\documentclass[fleqn,usenatbib]{mnras}

\usepackage{newtxtext,newtxmath}
\usepackage[T1]{fontenc}
\usepackage{ae,aecompl}

\usepackage{graphicx}	% Including figure files
\usepackage{amsmath}	% Advanced maths commands
\usepackage{amssymb}	% Extra maths symbols

\hypersetup{draft}

\newcommand{\beq}{\begin{equation}}
\newcommand{\eeq}{\end{equation}}
\newcommand{\kms}{\ensuremath{~\mathrm{km \, s^{-1}}}}
\newcommand{\hubble}{\ensuremath{~\mathrm{km \, s^{-1} \, Mpc^{-1}}}}
\newcommand{\Msun}{\ensuremath{M_{\odot}}}

\def\farcmin{\hbox{$.\mkern-4mu^\prime$}}

\def\df2{\mbox{[KKS2000]04}}

\title[The distance to the galaxy "lacking" dark matter]{A distance of 13 Mpc resolves the claimed anomalies of the galaxy lacking dark matter}

\author[I. Trujillo et al.]{Ignacio Trujillo,$^{1,2}$\thanks{E-mail: trujillo@iac.es}\thanks{All the authors of this work have contributed in a similar and very significant way. For this reason, with the exception of the first author (who has led and coordinated this global effort), the remaining list of authors has been ordered alphabetically.}
Michael A. Beasley,$^{1,2}$
Alejandro Borlaff,$^{1,2}$
Eleazar R. Carrasco,$^{3}$
\newauthor
Arianna Di Cintio,$^{1,2}$
Mercedes Filho,$^{4,5}$
Matteo Monelli,$^{1,2}$
Mireia Montes,$^{6}$
\newauthor
Javier Rom\'an,$^{1,2}$
Tom\'as Ruiz-Lara,$^{1,2}$
Jorge S\'anchez Almeida,$^{1,2}$
David Valls-Gabaud,$^{7}$
\newauthor
and Alexandre Vazdekis$^{1,2}$
\\
$^{1}$Instituto de Astrof{\'i}sica de Canarias (IAC), La Laguna, 38205, Spain\\
$^{2}$Departamento de Astrof{\'i}sica, Universidad de La Laguna (ULL), E-38200, La Laguna, Spain\\
$^{3}$Gemini Observatory/AURA, Southern Operations Center, Casilla 603, La Serena, Chile\\
$^{4}$Center for Astrophysics and Gravitation - CENTRA/SIM, Departamento de F\'\i sica, Instituto Superior T\'ecnico, Universidade de Lisboa,\\ Av. Rovisco Pais 1, P-1049-001 Lisbon, Portugal\\
$^{5}$Departamento de Engenharia F\'\i sica, Universidade do Porto, Faculdade de Engenharia, Universidade do Porto, Rua Dr. Roberto Frias, s/n,\\ P-4200-465, Oporto, Portugal\\
$^{6}$School of Physics, University of New South Wales, Sydney, NSW 2052, Australia\\
$^{7}$LERMA, CNRS, PSL, Observatoire de Paris, 61 Avenue de l'Observatoire, 75014 Paris, France\\
}

\date{Accepted XXX. Received YYY; in original form ZZZ}

\pubyear{2015}

\begin{document}
\label{firstpage}
\pagerange{\pageref{firstpage}--\pageref{lastpage}}
\maketitle

\begin{abstract}

The claimed detection of a diffuse galaxy lacking dark matter represents
a possible challenge to our understanding of the properties of these galaxies and galaxy formation in general. The galaxy, already identified in photographic plates taken in the summer of 1976 at the UK 48-in Schmidt telescope, presents normal distance-independent properties (e.g. colour, velocity dispersion of its globular clusters). However, distance-dependent quantities are at odds with those of other similar galaxies, namely the luminosity function and sizes of its globular clusters, mass-to-light ratio and dark matter content. Here we carry out a careful analysis of all extant data and show that they consistently indicate a much shorter distance (13 Mpc) than previously indicated (20 Mpc). With this revised distance, the galaxy appears to be a rather ordinary low surface brightness galaxy (R$_e$=1.4$\pm$0.1 kpc; M$_\star$=6.0$\pm$3.6$\times$10$^7$ M$_{\odot}$) with plenty of room for dark matter (the fraction of dark matter inside the half mass radius is $>$75\% and M$_{halo}$/M$_\star$$>$20) corresponding to a minimum halo mass $>$10$^9$ M$_{\odot}$. At 13 Mpc, the luminosity and structural properties of the globular clusters around the object are the same as those found in other galaxies.

\end{abstract}


%%%%%%%%%%%%%%%%%%%%%%%%%%%%%%%%%%%%%%%%%%%%%%%%%%

%%%%%%%%%%%%%%%%% BODY OF PAPER %%%%%%%%%%%%%%%%%%
\vspace{10 mm}

\section{Introduction} \label{sec:intro}

\citet{2018Natur.555..629V} claim the detection of a galaxy lacking dark matter "consistent with zero" dark matter content. If confirmed, this could be one of the most important discoveries in Extragalactic Astrophysics in decades. The galaxy (popularised as NGC1052-DF2\footnote{This galaxy was already catalogued before the publication by \citet{2018Natur.555..629V}. It has been identified with alternate names as KKSG04, PGC3097693 and [KKS2000]04 \citep[see e.g.][]{2000A&AS..145..45K}.  In this sense, we consider that the correct way of identifying the system is to use the name that follows the IAU rule (i.e. [KKS2000]04). This, in addition, is neutral regarding its possible association (or otherwise, as argued here) to NGC1052.}) is an extended (R$_e$=22.6 arcsec), very low central surface brightness ($\mu$(V$_{606}$,0)=24.4 mag arcsec$^{-2}$) object whose discovery can be traced back to at least 1978 \citep[see Plate 1 in][]{1978MNRAS.183..549F}\footnote{The plates were taken mostly during the summer of 1976.}. The mean velocity of the galaxy is 1793$\pm$2 km s$^{-1}$ \citep{2018arXiv181207345E}. Under the assumption that [KKS2000]04  is at the same distance ($\sim$20 Mpc) as the closest (in projection) massive galaxy in its vicinity (NGC1052; cz=1510 km s$^{-1}$), the effective radius of the object would be 2.2 kpc and the projected distance between the two galaxies only 84 kpc. With this effective radius and central surface brightness, the galaxy  would fall into the category of objects that are currently labeled as ultra-diffuse galaxies \citep[UDGs;][]{2015ApJ...798L..45V}. Assuming a distance of 20 Mpc, the estimated stellar mass of the object would be 2$\times$10$^8$ M$_{\odot}$ according to the mass-to-light ratio (M/L) derived from its colour V$_{606}$-I$_{814}$=0.37$\pm$0.05 (AB system). The system is also compatible with lacking neutral gas \citep[HI Parkes All-Sky Survey (HIPASS); 3$\sigma$ N$_{\rm HI}$$<$6$\times$10$^{18}$ cm$^{-2}$; ][]{2004MNRAS.350.1195M}. The total mass of the system has been obtained using the measured velocities of 10 compact sources (thought to be globular clusters) spatially located close to the galaxy. This group of compact objects has a very narrow velocity distribution, with a central peak at 1803$\pm$2 km s$^{-1}$ \citep[but see ][for a potential offset on this measurement]{2018arXiv181207345E} and a range of possible intrinsic velocity dispersions 8.8 km s$^{-1}$ $<$ $\sigma_{int}$ $<$ 10.5 km s$^{-1}$. Such a narrow dispersion would imply a dynamical mass of only 10$^8$ M$_{\odot}$ (i.e. fully compatible with the absence of dark matter in this system). 

In addition to the lack of dark matter, another intriguing result is that the luminosity of the compact sources (assuming that they are located at 20 Mpc) are much larger than those of typical globular clusters. In a follow up paper, \citet{2018ApJ...856L..30V} find the peak of the absolute magnitude distribution of the compact sources to be at M$_{V,606}$=-9.1 mag, significantly brighter than the canonical value for globular clusters of M$_{V}$=-7.5 mag \citep{2012Ap&SS.341..195R}. In this sense, [KKS2000]04  is doubly anomalous; if the compact sources are indeed globular clusters associated with the galaxy, it not only has an unexpected lack of dark matter (compatible with being a "baryonic galaxy"), but also a highly unusual population of globular clusters (some of them having absolute magnitudes similar to $\omega$ Centauri).

Both the absence of dark matter and the anomalous bright population of compact sources around [KKS2000]04 fully rely on the assumption that the galaxy is at a distance of 20 Mpc. In fact, if the galaxy were located much closer to us, for instance a factor of two closer (as motivated by the apparent magnitudes of the compact sources around the galaxy) then its stellar mass would go down significantly \citep[a fact already mentioned by ][]{2018Natur.555..629V}. A closer distance would make the properties of [KKS2000]04  perfectly ordinary. The question, then, is how accurate the adopted galaxy distance is? \citet{2018Natur.555..629V} use the surface brightness fluctuation (SBF) technique to infer a distance of 19.0$\pm$1.7 Mpc to the galaxy. In addition, the heliocentric velocity of the system (c$z$=1803$\pm$2 km s$^{-1}$) is also used as another argument to favor a large distance for this object. However, the validity of the SBF technique in the case of [KKS2000]04  should be taken with caution, since \citet{2018Natur.555..629V} extended the \citet{2010ApJ...724..657B} calibration to a range (in colour) where its applicability is not tested. Moreover, the use of heliocentric velocities to support a given distance in the nearby Universe should be done with care, as large departures from the Hubble flow are measured in our local Universe. For all these reasons, through a systematic analysis of different and independent distance indicators, we readdress the issue of the distance to [KKS2000]04. We  show that a distance of 13 Mpc not only resolves all the anomalies of the system but also is favoured by the colour - magnitude diagram of the system, the comparison of the luminosity function  of its stars with galaxies with similar properties, the apparent luminosity and size of its globular clusters and a revision of the SBF distance.

The structure of this paper is as follows. In Section \ref{sec:data} we describe the data used. Section \ref{sec:sed} shows the spectral energy distribution of the galaxy from the \textsl{FUV} to IR and the stellar populations properties derived from its analysis. In Section \ref{sec:distance} we discuss up to five different redshift independent distance estimations of the galaxy [KKS2000]04. We explore the velocity field of the galaxies around the system in Sec. \ref{sec:velfield} and the possibility that the galaxy is associated with other objects in Sec. \ref{sec:group}. Section \ref{sec:mass} readdresses the estimation of the total mass of the system. Finally, in Section \ref{sec:discussion} we put \df2 in context with other Local Group galaxies and give our conclusions in Section \ref{sec:conclusions}. We assume the following cosmological model: $\Omega_m$=0.3, $\Omega_\Lambda$=0.7 and H$_0$=70 km s$^{-1}$ Mpc$^{-1}$ (when other values are used this is indicated). All the magnitudes are given in the AB system unless otherwise explicitly stated.

\section{Data} \label{sec:data}

Given the potential relevance of the galaxy if the lack of dark matter is confirmed, we have made a compilation of all the public (SDSS, DECaLS, GALEX, WISE, Gemini, HST) data currently available for the object  with the aim of exploring its stellar population properties, its structure, and the properties of the compact sources around the system.

\subsection{SDSS, DECaLS, GALEX and WISE}

\textsl{SDSS u, g, r, i and z} band imaging data were retrieved from the DR14 SDSS \citep{2018ApJS..235...42A} Sky Server. The magnitude zero-point for all the data set is the same: 22.5 mag. The exposure time of the images is 53.9s and the pixel size 0.396 arcsec. Deeper optical data in the \textsl{g, r, and z} bands were obtained from the Dark Energy Camera Legacy Survey (DECaLS) archive\footnote{\url{http://legacysurvey.org/decamls/}}. The DECaLS survey is obtained with the Dark Energy Camera (DECam) on the Blanco 4m telescope. The zero-point of the images is 22.5 mag and the pixel scale 0.263 arcsec. At the position of the galaxy, the exposure time of the images is 450s (\textsl{r}) and 540s (\textsl{g and z}). We have used the DECaLS DR7 data for brick 0405m085.

Galaxy Evolution Explorer (GALEX) \textsl{FUV} and \textsl{NUV} data \citep{2005ApJ...619L...1M} were obtained from GALEX Mikulski Archive for Space Telescopes (MAST) archive\footnote{\url{http://galex.stsci.edu/GR6/?page=tilelist\&survey=allsurveys}}. The exposure times in each band are 2949.55s (\textsl{FUV}) and 3775.7s (\textsl{NUV}). The pixel size is 1.5 arcsec and the zero-points  18.82 mag (\textsl{FUV}) and 20.08 mag (\textsl{NUV}). Finally, Wide-field Infrared Survey Explorer \citep[WISE; ][]{2010AJ....140.1868W} data was downloaded from the WISE archive in IRSA\footnote{\url{http://wise2.ipac.caltech.edu/docs/release/allsky/}}. The total exposure time is 3118.5s. The WISE pixel scale is 1.375 arcsec and we use the following two channels \textsl{W1} (3.4 $\mu$m) and \textsl{W2} (4.6 $\mu$m) whose zeropoints are 23.183 and 22.819 mag, respectively.

\subsection{Gemini}

Very deep (3000s in \textsl{g and i} band) and good quality seeing (0.75 arcsec in \textsl{g} and 0.71 arcsec in \textsl{i}) data using the instrument GMOS-N \citep{2004PASP..116..425H} were obtained with the Gemini North telescope (program ID: GN-2016B-DD-3, PI: van Dokkum). Unfortunately, only the \textsl{g-band} data was useful for further analysis as the \textsl{i}-band was taken during non-photometric conditions with clouds passing during the observation. This affected the depth of the data as well as the photometry of the image. For this reason, we decided to use only the \textsl{g}-band in this work. The data was downloaded from the Gemini Observatory\footnote{\url{https://archive.gemini.edu/searchform}} archive and reduced using a minimal aggressive sky subtraction with the aim of keeping the low surface brightness features of the image. The images were processed with the reduction package \texttt{THELI} (Schirmer 2013). All images were bias, over-scan subtracted and flatfielded. Flatfields were constructed from twilight flats obtained during the evening and the morning. The reduced images present gradients across the entire GMOS-N FoV, especially in \textsl{i}-band. To remove the gradients, we modeled the background using a two pass model, following the \texttt{THELI} recipe. A superflat was constructed as follows: in the first pass, a median-combined image is created without object detection, to remove the bulk of the background signal. In the second pass, \texttt{SExtractor} \citep{1996A&AS..117..393B} is used to detect and mask all objects with a detection  threshold of 1.3 sigma above the sky and with a minimum area of 5 pixels. Because we did not want to oversubtract the outer part of the galaxy, a mask expansion factor of 5--6 was used to enlarge the isophotal ellipses. The resulting images, after the background modeling subtraction, are flat within 0.5\% or less. All images were registered to common sky coordinates and pixel positions using the software \texttt{SCAMP} \citep{2006ASPC..351..112B}. The astrometric solution was derived using the SDSS DR9 as a reference catalogue. The internal (GMOS) astrometric residual from the solution is ~0.05\arcsec and the external astrometric residual is ~0.2\arcsec. Before the co-addition, the sky of each single exposure was subtracted using constant values. Finally, the images were re-sampled to a common position using the astrometric solution obtained with \texttt{SCAMP} and then stacked using the program \texttt{SWARP} \citep{2002ASPC..281..228B}. The final co-added images were normalised to 1 sec exposure. The pixel size of the data is 0.1458 arcsec and the zero-point used 26.990 mag (\textsl{g}-band).

\subsection{Hubble Space Telescope}

Charge-transfer efficiency (CTE) corrected data were obtained from the MAST  archive webpage\footnote{\url{https://mast.stsci.edu/portal/Mashup/Clients/Mast/Portal.html}}. The data was taken as part of the programme GO-14644  (PI van Dokkum) and consist on two Advanced Camera for Survey (ACS) orbits: one  in \textsl{F606W} (\textsl{V$_{606}$}; 2180s) and one in \textsl{F814W} (\textsl{I$_{814}$}; 2320s). Each orbit is composed of 4 images whose individual exposures were of 545s (\textsl{F606W}) and 580s (\textsl{F814W}). The pixel scale is 0.05\arcsec and the AB magnitude zero-points used are 26.498 (\textsl{V$_{606}$}) and 25.944 (\textsl{I$_{814}$})\footnote{\url{http://www.stsci.edu/hst/acs/analysis/zeropoints}}.

To build final drizzled mosaics we used the flc files. Following the Drizzlepac and ACS documentation, we applied the Lanzcos3 kernel for the drizzling and the "iminmed" combine algorithm (which is recommended for a low number of images). This corrects satellite trails and cosmic rays. The pixfrac parameter was set to 1 (the whole pixel). The sky correction was set to manual. We masked foreground and background sources using  \texttt{NoiseChisel} \citep{2015ApJS..220....1A}. For masking, we set \texttt{NoiseChisel} with a tile size of 60x60 pix, which allows for a higher S/N of the low surface brightness features in the masking process. To mask the [KKS2000]04 galaxy, we used a wide manual (radial size larger than 50\arcsec) mask over the [KKS2000]04 galaxy. To estimate the sky, we applied the remaining (non-masked) pixels. The centroid of the sky distribution was calculated using bootstrapping (using Bootmedian, a code made publicly available at \url{https://github.com/Borlaff/bootmedian}).  After performing the reduction, we have been able to decrease the strong wavy pattern that appears in the publicly available reduction of the [KKS2000]04 dataset. Finally, to remove large scale gradients on the sky background we used again \texttt{NoiseChisel}. In this case, the underlying shape of the sky was model using large tile sizes (200x200 pixels).

\begin{figure*}

\includegraphics[width=\textwidth]{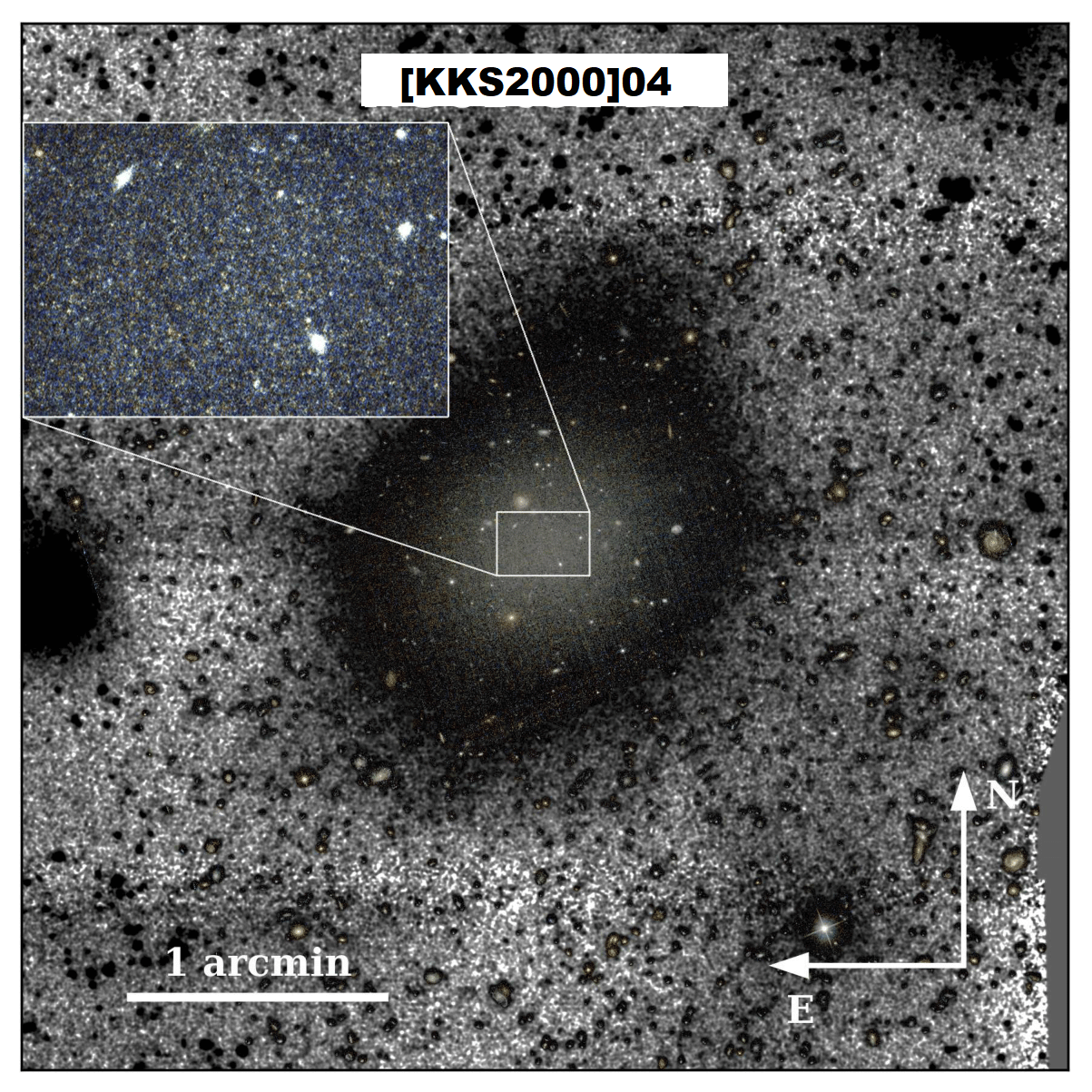}
  \caption{Colour composite image of [KKS2000]04 combining \textsl{F606W} and \textsl{F814W} filters with black and white background using \textsl{g}-band very deep imaging from Gemini. The ultra-deep \textsl{g}-band Gemini data reveals a significant brightening of the galaxy in the northern region. An inset with a zoom into the inner region of the galaxy is shown. The zoom shows, with clarity, the presence of spatially resolved stars in the HST image.}
\label{fig:hstgemini}
\end{figure*}

All the photometric data has been corrected by foreground Galactic extinction. We use (see Table \ref{table:sed}) the values provided by the NED web-page at the spatial coordinates of the galaxy. A colour composite figure showing the HST data combined with the ultra-deep Gemini \textsl{g}-band data is shown in Fig.~\ref{fig:hstgemini}.

\section{The spectral energy distribution}
 \label{sec:sed}

\begin{figure*}

\includegraphics[width=\textwidth]{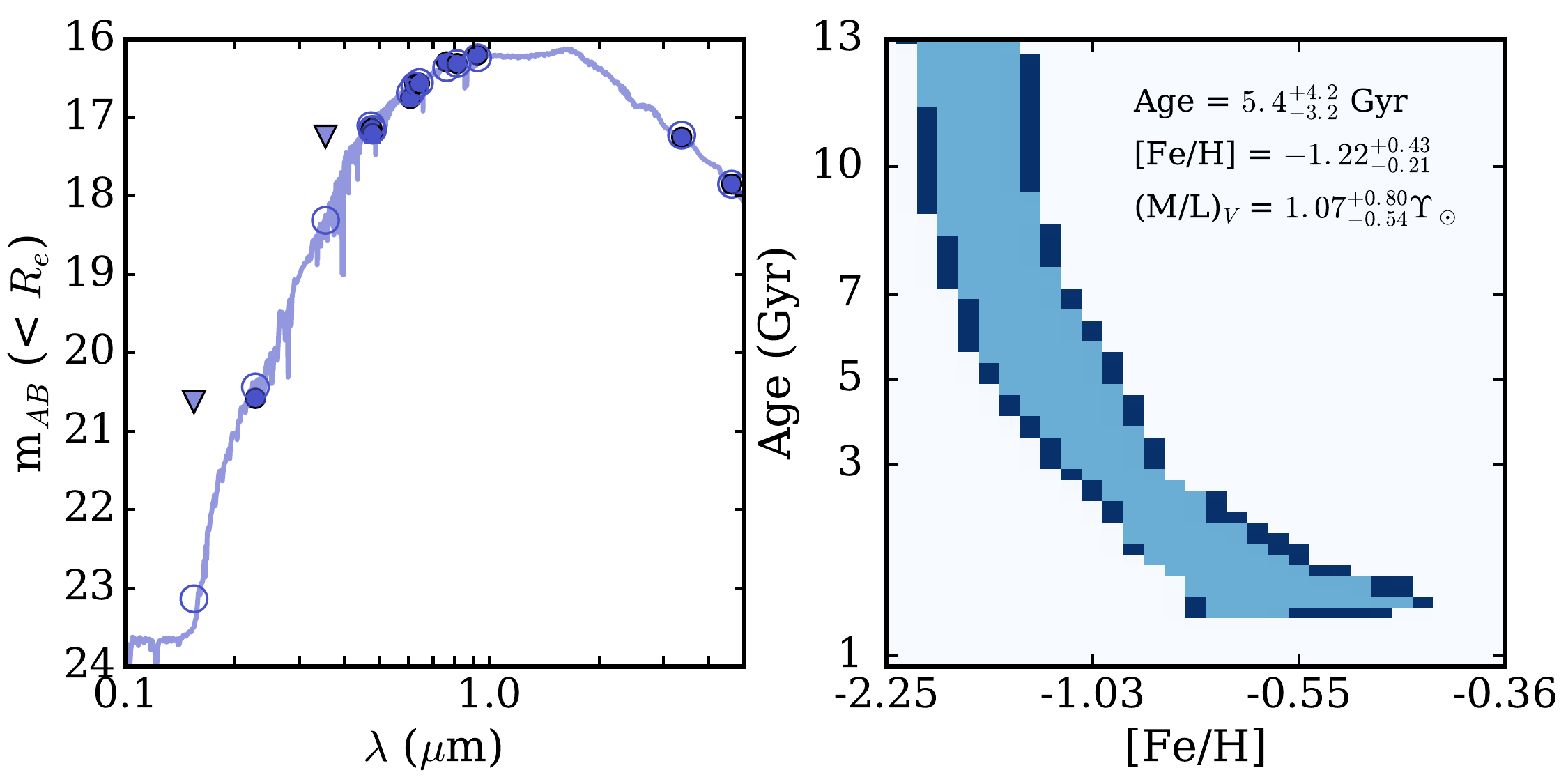}
  \caption{Left panel: Spectral Energy Distribution of the galaxy [KKS2000]04. The figure shows the magnitude of the object within a radial aperture of 1 R$_e$ (22.6\arcsec). Inverted triangles correspond to upper limits in those bands  were the signal was not sufficient to get a proper estimate of the flux (\textsl{FUV and u}). Blue circles correspond to the flux measured in the rest of the bands. Open circles are the expected flux after convolving the best fit model with the filter responses. The best fit model is a \citet{2003MNRAS.344.1000B} model with 5.4$^{+4.2}_{-3.2}$ Gyr, [Fe/H]=-1.22$^{+0.43}_{-0.21}$ and Chabrier IMF (solid line overplotted). Right panel: Age and metallicity plane. The contours correspond to fits compatible with the observations at 68\% (light blue) and 95\% (dark blue) confidence level.}
\label{fig:sed}
\end{figure*}

Lacking a spectroscopic analysis of the galaxy, our best way to  constrain the age, metallicity and stellar mass to light ratio of the stellar population of the object is by analysing its spectral energy distribution (SED) from broad band photometry. We measure the SED of the system from FUV to IR (Fig. \ref{fig:sed}). The photometry (after correcting for Galactic reddening) was derived in a common circular aperture of R=1R$_e$=22.6\arcsec (i.e. containing half of the total brightness of the object) as indicated on the vertical axis.   We use such radial aperture to guarantee  that we have enough signal to produce a reliable characterisation of the SED in all the photometric bands. The images in each filter were masked to avoid the contamination from both foreground and background objects. For those filters with poorer spatial resolution, we use the information provided by the HST data to see whether the addition of extra masking was needed. 

The inverted triangles in the left panel of Fig. \ref{fig:sed} correspond to the magnitude detection limits in those images were the galaxy was not detected (\textsl{FUV} and \textsl{u}). These upper limits were estimated as the 3$\sigma$ fluctuations of the sky (free of contaminating sources) in circular apertures of radius 1 R$_e$.

In order to characterise the stellar population properties of the galaxy we have fitted its SED using \citet{2003MNRAS.344.1000B} single stellar population (instantaneous burst) models. We used a Chabrier Initial Mass Function \citep[IMF;][]{2003PASP..115..763C}. For the fitting we use $\chi^2$ minimisation approach \citep[see][for the details of the fitting procedure]{2014MNRAS.439..990M}. From this fitting we derive a most likely age of 5.4$^{+4.2}_{-3.2}$ Gyr, a metallicity of [Fe/H]=-1.22$^{+0.43}_{-0.21}$ and a mass-to-light ratio in the \textsl{V}-band (M/L)$_V$=1.07$^{+0.80}_{-0.54}$ $\Upsilon$$_\odot$ (see right panel in Fig. \ref{fig:sed}). The uncertainties on the above parameters have been estimated by marginalising the 1D probability distribution functions.  The metallicity we obtain is  similar to the average metallicity found by \citet{2018ApJ...856L..30V} for the GCs surrounding this object. Very recently, using MUSE spectroscopy, \citet{2018arXiv181207346F} has found the following age and metallicity for the galaxy: 8.9$\pm$1.5 Gyr and [Fe/H]=-1.07$\pm$0.12. Our results are compatible with such spectroscopic determination within the error bars. In what follows we assume that these properties obtained within 1 R$_e$ are representative of the whole galaxy. In Table \ref{table:sed}, we provide the total magnitudes of the galaxy in the different filters. The total magnitudes were obtained from the 1 R$_e$ aperture magnitudes shown in Fig. \ref{fig:sed} and subtracting 2.5$\times$$\log$(2) to that magnitudes to account for the flux beyond 1 R$_e$. We also include an extra correction on the total magnitudes by adding back the flux of the galaxy that is under the masks used to avoid the contamination by both foregrounds and backgrounds objects. The correction is done assuming that the flux that is masked is not in a particular place of the galaxy but randomly distributed over the image (which is in fact the case). Under such assumption the correction is estimated by calculating which area of the galaxy is masked and multiplying the observed flux by the amount of area that is not observed. The values of the parameters provided above corresponds  to the ones obtained with this correction. The correction of the flux under the masks is especially relevant on the low spatial resolution imaging as the ones provided by Spitzer. We have fitted the SEDs using both the observed (masked) magnitudes and the magnitudes applying the flux correction below the mask. The age, metallicity and mass-to-light correction obtained in both cases are compatibles within the uncertainties.

\begin{table*}
\centering
\caption{Half m$_{AB}$($<$R$_e$), total m$_{AB}$ and total m$_{AB}$ mask corrected magnitudes (all corrected of foreground Galactic extinction) of [KKS2000]04 in different bands. The applied foreground extinction is indicated in the last column.}
\begin{tabular}{cccccc} %\toprule
\hline
\label{table:sed}
Filter & $\lambda_{eff}$ &  m$_{AB}$($<$R$_e$)  & m$_{AB}$ & m$_{AB,mc}$ & A$_\lambda$  \\
       &        \AA         & (mag)  & (mag)  &  (mag) \\

\hline
\textsl{FUV}       & 1542.3 &  $<$20.63  & $<$19.88          & $<$19.88  & 0.171 \\ 
\textsl{NUV}       & 2274.4 & 20.77 $\pm$  0.05& 20.02 $\pm$  0.05  & 19.83$\pm$0.04  & 0.192 \\
\textsl{SDSS u}    & 3543   & $<$ 17.24 & $<$ 16.49           & $<$ 16.49 & 0.104 \\
\textsl{DECAM g}   & 4720   & 17.29 $\pm$ 0.04  & 16.53 $\pm$ 0.04  &  16.41 $\pm$ 0.04     & 0.081 \\
\textsl{Gemini g}  & 4750   & 17.28 $\pm$  0.012 & 16.52 $\pm$  0.012 & 16.39 $\pm$  0.011  & 0.081 \\
\textsl{SDSS g}    & 4770   & 17.36 $\pm$  0.10 & 16.60 $\pm$  0.10   & 16.45 $\pm$  0.09 & 0.081 \\
\textsl{ACS F606W} & 6060   & 16.87 $\pm$  0.013 & 16.12 $\pm$  0.013 & 16.00 $\pm$  0.013 & 0.060 \\
\textsl{SDSS r}    & 6231   & 16.73 $\pm$  0.08 & 15.97 $\pm$  0.08   & 15.82 $\pm$  0.07  & 0.056 \\
\textsl{DECAM r}   & 6415   & 16.69 $\pm$  0.03 & 15.94 $\pm$  0.03   & 15.81 $\pm$  0.03 & 0.056 \\
\textsl{SDSS i}    & 7625   & 16.44 $\pm$  0.07 & 15.69 $\pm$  0.07  & 15.54 $\pm$  0.06  & 0.042 \\
\textsl{ACS F814W} & 8140   & 16.42 $\pm$ 0.014 & 15.67 $\pm$ 0.014  & 15.55 $\pm$ 0.012 & 0.037 \\
\textsl{SDSS z}    & 9134   & $<$15.38          & $<$ 14.63          & $<$ 14.63  & 0.031 \\
\textsl{DECAM z}   & 9260   & 16.32 $\pm$  0.03 & 15.57 $\pm$  0.03  & 15.45 $\pm$  0.03 & 0.031 \\
\textsl{WISE W1}   & 33680  & 17.61 $\pm$ 0.07  & 16.86 $\pm$ 0.07   & 16.50 $\pm$ 0.07 & 0.004 \\
\textsl{WISE W2}   & 46180  & 18.21 $\pm$ 0.12  & 17.46 $\pm$ 0.12   & 17.09 $\pm$ 0.12 & 0.003 \\
\hline
\end{tabular}
\end{table*}

\section{The distance to [KKS2000]04}
 \label{sec:distance}

Up to five different  redshift-independent distance measurements converge to a distance of 13 Mpc for the galaxy [KKS2000]04. In the following we will describe each one of them.

\subsection{The colour - magnitude diagram distance}

When available, the analysis of the colour-magnitude diagram (CMD) of resolved stars is one of the most powerful techniques to infer a redshift-independent distance to a galaxy. The data we have used to create the CMD of [KKS2000]04 is the HST optical data described above. Data treatment for creating the CMD was carried out following essentially the prescriptions of \citet{2010ApJ...720.1225M}. Photometry was performed on the individual $flc$ images using the \texttt{DAOPHOT/ALLFRAME} suite of programs \citep{stetson87,stetson94}. Briefly, the code performs a simultaneous data reduction of the images of a given field, assuming individual Point Spread Functions (PSF) and providing an input list of stellar objects. The star list was generated on the stacked median image obtained by registering and co-adding the eight individual available frames, iterating the source detection twice. To optimise the reduction, only the regions around the centre of [KKS2000]04, within 5 R$_e$ (113\arcsec), were considered. We explore different apertures and the main result did not change. The photometry was calibrated to the AB system. The photometric catalogue was cleaned using the {\itshape sharpness} parameter provided by \texttt{DAOPHOT}, rejecting sources with $|sha| >$ 0.5.
	
\begin{figure}
  \includegraphics[width=\columnwidth]{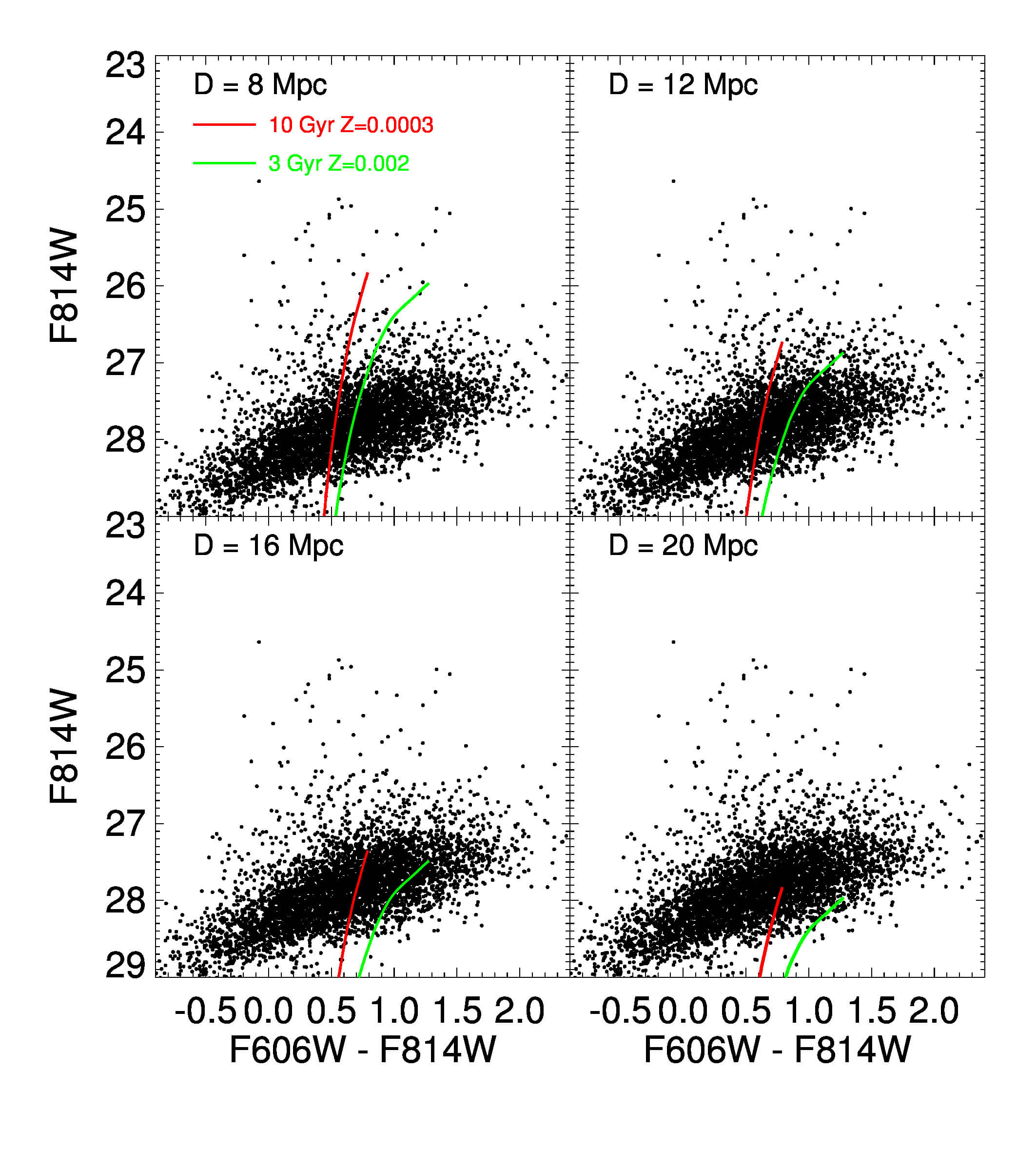}
  \caption{CMD in AB mag of the 5150 point sources measured in the central regions of [KKS2000]04 (R$<$1.5 R$_e$). The four panels show the comparison with theoretical isochrones from the BaSTI database assuming different distances, from 8 to 20 Mpc in steps of 4 Mpc.}
     \label{fig:cmd}
  \end{figure}

Fig. \ref{fig:cmd} presents the obtained (\textsl{F606W-F814W, F814W}) colour-magnitude diagram. The four panels show a comparison with selected isochrones from the BaSTI database\footnote{\url{http://basti-iac.oa-abruzzo.inaf.it/}}  spanning a wide range of ages and metallicities. Isochrones were moved according to different assumptions of the distance, from 8 Mpc (top left) to 20 Mpc (bottom right). The vast majority of  detected sources are compatible with being bright Red Giant Branch (RGB) stars, though a small contamination from asymptotic giant branch stars cannot be excluded with the present data. There is no strong evidence of bright main sequence stars, corresponding to a population younger than few hundred million years. The comparison with isochrones seems to exclude distances as close as 8 Mpc or as distant as 20 Mpc. In fact, a qualitative comparison seems to favour an intermediate distance.

\begin{figure}
  \includegraphics[width=\columnwidth]{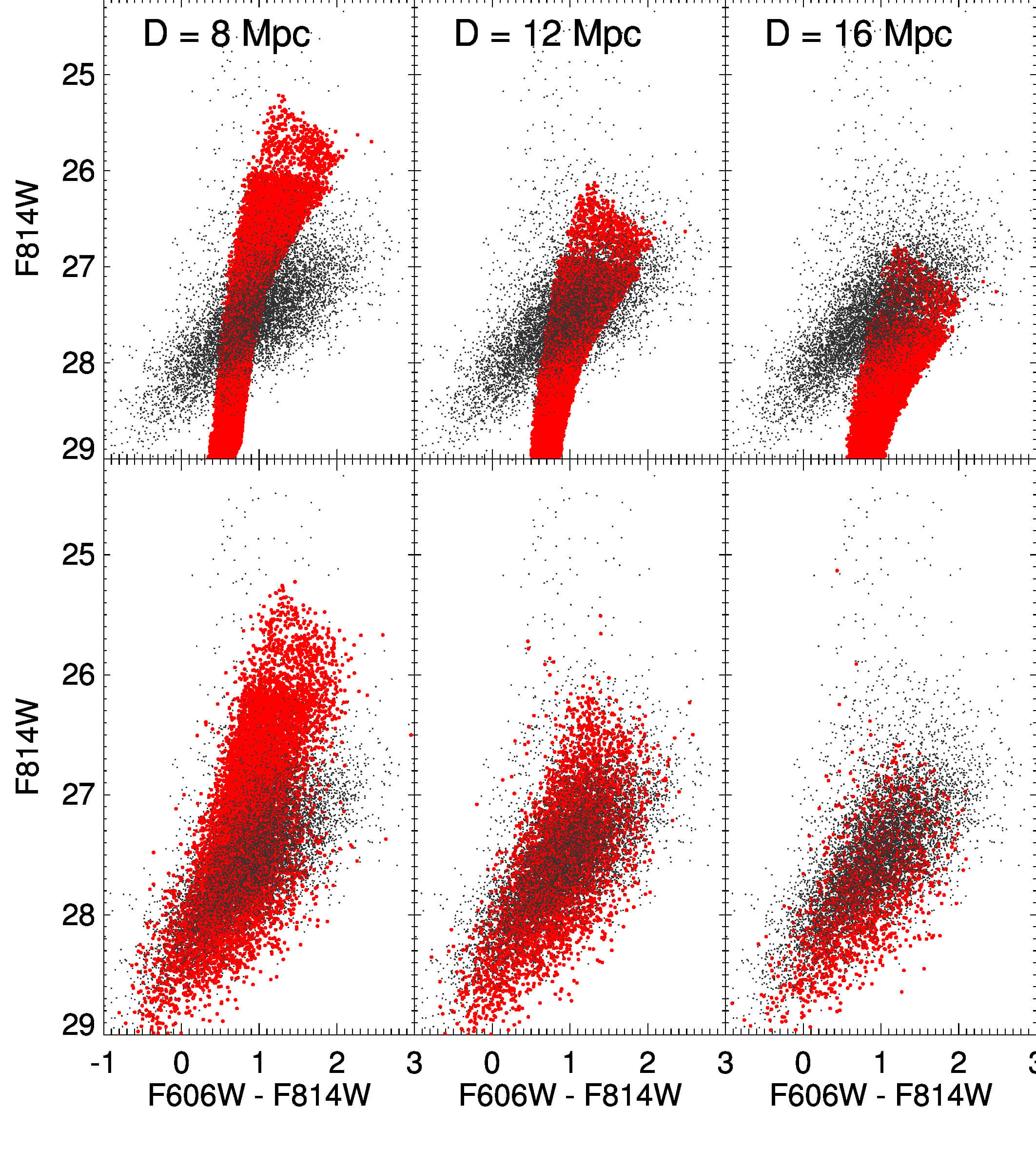}
  \caption{Top panels: CMD of [KKS2000]04 (grey points) compared to the CMD of a synthetic population located at 8, 12, and 16 Mpc (red points, top panels from left to right). Bottom panels: The three synthetic CMDs were used to perform artificial stars tests, and the recovered CMDs are shown in the bottom panels. Magnitudes are shown in the AB system. See text for details.}
     \label{fig:crowd}
  \end{figure}

To make a quantitative analysis we have conducted specific artificial star tests. In particular, we created a mock population of 50,000 stars covering a range of ages between 10 and 13.5 Gyr and metallicity between Z=0.001 and Z=0.008. The top panels of Fig. \ref{fig:crowd} compares the CMD of this mock population (red points) with that of [KKS2000]04 (dark grey), for three different assumptions of the distance: 8 Mpc (left), 12 Mpc (center) and 16 Mpc (right). Synthetic stars were injected in individual images and the photometric process was repeated for the three cases. The bottom panels of Fig. \ref{fig:crowd} present the superposition of the [KKS2000]04  CMD and the recovered mock stars (red points). Clearly, the distribution of the recovered stars in the 8 Mpc and the 16 Mpc cases is not compatible with the observed CMD of [KKS2000]04. In fact, in the case of the 8 Mpc distance the brightest portion of the RGB would be clearly detected between \textsl{F814W}= 26 mag and \textsl{F814W}= 27 mag. On the other hand, most of the injected stars assuming a 16 Mpc distance have been lost in the photometric process, resulting in a very sparsely recovered CMD.

A more precise estimate of the distance to [KKS2000]04 can be derived using the tip of the red giant branch (TRGB). This is a well-established distance indicator for resolved stellar populations \citep{1993ApJ...417..553L}. From the stellar evolutionary point of view, the TRGB corresponds to the end of the red giant branch phase, when the helium core reaches sufficient mass to explosively ignite He in the centre. Observationally, this corresponds to a  well-defined discontinuity in the luminosity function, which can be easily identified from photometric data. This method presents two strong advantages. First, the luminosity of the TRGB in the \textsl{I/F814W} filters has a very mild dependence on the age and the metallicity over a broad range \citep{1997MNRAS.289..406S}. Second, since the TRGB is an intrinsically bright feature (M$_{I}$$\sim$ -4), this method can be reliably used  beyond 10 Mpc.

\begin{figure}
  \includegraphics[width=\columnwidth]{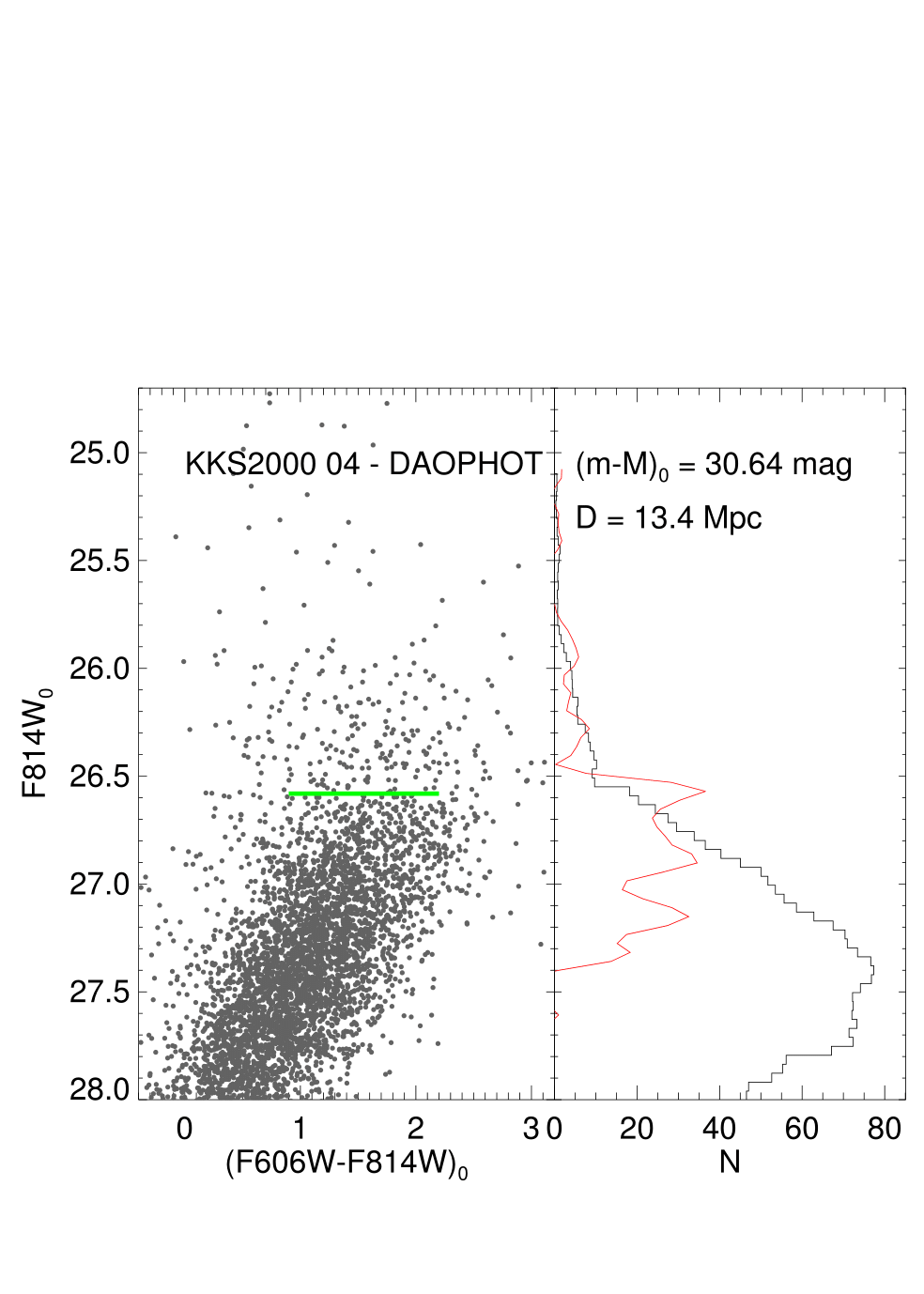}
  \caption{{\em Left panel-} De-reddened CMD of [KKS2000]04 stars. The green line shows the  estimated position of the TRGB. {\em Right panel- }Luminosity function (black line), and the response filter (red curve). Magnitudes are shown in the Vega system.}
     \label{fig:tip}
  \end{figure}

Fig. \ref{fig:tip} shows the TRGB distance estimate in the case of [KKS2000]04 based on the current data. We adopted the calibration by \citet{2007ApJ...661..815R}

\begin{equation}
M_{F814W}^{ACS} = -4.06+ 0.20\times[(F606W-F814W) - 1.23] 
\label{eq:trgb}
\end{equation}

which accounts for the mild dependency on the metallicity by taking into account a colour term. Following \citet{mcquinn17a}, we applied a colour correction to the \textsl{F814W} photometry, as it results in a steeper RGB and therefore a better defined TRGB. The resulting modified and de-reddened CMD (corrected after taking into account the different zero points between the AB and the Vega mag systems) is shown in the left panel of Fig. \ref{fig:tip}. The right panel presents the luminosity function (in black) and the filter response after convolving with a Sobel kernel \citep{1996ApJ...461..713S,1997ApJ...478...49S} K=[-2,-1,0,1,2] (red). The filter presents a well defined peak which we fitted with a Gaussian curve, obtaining \textsl{F814W$_{0, Vega}$} = 26.58$\pm$0.08 mag, marked by the green line on the CMD.  The distance modulus was derived with the \citet{2007ApJ...661..815R} zero points, obtaining $(m-M)_0$ = 30.64 $\pm$ 0.08 mag, corresponding to a distance D = 13.4 $\pm$ 1.14 Mpc. The error budget includes the uncertainty of the calibration relation by \citet{2007ApJ...661..815R} and the error on the determination of the TRGB position. 

Interestingly, using the different photometric code \texttt{DOLPHOT} \citep{2000PASP..112.1383D}, \citet{2018ApJ...868...96C} confirm the presence of the TRGB of the galaxy at a very similar location as we have found, finding a distance of 12.6 Mpc. However, they discard such measurement arguing that the TRGB feature found in their CMD is produced by the presence of blends of stars \citep{2018ApJ...864L..18V}.  A straightforward to test whether blends are an issue or not is to remeasure the distance to [KKS2000]04 avoiding the central region of the object. If blends were causing an artificial TRGB, the derived TRGB using the outer (significantly less crowded) part of the galaxy would produce a larger distance. We, consequently, have estimated the location of the TRGB using only stars within 0.8R$_e$ and outside this radial distance. We find the same distance modulus: 30.64$\pm$0.08 mag (R$<$0.8R$_e$) and 30.65$\pm$0.08 mag (R$<$0.8R$_e$). We have chosen 0.8R$_e$ to have a similar number of stars in both radial bins. We conclude, consequently, that the claim that the location of TRGB of this galaxy is affected/confused by the presence of groups of stars is ruled out by this simple test, and the resulting TRGB distance is 13.4$\pm$1.14 Mpc.

\subsection{The comparison with the nearby analogous galaxy DDO44}

A robust differential measure of the distance to [KKS2000]04 is also possible by comparing its colour - magnitude diagram with that of a galaxy with similar characteristics to our object, but which has a well-established distance. The idea is the following. The CMD of the reference galaxy is placed at different distances and its photometry degraded accordingly in order to satisfy both the photometric errors and completeness of the CMD of [KKS2000]04. Then, a quantitative comparison between both the CMD of the reference galaxy and the CMD of our target can indicate which distance is the most favoured.

\begin{figure*}
  \includegraphics[width=\textwidth]{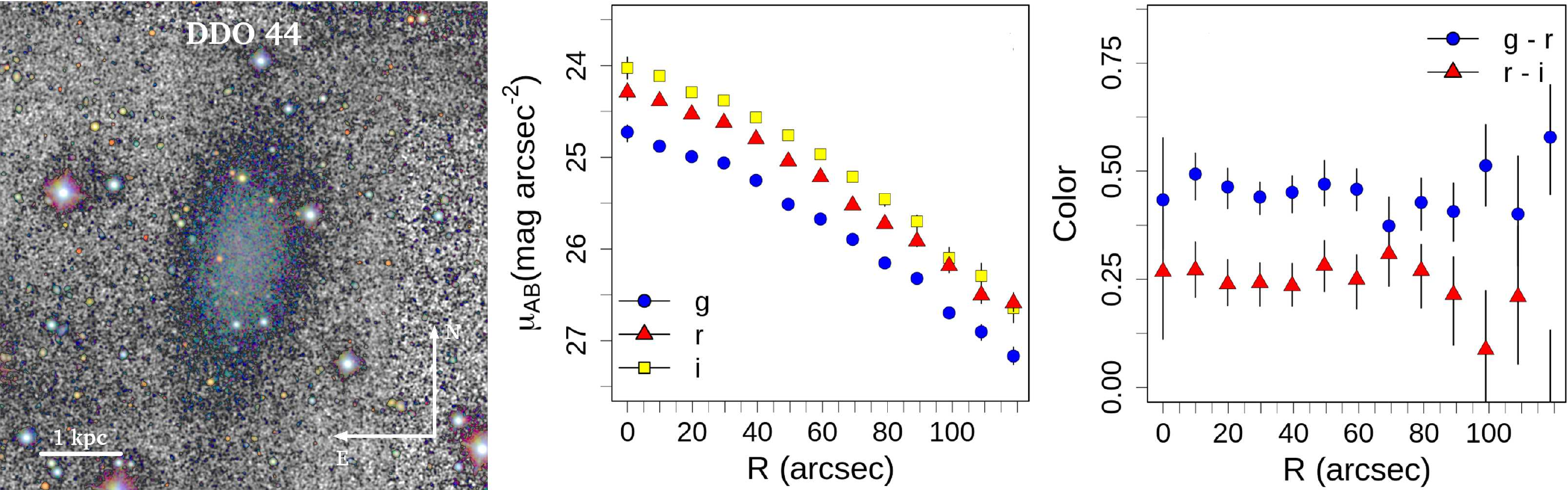}
  \caption{DDO44 SDSS colour image and its surface brightness and radial colour profiles. The left panel shows a colour composite image based on the SDSS \textsl{g}, \textsl{r}  and \textsl{i} filters. The galaxy, located at 3.1 Mpc, has an extended and smooth appearance typical of  UDG galaxies. The middle panel presents the surface brightness profiles of DDO44 in the SDSS \textsl{g}, \textsl{r}  and \textsl{i} filters . The right panel shows the \textsl{g-r} and \textsl{r-i} radial colour profiles. }
     \label{fig:ddo44}
  \end{figure*}

The galaxy we have chosen to compare with [KKS2000]04 is the dSph DDO44 \citep[R.A.: 07:34:11.3, Dec: +66:53:10;][]{1959PDDO....2..147V, 1985A&AS...60..213K}. This galaxy is a member of the M81 group and is located at a distance of 3.10$\pm$0.05 Mpc \citep{2009ApJS..183...67D}. Based on its stellar and structural properties, DDO44  is almost a perfect analogy to [KKS2000]04. In Fig. \ref{fig:ddo44} we show a colour image of the galaxy based on the combination of the SDSS filters \textsl{g}, \textsl{r}  and \textsl{i}.  We have used the SDSS data of DDO44 to explore the structural properties of the galaxy. This allows for a direct comparison with the structural characteristics of [KKS2000]04. The surface brightness profiles in the  \textsl{g}, \textsl{r}  and \textsl{i} filters are shown in Fig. \ref{fig:ddo44}. The central surface brightness of DDO44 is 24.3$\pm$0.1 mag/arcsec$^2$ (\textsl{r}-band; AB system). This value is almost identical to the central surface brightness of [KKS2000]04 (i.e. $\mu$(V$_{606}$,0)=24.4 mag arcsec$^{-2}$; see Table \ref{table:sed} for finding the colour difference between \textsl{r} and \textsl{V$_{606}$}). The structural parameters of DDO44 were derived using \texttt{IMFIT} \citep{2015ApJ...799..226E} assuming a S\'ersic model for the light distribution of the galaxy. We obtained the following values R$_e$=68$\pm$1 arcsec, axis ratio b/a=0.52$\pm$0.02 and S\'ersic index $n$=0.64$\pm$0.02. The S\'ersic index of DDO44 is very similar to that obtained for [KKS2000]04 \citep[i.e. $n$=0.6;][]{2018Natur.555..629V}. At the distance of DDO44, the effective radius measured is equivalent to 1.00$\pm$0.02 kpc. Our structural parameters are also in agreement with those reported in the literature \citep{1999A&A...352..399K}. The apparent total magnitude of DDO44 in the \textsl{r}-band is 14.3 mag, which implies M$_r$=-13.2 mag and a total stellar mass of $\sim$2$\times$10$^7$ M$_{\odot}$ according to the age and metallicity of this galaxy (see below).

DDO44 has \textsl{g-r} and \textsl{r-i} colours which are also similar to those measured for [KKS2000]04  (see Fig. \ref{fig:ddo44}): \textsl{g-r}=0.48$\pm$0.05 (vs 0.6$\pm$0.1 for [KKS2000]04) and \textsl{r-i}=0.26$\pm$0.05 (vs 0.29$\pm$0.08 for [KKS2000]04). The age of the DDO44 stellar population is mainly old ($>$8 Gyr) with only 20\% of its stars being compatible with intermediate ages (2-8 Gyr). Its metallicity is [Fe/H]=-1.54$\pm$0.14 \citep{2006PASP..118..580A}. These values are comparable to the ones found in \citet{2018arXiv181207346F} for [KKS2000]04: 8.9$\pm$1.5 Gyr and [M/H]=-1.1$\pm$0.1. Moreover, its stellar population is found to be very similar along the radial structure of the object. 

The fact that stellar population properties and stellar density of DDO44 is so close to  [KKS2000]04 makes this galaxy a perfect target to explore the effect of the distance in its CMD diagram. Luckily,  DDO44 was also observed by HST (Program GO 8192; PI: Seitzer) using the same set of filters as the one we have used here for the CMD of  [KKS2000]04 (i.e. \textsl{F606W} and \textsl{F814W}). The exposure time on each filter was 600s. The camera, however, was not the ACS but the WFPC2. DDO44 was centred on the WF3 chip covering a region of 80$\times$80 arcsec. Considering the effective radius of the galaxy, the region of DDO44 that was probed corresponds to $\sim$0.8 R$_e$. In order to make a sensible comparison between the photometry used in [KKS2000]04 (ACS) and DDO44 (WFPC2), we have explored whether using the same filters but different cameras  introduces any bias on the photometry. We find the differences to be negligible:  \textsl{F606W$_{ACS}$}-\textsl{F606W$_{WFPC2}$}=-0.015 mag and \textsl{F814W$_{ACS}$}-\textsl{F814W$_{WFPC2}$}=0.01 mag \citep[for an age of 8 Gyr and metallicity of Fe/H=-1.3; based on ][]{2016MNRAS.463.3409V}. As the expected difference is so small, we did not correct for this effect in what follows.  

\begin{figure*}
 \includegraphics[width=0.95\textwidth]{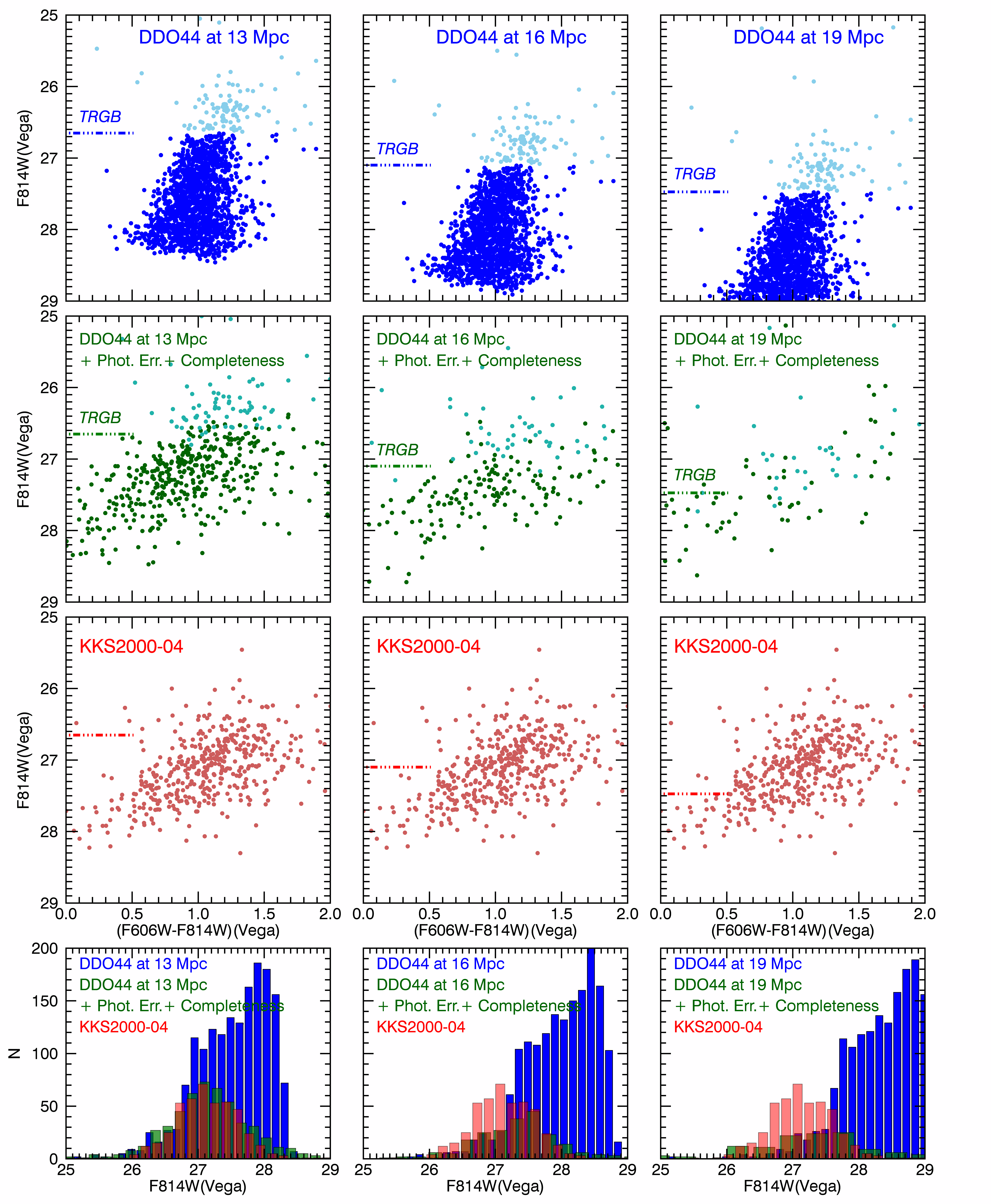}
  \caption{The CMD of the well-established distance galaxy DDO44 \citep[3.1$\pm$0.05 Mpc;][]{2009ApJS..183...67D} placed at different distances compared with the CMD of [KKS2000]04. The first row shows the observed CMD of DDO44 as produced by the ANGST \citep{2009ApJS..183...67D} team simulated at 13, 16 and 19 Mpc. The second row shows the same CMD taking into account the photometric errors and the completeness of the observations of [KKS2000]04 (see text for more details). AGB stars have been plotted with a light colour to facilitate their identification. The third row illustrates the CMD of [KKS2000]04 using \texttt{DOLPHOT} and applying the same quality cuts as the ones used in the  ANGST survey. The expected location of the TRGB at the different distances is indicated with a horizontal dashed line. The fourth row shows the luminosity functions of the stars of the CMD of DDO44 (as observed; blue colour) and considering photometric error and completeness (green colour), together with the observed luminosity function of  [KKS2000]04. All the magnitudes shown in this plot are in the Vega system and have been obtained using the software \texttt{DOLPHOT} and  applying identical photometric quality cuts.}
     \label{fig:ddo44cmd1}
 \end{figure*}

Fig. \ref{fig:ddo44cmd1} shows the CMD of DDO44 assuming three different distances: 13, 16 and 19 Mpc. To create the first row of Fig. \ref{fig:ddo44cmd1} we have used the publicly available CMD of DDO44 from the ANGST (ACS Nearby Galaxy Survey Treasury program) webpage \citep[https://archive.stsci.edu/prepds/angst/;][]{2009ApJS..183...67D}. The file we retrieved is \verb hlsp_angst_hst_wfpc2_8192-kk061_f606w-f814w_v1.phot. We  applied the following Galactic extinction corrections: 0.1 mag (\textsl{F606W}) and 0.064 (\textsl{F814W}). The stellar photometry of the DDO44 catalogue was created using \texttt{DOLPHOT} \citep{2000PASP..112.1383D} and we have used the output in Vega magnitudes.  As  explained in the previous section,  we applied a colour correction to the published \textsl{F814W} photometry \citep{mcquinn17a}, as this results in a steeper RGB and therefore a better defined TRGB. Following standard procedures \citep[see e.g.][]{2009ApJS..183...67D}, to avoid  stars with uncertain photometry we applied the cut (sharp$_{606}$+sharp$_{814}$)$^2$$\leq$0.075 and (crowd$_{606}$+crowd$_{814}$)$\leq$0.1. To model the different distances shown in the first row of Fig. \ref{fig:ddo44cmd1} we have simply applied an additive term to the observed DDO44 \textsl{F814W} photometry to mimic the expected distance modulus at 13, 16 and 19 Mpc (i.e. +3.113,  +3.563 and +3.937 mag). The dashed lines in the panels of the two upper rows of Fig. \ref{fig:ddo44cmd1} correspond to the expected location of the TRGB at those distances.

As mentioned above, the published photometry of DDO44 was obtained using the software  \texttt{DOLPHOT}. With the aim of making a direct comparison between the CMDs of DDO44 and [KKS2000]04 we have also applied \texttt{DOLPHOT} to the HST images of [KKS2000]04 to obtain its CMD. Again, we follow the same procedure as the one described in  \citet[][]{2009ApJS..183...67D}. In order to remove false detections and blends we take advantage of the various \texttt{DOLPHOT}  parameters (such as sharpness and crowding). The CMD of [KKS2000]04 shown in the third row of Fig. \ref{fig:ddo44cmd1} corresponds to all the detections within 1 R$_e$ applying the (above described) quality cuts (i.e. (sharp$_{606}$+sharp$_{814}$)$^2$$\leq$0.075 and (crowd$_{606}$+crowd$_{814}$)$\leq$0.1) and with simultaneous photometry in both HST ACS bands. This resulted in 447 objects. As a sanity check we compare the difference in magnitude between the photometry of individual stars obtained both using  \texttt{DAOPHOT} and \texttt{DOLPHOT}. We find no offsets between the codes.

To permit a direct comparison between the CMDs of DDO44 and [KKS2000]04 is necessary to model in the published photometry of DDO44 the photometric errors and completeness of the HST observations of [KKS2000]04.  To be as precise as possible, we have also taken into account the photometric errors and the completeness of the published DDO44 photometry. As expected, this is a small correction considering that the stars observed in DDO44 are significantly brighter than those explored in  [KKS2000]04. To help the reader compare between the two photometries, these are the typical photometric errors of the DDO44 catalogue: $\pm$0.05 mag at \textsl{F606W(Vega)}=24 mag and $\pm$0.11 mag at \textsl{F814W(Vega)}=24 mag,  and their 50\% completeness: \textsl{F606W(Vega)}=26.04 and \textsl{F814W(Vega)}=24.86 mag  \citep{2009ApJS..183...67D}. Note that in order to see where this affects the observed CMDs shown in the first row of Fig.  \ref{fig:ddo44cmd1} it is necessary to add to the previous quantities  +3.113,  +3.563 and +3.937 mag to take into account the difference in the distance moduli. Similarly, the typical photometric errors and the completeness of the \texttt{DOLPHOT} photometry of [KKS2000]04 are: $\pm$0.12 mag at \textsl{F606W(Vega)}=27 mag and $\pm$0.22 mag at \textsl{F814W(Vega)}=27 mag,  and the 50\% completeness: \textsl{F606W(Vega)}=28.14 and \textsl{F814W(Vega)}=27.61 mag. Once the completeness and photometric errors of both datasets are accounted for, we can start to simulate the observational effects on the CMDs of DDO44 at different distances. This is illustrated in the second row of Fig. \ref{fig:ddo44cmd1}. 
 
In order to show which distance is the most favoured by comparing the CMDs of the simulated CMDs of DDO44 (second row of Fig. \ref{fig:ddo44cmd1}) and  [KKS2000]04 (third row of Fig. \ref{fig:ddo44cmd1}), we plot in the fourth row of Fig. \ref{fig:ddo44cmd1} the luminosity functions of the stars in the CMD of DDO44 (as observed; blue colours) and considering both photometric errors and completeness (green colours). For comparison, we also plot the luminosity function of the stars of [KKS2000]04. The luminosity functions of the stars of both galaxies are reasonably similar when DDO44 is located at a distance of 13 Mpc and start to significantly deviate when DDO44 is placed at distances beyond 16 Mpc.  

\begin{figure*}
 \includegraphics[width=\textwidth]{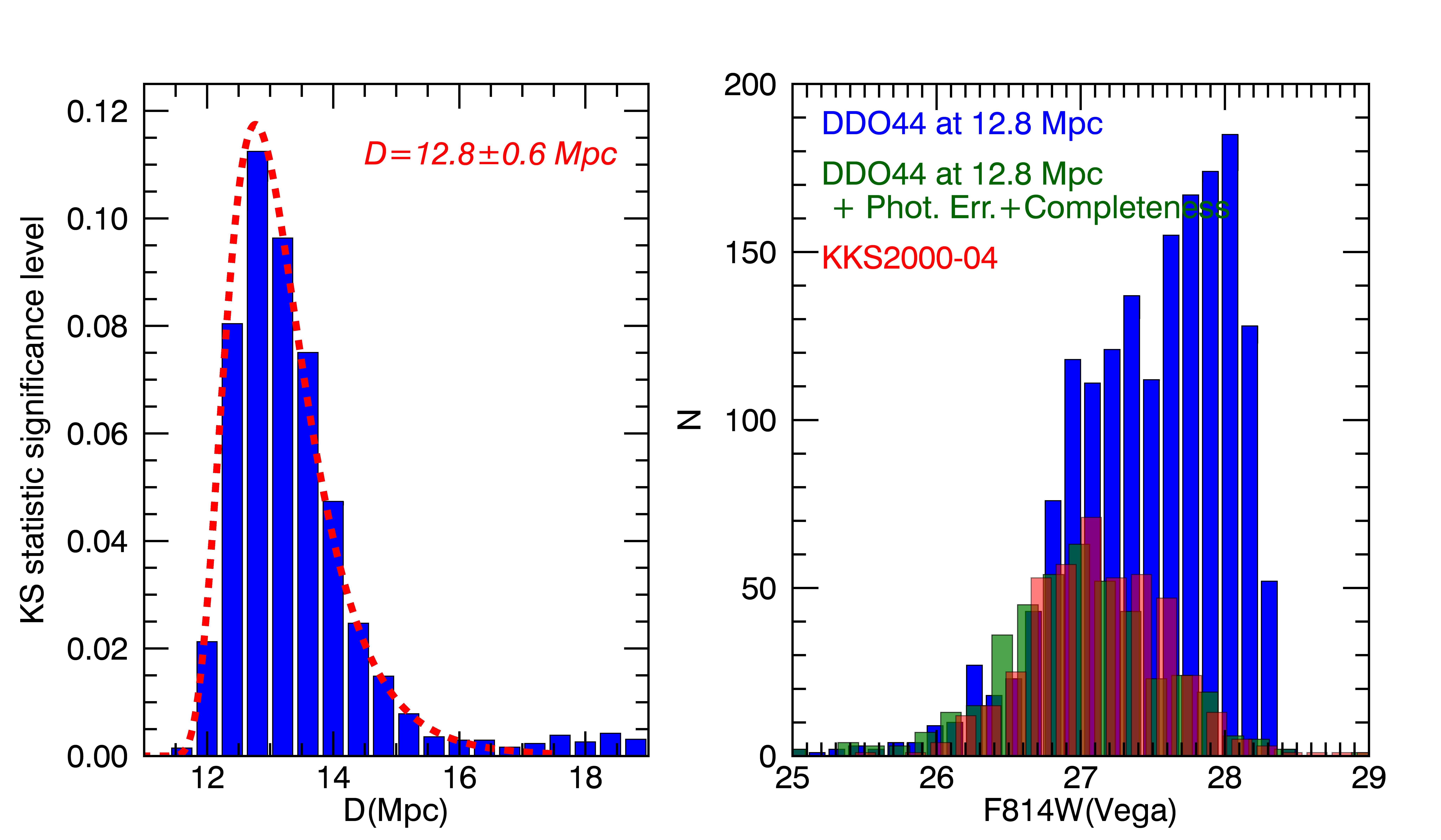}
  \caption{{\it Left Panel:} Kolmogorov-Smirnov (KS) statistic significance level at comparing the F814W luminosity function of the stars of DDO44 at different distances and [KKS2000]04. The peak of the maximum similarity is reached at a distance of D=12.8$\pm$0.6 Mpc. Distances for [KKS2000]04 larger than 16 Mpc are rejected at 99.5\% level. Overplotted in red is the shape of an f-distribution. {\it Right Panel:} The luminosity functions of the CMD of DDO44 at 12.8 Mpc (as it would be observed using the ANGST photometry; blue colour) and considering photometric errors and completeness (green colour), together with the observed luminosity function of  [KKS2000]04. All  magnitudes shown in this plot are in the Vega system and have been obtained using the  software \texttt{DOLPHOT} and  applying the same photometric quality cuts.}
     \label{fig:ddo44cmd2}
\end{figure*}

To have a quantitative estimation of which distance is most favoured, we have simulated the appearance of the DDO44 CMD from 11 to 19 Mpc. We  created 3000 mock CMDs to densely cover this interval in distance. For every  simulated DDO44 CMD we  obtained the F814W luminosity function of the stars taking into account the photometric errors and completeness as explained above. Then, we  compared the simulated luminosity function of DDO44 at a given distance D with the observed luminosity function of  [KKS2000]04. To explore the similarities between both luminosity functions, we  used the Kolmogorov-Smirnov (KS) test. The result of doing this is shown in Fig. \ref{fig:ddo44cmd2}.

We estimated the best-matching distance using the KS test whereby the cumulative luminosity function of the stars in DDO44 placed at different distances is compared with the one (within the same cuts) of [KKS2000]04. Fig. \ref{fig:ddo44cmd2} shows that at the 99.5\% confidence level, a distance larger than 16 Mpc is rejected. This non-parametric technique yields a distance of 12.8$\pm$0.6 Mpc (where the significance level for rejecting that both luminosity functions come from the same parent distribution can not be rejected at $>$88\% level). Also in Fig.  \ref{fig:ddo44cmd2} we show an example of how the luminosity function of the stars of DDO44 at a distance of 12.8 Mpc would look like when compared with the observed luminosity of [KKS2000]04. The similarity between both distributions is rather obvious.

\subsection{The size and magnitudes of the globular clusters as distance indicators}

The globular clusters around [KKS2000]04  can be used to give two independent distance estimators. The first is based on the fact that the peak of the luminosity function of globular clusters is rather invariant from galaxy to galaxy with a value of  M$_V$$\sim$-7.5$\pm$0.2 \citep{2012Ap&SS.341..195R}. The second takes advantage of the fairly constant (and independent from magnitude) half-light radii of globular clusters \citep{2005ApJ...634.1002J}. 

\begin{figure}
\includegraphics[width=\columnwidth]{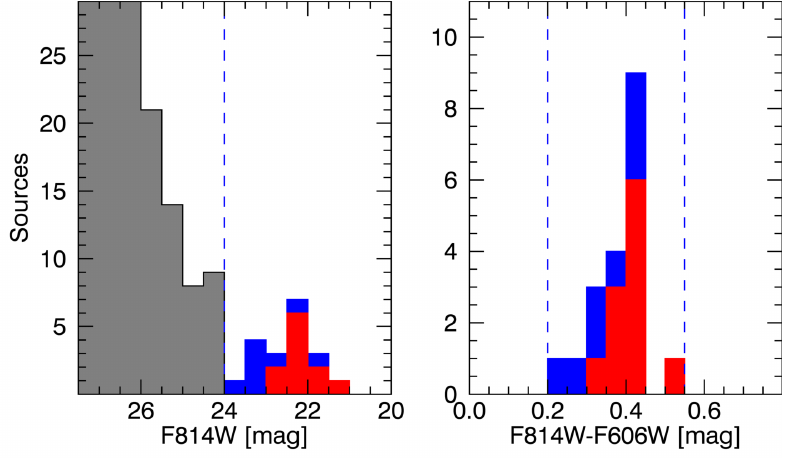}
\caption{Left panel: \texttt{SExtractor} AUTO magnitudes (corrected by foreground Galactic extinction) of all the sources (grey, blue and red histograms) in the \textsl{F814W(AB)} image with FWHM$<$5 pixels and colour 0.2$<$\textsl{F606W(AB)-F814W(AB)}$<$0.55 mag. The vertical dashed line corresponds to \textsl{F814W(AB)}=24 mag. Right panel: Colour distribution of the GC candidates around [KKS2000]04. The vertical dashed lines encircle our colour cut in the sample selection. In both panels, red corresponds to the GCs detected by \citet{2018ApJ...856L..30V} whereas blue indicates the new candidates found in this paper.}
\label{fig:histosgc}
\end{figure}

\citet{2018ApJ...856L..30V} have explored 11 spectroscopically confirmed clusters around [KKS2000]04. These authors acknowledge the possibility that further clusters may exist around the galaxy but lacking a spectroscopic confirmation they refrain to include new objects. Note that \citet{2018ApJ...856L..30V} target selection gave priority to compact objects with \textsl{F814W}(AB)$<$22.5 mag. We have probed whether new GCs can be found around the galaxy. To do that we created a \texttt{SExtractor} catalogue with all the sources in the \textsl{F814W} image satisfying the following: a FWHM size less than 5 pixels \citep[their spectroscopically confirmed GCs have FWHM$<$4.7 pixels;][]{2018ApJ...856L..30V} and a range in colour 0.2$<$\textsl{F606W(AB)}-\textsl{F814W(AB)}$<$0.55 mag\footnote{This colour range is a compromise between maximising the detection of metal-poor GCs  and minimising the contamination of background red sources \citep{2012ApJ...746...88B}.} (the range in colour for their spectroscopically confirmed GCs is 0.28--0.43). We used \texttt{SExtractor} AUTO magnitudes to build this catalogue. The distribution of magnitudes of all the sources in the \textsl{F814W} image satisfying the colour and size cut are shown in Fig. \ref{fig:histosgc}. Motivated by the shape of the luminosity distribution shown in Fig. \ref{fig:histosgc}, we added another restriction to create our final sample of GC candidates around the galaxy: \textsl{F814W}(AB)$<$24 mag. Note that this magnitude is well above the completeness magnitude (\textsl{F814W}(AB)=25.4 mag) for point-like sources in the HST image.

\begin{figure*}
\includegraphics[width=\textwidth]{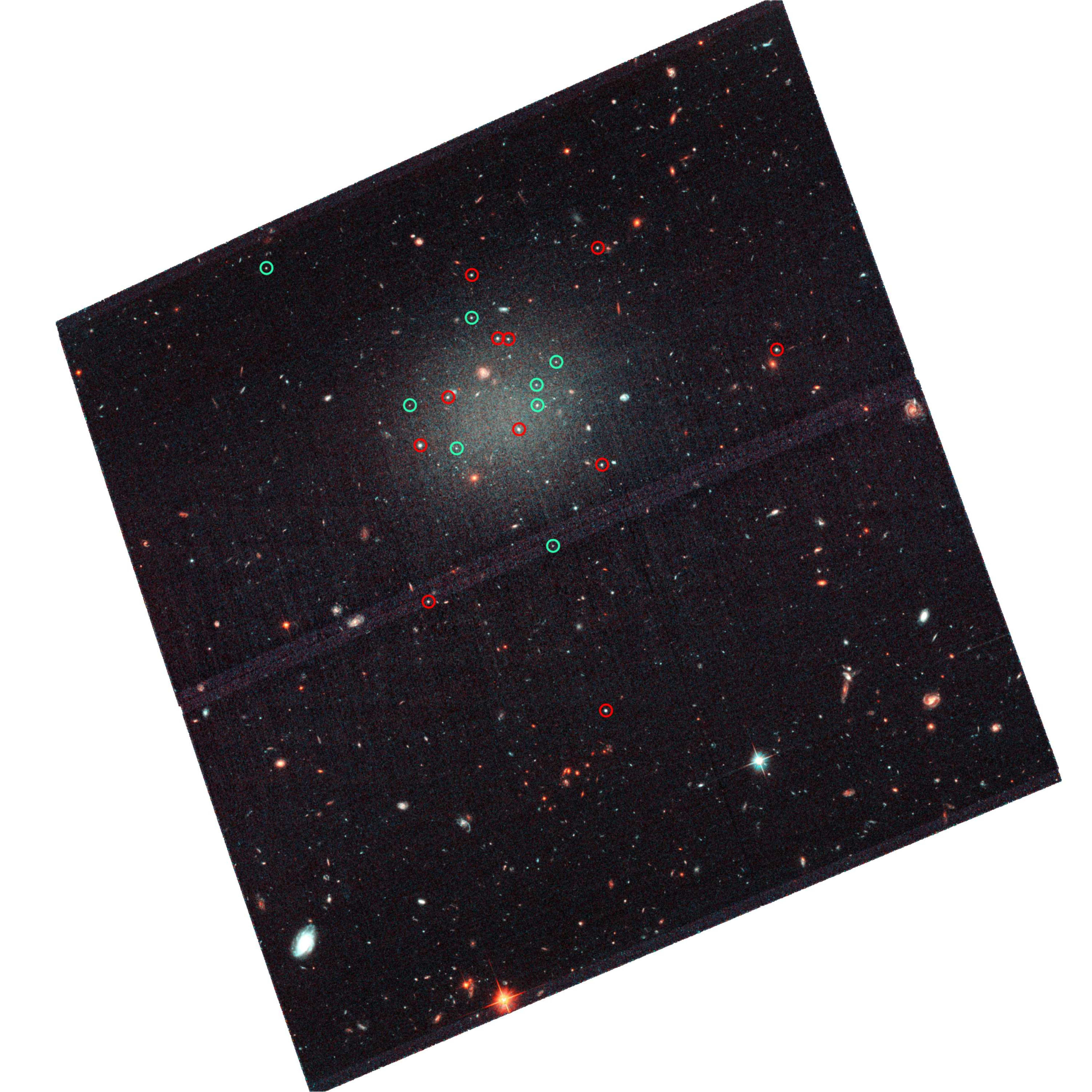}
\caption{Globular clusters surrounding [KKS2000]04. The red circles correspond to the GCs detected in \citet{2018ApJ...856L..30V}, whereas the blue circles indicate the location of the new GC candidates proposed in this work.}
\label{fig:newgc}
\end{figure*}

As expected, our criteria recovers the 11 clusters found by \citet{2018ApJ...856L..30V} but also adds another 8 new candidates. The final sample of GCs explored in this paper is shown on Fig. \ref{fig:newgc}. It is worth noting that the majority of the new GC candidates added in this work are spatially located close to the galaxy, suggesting a likely association with this object.

Once the sample of GCs is built, to use both distance estimators based on the GCs properties, we need to quantify the apparent magnitudes, sizes and ellipticities of the GC population around [KKS2000]04. As mentioned before, the magnitudes we used were those produced by \texttt{SExtractor}. Following \citet{2018ApJ...856L..30V}, to derive the size and ellipticities of the GCs we use PSF-convolved  \citet{1962AJ.....67..471K} and   \citet{1968adga.book.....S} models. The PSFs used were synthetic PSFs from Tiny Tim \citep{1993ASPC...52..536K}. The model fitting of the globular clusters was conducted using \texttt{IMFIT} \citep{2015ApJ...799..226E}. Note that for this we account for the spatial distortions of the HST PSF along the field-of-view. This is key due to the small sizes of the GCs (subtending only a few pixels) and necessary for a correct fitting. The results of the model fitting were comparable in both \textsl{F606W} and \textsl{F814W}, but slightly more accurate in the \textsl{F814W} band. For this reason, we use this band to characterise the structural parameters of the objects. The values measured (with their typical errors) are shown in Table \ref{table:gcs}. The magnitudes in the table correspond to \texttt{SExtractor} magnitudes (corrected by foreground Galactic extinction: 0.06 mag for \textsl{F606W} and 0.037 mag for \textsl{F814W}). The structural parameters provided are those obtained adopting a S\'ersic model.

\begin{table*}
\centering
\caption{Structural parameters of the globular clusters surrounding [KKS2000]04.}
\begin{tabular}{ccccccc} %\toprule
\hline
\label{table:gcs} 
   ID & R.A.  & Dec.  & \textsl{V$_{606}$} & \textsl{I$_{814}$} & R$_{e,F814W}$ & $\epsilon_{F814W}$  \\ 
      &(J2000)&(J2000)&  (AB mag)    & (AB mag)     & (arcsec)      &  \\
      &  &  & ($\pm$0.05) &  ($\pm$0.05) & ($\pm$0.020) & ($\pm$0.1) \\
\hline
\multicolumn{7}{c}{Globular clusters detected in \citet{2018Natur.555..629V}} \\
   \hline
GC39  & 40.43779 & -8.423583   &   22.38   & 22.02   &  0.083 &  0.05 \\
GC59  & 40.45034 & -8.415959   &   22.87   & 22.44   &  0.107 &  0.27 \\
GC71  & 40.43807 & -8.406378   &   22.70   & 22.30   &  0.083 &  0.15 \\
GC73  & 40.45093 & -8.405026   &   21.52   & 21.19   &  0.077 &  0.14 \\
GC77  & 40.44395 & -8.403900   &   22.03   & 21.66   &  0.118 &  0.25 \\
GC85  & 40.44896 & -8.401659   &   22.42   & 22.01   &  0.060 &  0.31 \\
GC91  & 40.42571 & -8.398324   &   22.49   & 22.08   &  0.114 &  0.20 \\
GC92  & 40.44544 & -8.397534   &   22.39   & 21.88   &  0.055 &  0.13 \\
GC93  & 40.44469 & -8.397590   &   22.97   & 22.59   &  0.046 &  0.26 \\
GC98  & 40.44728 & -8.393103   &   22.90   & 22.49   &  0.056 &  0.13 \\
GC101 & 40.43837 & -8.391198   &   23.01   & 22.58   &  0.050 &  0.06 \\

\hline
\multicolumn{7}{c}{New globular clusters proposed in this work} \\
\hline
GC$_{new}$1  & 40.44153 & -8.412058   &   23.46   & 23.15   &  0.048 &  0.22\\
GC$_{new}$2  & 40.44836 & -8.405235   &   23.81   & 23.39   &  0.083 &  0.12\\
$^a$GC$_{new}$3 & 40.44265 & -8.402213 &   22.61   & 22.21   &  0.077 & 0.06\\
GC$_{new}$4  & 40.45166 & -8.402218   &   24.29   & 23.87   &  0.048 &  0.17\\
GC$_{new}$5  & 40.44270 & -8.400788   &   23.84   & 23.49   &  0.100 &  0.18\\
GC$_{new}$6  & 40.44131 & -8.399180   &   23.86   & 23.46   &  0.064 &  0.05\\
$^b$GC$_{new}$7 & 40.44729 & -8.396096  &  22.21   & 21.92   &  0.048 & 0.10\\
GC$_{new}$8  & 40.46181 & -8.392609   &   22.82   & 22.58   &  0.046 &  0.01\\
\hline
\multicolumn{7}{c}{$^a$This candidate has been spectroscopically confirmed \citep{2018arXiv181207345E}.} \\
\multicolumn{7}{c}{$^b$This candidate has been spectroscopically ruled out \citep{2018arXiv181207345E}.} \\
\end{tabular}
\end{table*}

We have compared the difference in the magnitudes using \texttt{SExtractor} with those retrieved using King and S\'ersic models. The mean difference between the \texttt{SExtractor} AUTO magnitudes and the King model magnitudes is 0.06 mag with a standard deviation of 0.05 mag. Similarly, the difference between the King and S\'ersic models is 0.06 mag with a standard deviation of 0.12 mag. As the model-based magnitudes are prone to overestimation (particularly if one of the fitting parameters as the S\'ersic index or the tidal King radius is wrong) we conservatively decided to continue our analysis with the \texttt{SExtractor} magnitudes. Nonetheless, we will show how using the model magnitudes barely affects the results.

\begin{figure}

\includegraphics[width=\columnwidth]{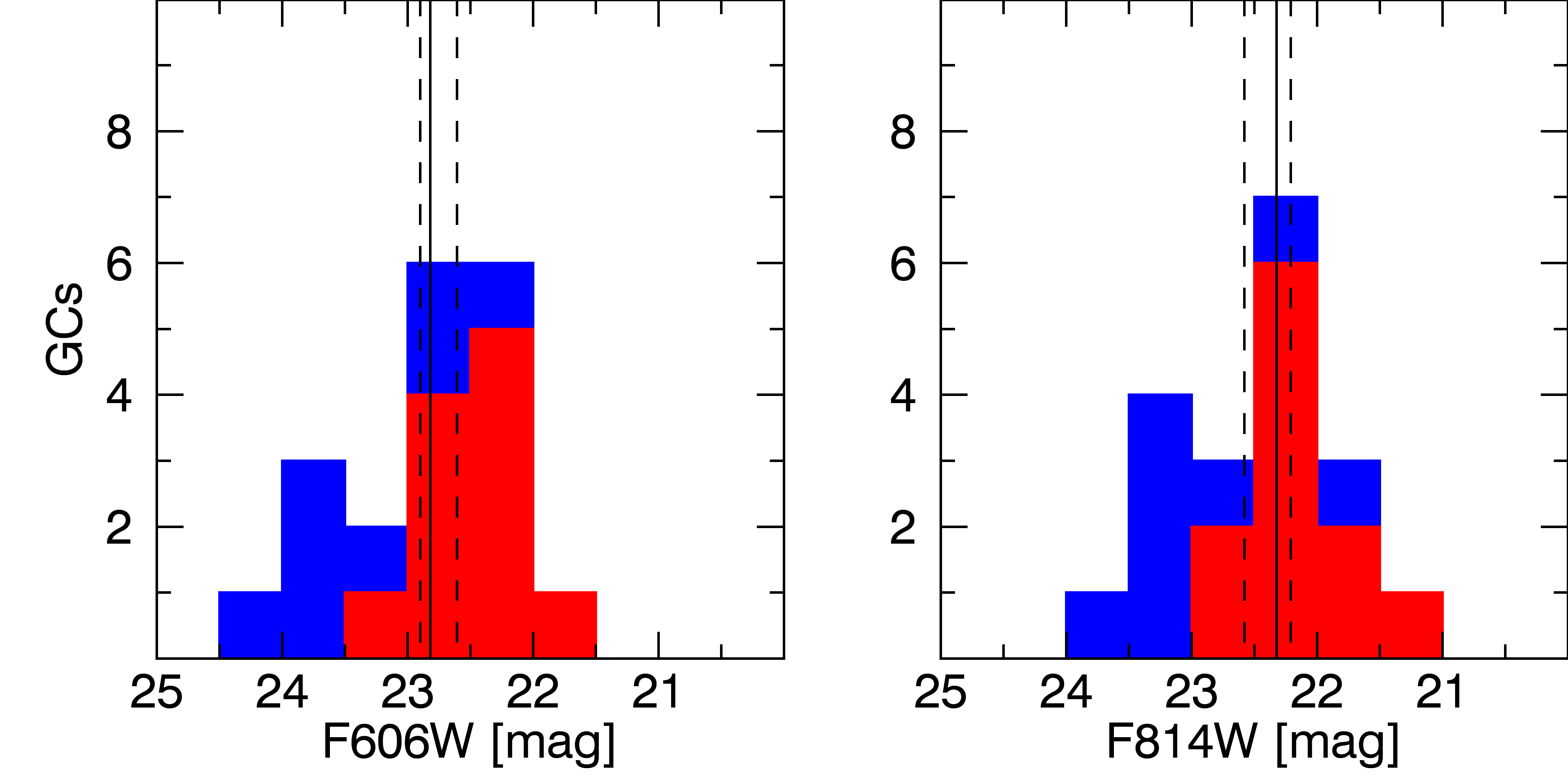}
\caption{The luminosity functions of the globular clusters surrounding [KKS2000]04. The magnitudes have been corrected for foreground Galactic extinction. The left panel shows the magnitude distribution in the \textsl{F606W(AB)} band whereas the right panel shows \textsl{F814W(AB)}. In red we show the location of the globular clusters identified by \citet{2018ApJ...856L..30V} and in blue the new sources found in this work. The vertical solid lines represent the location of the median values of the distributions for the entire sample of 19 clusters. The dashed lines show the 1$\sigma$ uncertainties on the location of the median values using bootstrapping. }
\label{fig:lumgc}
\end{figure}

The luminosity functions of the globular clusters  in each HST band are shown in Fig. \ref{fig:lumgc}. The location of the peaks of the luminosity functions and their errors were obtained using a bootstrapping median. For the 19 GCs proposed in this work, we derive the following peak locations: 22.82$^{+0.08}_{-0.21}$ mag (\textsl{F606W}(AB)) and 22.43$^{+0.14}_{-0.23}$ mag (\textsl{F814W}(AB)). Using only the 11 \citet{2018ApJ...856L..30V} GCs we get 22.48$^{+0.38}_{-0.06}$ mag (\textsl{F606W}(AB)) and 22.08$^{+0.35}_{-0.08}$ mag (\textsl{F814W}(AB)). This last value is also provided by \citet{2018ApJ...856L..30V} and we find here a good agreement with their measurement. Note how the spectroscopic completeness criteria introduced by \citet{2018ApJ...856L..30V} for selecting their GC sample moves the location of the peak of the luminosity function towards brighter values than in our case.

\begin{figure}

\includegraphics[width=\columnwidth]{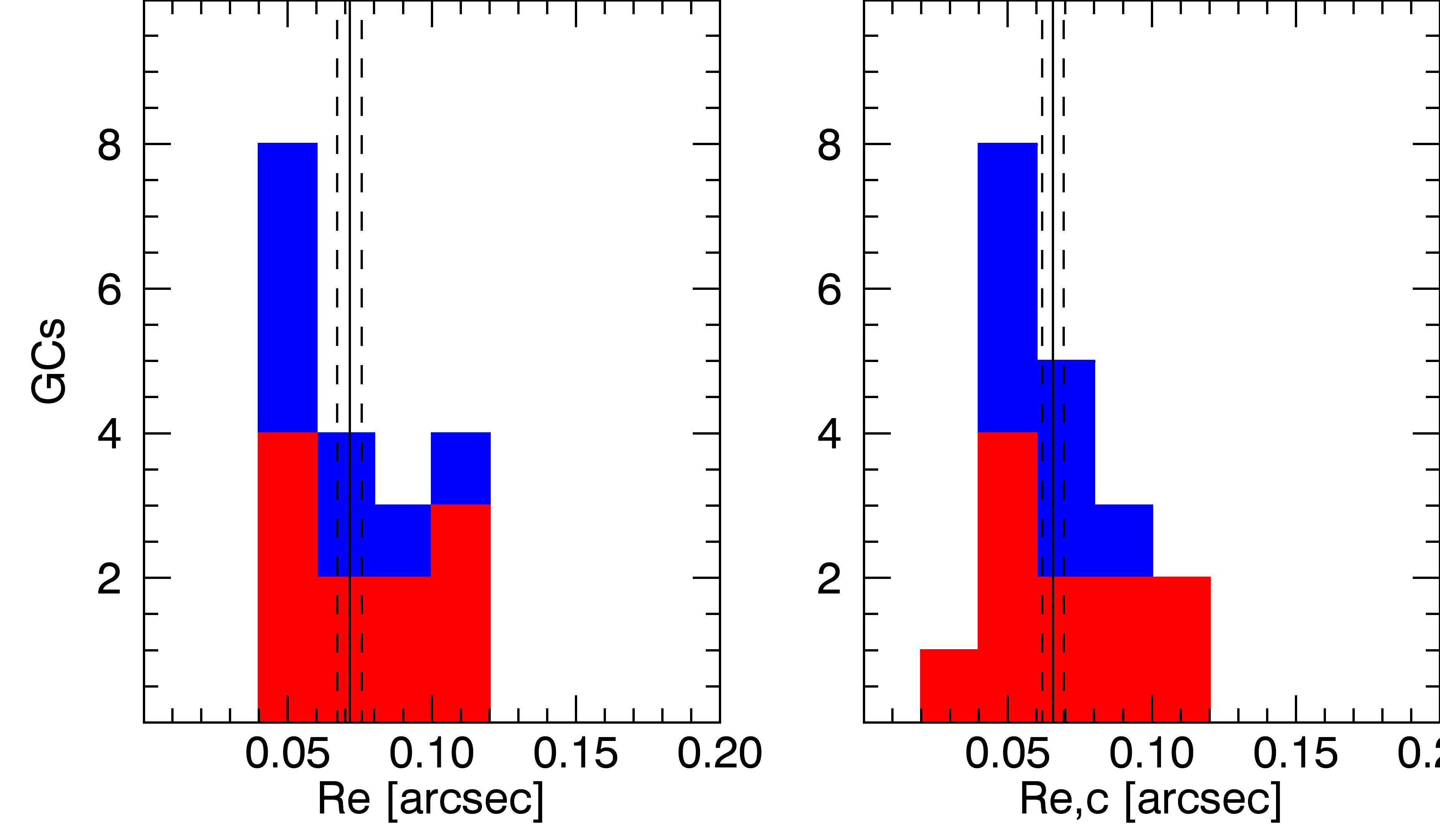}
\caption{{\it Left panel.} Effective radii using S\'ersic models for the sample of GCs analysed in this work. {\it Right panel.} Same as in the left panel but using circularised effective radii. Vertical solid lines correspond to the mean values whereas the dashed lines show the uncertainty region. The red histograms show the distribution for the GCs in \citet{2018ApJ...856L..30V} whereas the blue histograms show the new GCs proposed in this work.}
\label{fig:radiusgc}
\end{figure}

The distribution of \textsl{F814W} S\'ersic half-light radii and the circularised half-light radii are shown in Fig. \ref{fig:radiusgc}. The mean  size for all the GCs is $<$R$_e$(F814W)$>$ =0.071$\pm$0.004\arcsec \citep[0.077$\pm$0.006\arcsec using the 11 GCs studied in][]{2018ApJ...856L..30V} and the mean circularised size is $<$R$_{e,c}$(F814W)$>$=0.065$\pm$0.004\arcsec \citep[0.069$\pm$0.005\arcsec using the 11 GCs given in ][]{2018ApJ...856L..30V}. We do not find any difference between the average sizes of the GCs using \textsl{F814W} and \textsl{F606W} within their uncertainties. Finally, the mean ellipticity obtained in the \textsl{F814W} band is 0.15$\pm$0.02 \citep[0.18$\pm$0.02 using only the 11 GCs provided by][]{2018ApJ...856L..30V}. The mean ellipticities obtained from the two HST bands are consistent within their error bars. The ellipticity values obtained here are also compatible  with the value reported in \citet{2018ApJ...856L..30V}.

Once the structural parameters of the GC population are measured, we can derive a first distance estimate using the peak of the luminosity function of the GCs. \citet{2012Ap&SS.341..195R} suggested M$_V$=-7.66$\pm$0.09 mag for the absolute magnitude of the peak of the GC luminosity distribution. It is worth noting that this value is the one recommended for metal-poor clusters. The globular clusters of [KKS2000]04 have an average metallicity of [Fe/H]=-1.35 \citep{2018ApJ...856L..30V}. In this sense, this choice is well justified. To transform our \textsl{V$_{606}$} (\textsl{F606W} HST filter) into V magnitudes, we apply the following correction: \textsl{V=V$_{606}$}+0.118. The correction is obtained assuming a single stellar population model with age=9.3 Gyr and [Fe/H]=-1.35 \citep[the values measured spectroscopically by][]{2018ApJ...856L..30V}. Thus, the peak of the luminosity function has a magnitude of \textsl{V}=22.94$^{+0.08}_{-0.21}$ mag which corresponds to a distance modulus of (m-M)$_0$=30.60$^{+0.16}_{-0.30}$ mag (where the error bars of the distance modulus also account for the uncertainty on the location of the GC luminosity function peak given by \citet{2012Ap&SS.341..195R}). This distance modulus is equivalent to the following distance: D$_{GC,1}$=13.2$^{+1.1}_{-1.7}$ Mpc (using all the GCs)\footnote{Removing the GCnew7 candidate, the peak of the luminosity function moves to \textsl{V}=22.99$^{+0.09}_{-0.17}$ mag which corresponds to a distance D$_{GC,1}$=13.5$^{+1.2}_{-1.5}$ Mpc. This is fully consistent with the measurement we have estimated using all the GCs within the error bars.}. The same analysis but using the 11 GCs published in \citet{2018ApJ...856L..30V} provides D$_{GC,1}$=11.3$^{+2.8}_{-0.9}$ Mpc. 

It is worth exploring how this distance estimate  changes if instead of \texttt{SExtractor} AUTO magnitudes we would have used the model magnitudes described above. Using both the S\'ersic and the King models, the distances are: D$_{GC,1}$=12.0$^{+1.5}_{-1.3}$ Mpc (S\'ersic model magnitudes) and D$_{GC,1}$=12.3$^{+2.0}_{-0.9}$ Mpc (King model magnitudes). As expected, the distances using the model magnitudes are slightly smaller than the ones using the \texttt{SExtractor} magnitudes  (although compatible within the uncertainties).

Another independent measurement of the distance to [KKS2000]04 is based on the average size of its globular clusters \citep{2005ApJ...634.1002J}. The size of the globular clusters is almost independent of their absolute magnitude, with an average size of $<$R$_e$$>$=3.6$\pm$0.2 pc \citep[][2010 edition]{1996AJ....112.1487H} for MW GCs and $<$R$_e$$>$=4.3$\pm$0.2 pc for dwarf galaxies GCs \citep{2009MNRAS.392..879G}. A clear illustration of this independence of size on magnitude is shown in Fig. 2 of \citet[][]{2015ApJ...813L..15M} for a large sample of GCs. We take advantage of this observational fact to get another distance estimate to [KKS2000]04. From the mean measured sizes of the GCs we derive the following distances: D$_{GC,2}$=12.4$^{+1.4}_{-1.2}$ Mpc\footnote{Removing the GCnew7 candidate has a tiny effect on the distance determination using the size of the GCs:  D$_{GC,2}$=12.6$^{+1.5}_{-1.3}$ Mpc.} \citep[comparing with GCs in Dwarfs; D$_{GC,2}$=11.6$^{+1.5}_{-1.5}$ Mpc using the 11 GCs given in][]{2018ApJ...856L..30V} and D$_{GC,2}$=10.3$^{+1.3}_{-1.2}$ Mpc  \citep[comparing with GCs in the MW; D$_{GC,2}$=9.4$^{+1.5}_{-1.1}$ Mpc using the 11 GCs given in][]{2018ApJ...856L..30V}. In what follows we assume the determination of the distance to  [KKS2000]04 based on the comparison with the GCs in dwarf galaxies since [KKS2000]04 has all the characteristics of a dwarf galaxy.

\begin{figure}
\includegraphics[width=\columnwidth]{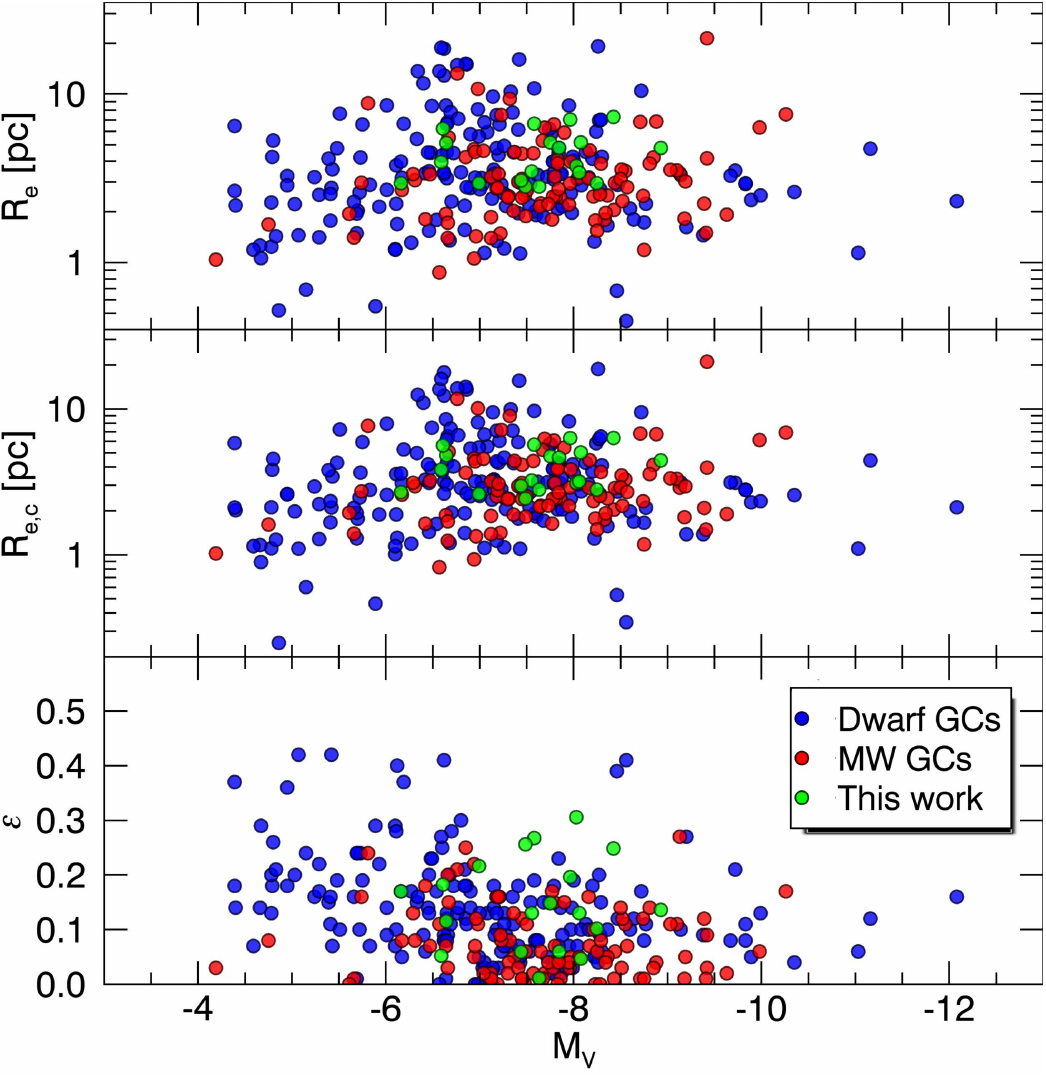}

  \caption{Structural properties (effective radii, circularised effective radii and ellipticities) of the globular clusters around [KKS2000]04 (green points) compared to the GCs of the Milky Way \citep[red points;][]{1996AJ....112.1487H} and a compilation of dwarf galaxies \citep[blue points;][]{2009MNRAS.392..879G}. At a distance of 13 Mpc, the properties of the GCs in [KKS2000]04 are consistent with the properties of GCs of regular dwarf galaxies.}
\label{fig:gcs}
\end{figure}

Finally, the mean ellipticity of the GCs of [KKS2000]04 is $<$$\epsilon$$>$=0.15$\pm$0.02 for all the GCs and $<$$\epsilon$$>$=0.18$\pm$0.02 for the 11 clusters provided by \citet[][]{2018ApJ...856L..30V}. This mean ellipticity  is anomalously high compared to the Milky Way GCs but rather normal in comparison with the ellipticity measured in dwarf galaxies such as in the Small and Large Magellanic Clouds \citep{1996A&AS..116..447S}. In this sense, [KKS2000]04 has a normal population of GCs if compared to GCs in dwarf galaxies.  Fig. \ref{fig:gcs} shows the location of the GCs of [KKS2000]04 in the size - absolute magnitude plane, and in the ellipticity - absolute magnitude plane. The sizes and ellipticities of the [KKS2000]04 are very similar to those found in regular dwarf galaxies. To quantify this statement, Table \ref{table:gcs2} shows the mean and median structural properties of the GCs of [KKS2000]04 (under the assumption this galaxy is at a distance of 13 Mpc) compared to both MW and dwarf galaxy GCs. Note the range of absolute magnitudes covered by MW GCs is broader than the one by [KKS2000]04. This decrease of the width of the luminosity distribution of the GCs as the host galaxy gets fainter has been already reported by \citet{2007ApJS..171..101J}.

\begin{table}
\centering
\caption{Globular cluster structural properties (effective radius R$_e$, circularised effective radius R$_{e,c}$ and ellipticity) for different galaxy hosts. The errors correspond to the 1$\sigma$ interval.}
\begin{tabular}{lccc} %\toprule
\hline
\label{table:gcs2} 
 & MW & Dwarfs & [KKS2000]04  \\
 &  &  & (at 13 Mpc) \\
 \hline
 Mean R$_e$ (pc)  & 3.6$_{-0.2} ^{+0.2}$ & 4.3$_{-0.2} ^{+0.2}$  & 4.4$_{-0.3} ^{+0.3}$ \\
 Median R$_e$ (pc) & 2.9$_{-0.2} ^{+0.1}$ & 3.2$_{-0.2} ^{+0.1}$  & 3.9$_{-0.5} ^{+0.8}$ \\ \hline
 Mean R$_{e,c}$ (pc)  & 3.4$_{-0.2} ^{+0.2}$ & 4.0$_{-0.2} ^{+0.2}$  & 4.1$_{-0.2}^{+0.2}$ \\
 Median R$_{e,c}$ (pc) & 2.8$_{-0.2} ^{+0.1}$ & 3.0$_{-0.2} ^{+0.1}$  & 3.8$_{-0.7}^{+0.8}$  \\
   \hline
 Mean $\epsilon$  & 0.08$_{-0.01} ^{+0.01}$ & 0.14$_{-0.01} ^{+0.01}$  & 0.15$_{-0.01}^{+0.02}$  \\
 Median $\epsilon$ & 0.06$_{-0.01} ^{+0.01}$ & 0.13$_{-0.02} ^{+0.01}$  & 0.14$_{-0.01} ^{+0.03}$  \\  
 \hline

\end{tabular}
\end{table}

\subsection{The surface brightness fluctuation (SBF) distance}
\label{sec:sbf}

\begin{figure*}

\includegraphics[width=\textwidth]{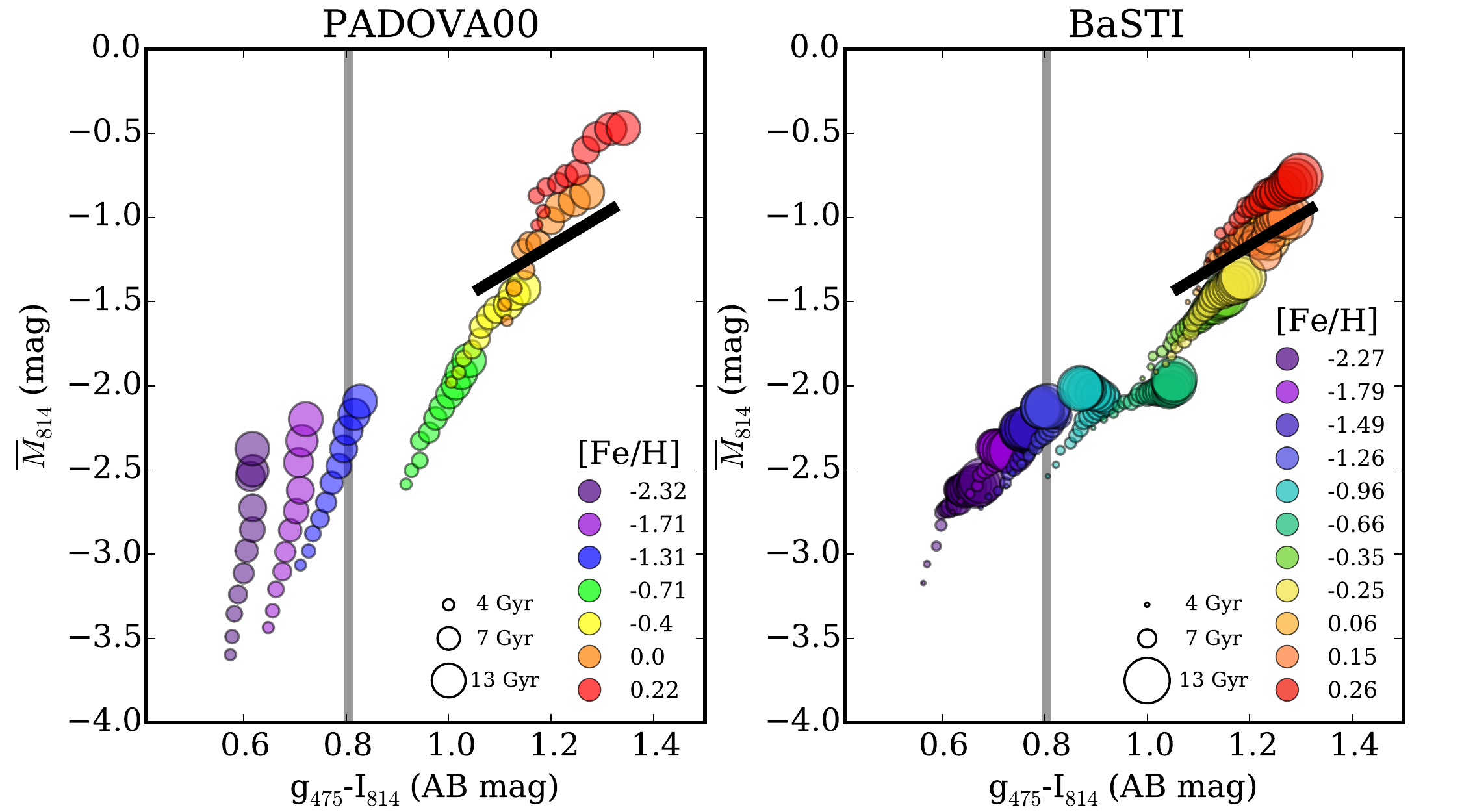}

\caption{The absolute fluctuation (AB) magnitude in the \textsl{F814W} filter as a function of the $g_{475} -I_{814}$ colour for a variety of SSP models using the E-MILES library \citep{2016MNRAS.463.3409V}, assuming a Kroupa  Universal IMF and two sets of stellar tracks: Padova00  \citep[left panel, ][]{2000A&AS..141..371G} and  BaSTI \citep[right panel,][]{2004ApJ...612..168P}. The symbols are colour-coded with metallicity, ranging from metal-poor (purple) to metal-rich (red) populations, while their sizes are proportional to their ages. The black line  is the \citet{2010ApJ...724..657B} calibration (Eq.~\ref{eq:blakeslee}) within
its validity range. The grey vertical line is the $g_{475}-I_{814}$ colour as inferred from the best-fit SED.}
\label{fig:sbf1}
\end{figure*}

\begin{figure*}

\includegraphics[width=\textwidth]{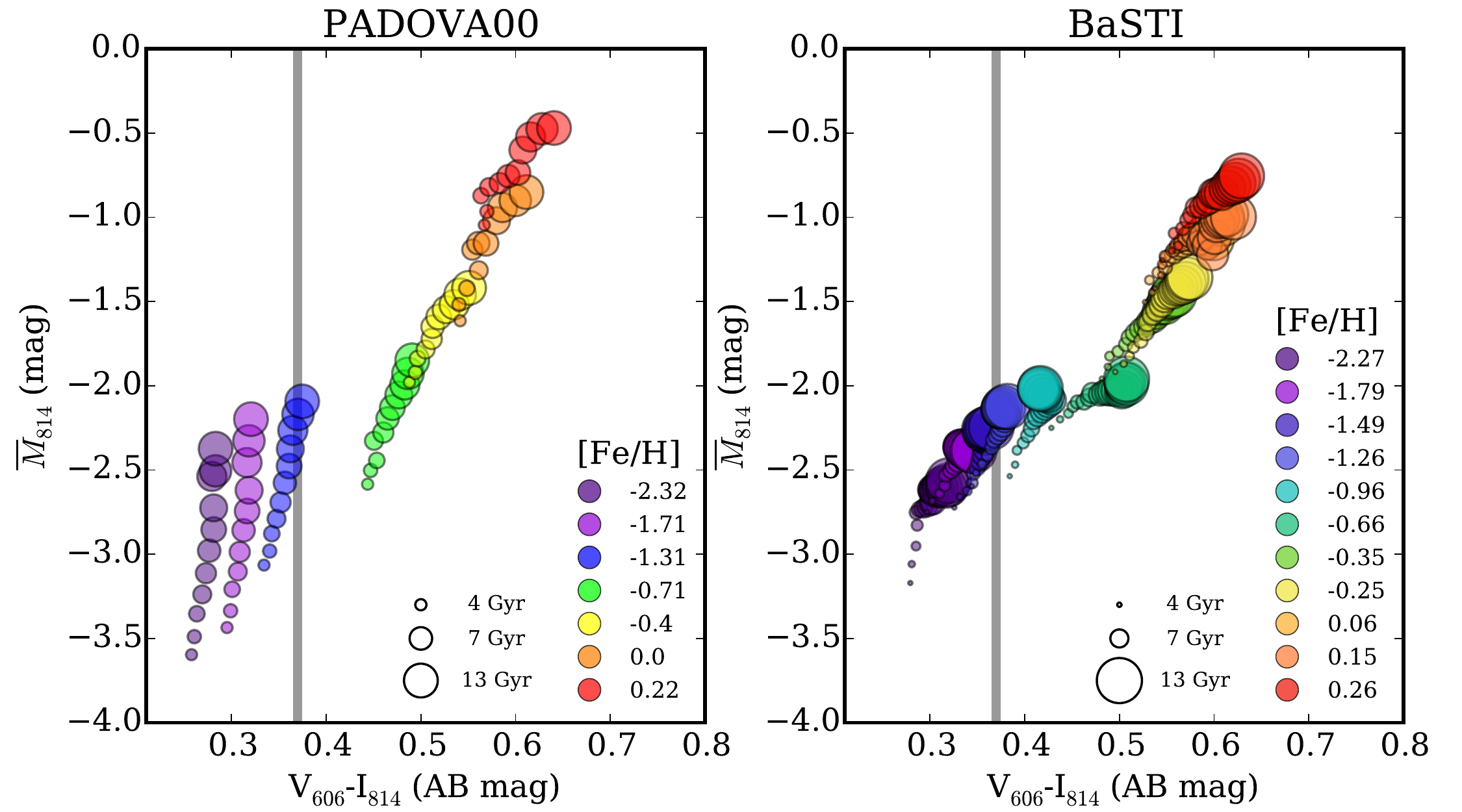}
\caption{The same as Fig.~\ref{fig:sbf1} but in terms  of the directly observed $V_{606}-I_{814}$ colour, which is independent of the assumed or fitted SED. In both cases, the theoretical predictions show a  non-linear behaviour at colours redder than the observed one, making unreliable any linear extrapolation of Eq.~\ref{eq:blakeslee} beyond its validity range. The grey vertical line is the $V_{606}-I_{814}$=0.37 colour given by \citet{2018Natur.555..629V}.}
\label{fig:sbf2}
\end{figure*}

A promising way to measure distances is the method put forward by \cite{1988AJ.....96..807T} using surface brightness fluctuations. The method relies on measuring the luminosity fluctuations that arise from the counting statistics of the stars contributing the flux in each pixel image of a galaxy. The amplitude of these fluctuations is inversely proportional to the distance of the galaxy. The SBF method is precise enough to resolve the depth of the Virgo cluster \citep{2007ApJ...655..144M} and is therefore ideal to measure the distance to \df2. \citet{2018Natur.555..629V} used this technique and measured a fluctuation magnitude of $\overline{m}_{814}=29.45\pm0.10$ to infer a distance of 19.0$\pm1.7$ Mpc for \df2. This determination relies on two key steps. First, they assume a colour transformation which does not use the actual SED of the galaxy, namely, 
\beq
g_{475} \; = \; I_{814} + 1.852 \left(V_{606} - I_{814}\right) \, + \, 0.096 \, , 
\eeq
and which, for their observed colour $V_{606} - I_{814} = 0.37\pm0.05$, yields
$g_{475} - I_{814} = 0.78 \pm 0.05$, while the actual value is 0.85 $\pm 0.02$ (see Table~1; where we have used the Gemini \textsl{g}-band as a proxy for \textsl{g$_{475}$}). Second, and much more importantly, they adopt the calibration by \citet[][their Eq.(2)]{2010ApJ...724..657B} 
\begin{multline}
 \overline{M}_{814} \; = \left( -1.168\pm0.013\pm0.092\right) \, + \\
 + \, \left(1.83\pm0.20\right)\times\left[(g_{475} - I_{814}) - 1.2\right] \; .
 \label{eq:blakeslee}
\end{multline}
This calibration is only valid for the colour range $1.06 \leq (g_{475} - I_{814}) \leq 1.32$, which corresponds to a range in absolute fluctuation magnitude $-1.4 \leq \overline{M}_{814} \leq -0.8$. \citet{2018Natur.555..629V} extrapolate this relation well outside its validity range, yielding an absolute fluctuation magnitude $\overline{M}_{814}=-1.94$ which is therefore highly unreliable. Not only a linear extrapolation to  bluer colours is not warranted, but both observations \citep{2007ApJ...655..144M,2009ApJ...694..556B,2015ApJ...808...91J} and theoretical models \citep{2001MNRAS.320..193B,2003AJ....125.2783C,2005AJ....130.2625R,2006A&A...458.1013M} predict a non-linear behaviour for bluer colours. This is illustrated in Figure~\ref{fig:sbf1}, where the 
absolute fluctuation magnitude is shown as a function of the $g_{475} - I_{814} $ colour for a variety of ages and metallicities (including the ones inferred from the fitting of the SED), and for two
sets of stellar evolutionary tracks, as predicted by the \citet{2016MNRAS.463.3409V} models. While the models agree rather well with the empirical calibration, they also show a dramatic drop
in the absolute fluctuation magnitude at colours bluer than $g_{475} - I_{814} \approx 1.0$, well below the values given by a linear extrapolation. This pattern does not depend strongly  on the filter set. Figure~\ref{fig:sbf2} shows  that, using the observed HST colour (without interpolations which depend upon the assumed or fitted SED), the absolute fluctuation magnitude is consistently brighter and would
therefore imply an even larger distance than the one derived by \citet{2018Natur.555..629V}, well beyond 25 Mpc (in tension with the spatially resolved stellar population of [KKS2000]04).

The predicted behaviour of the SBF magnitude at bluer colours depends on a number of factors which are difficult to assess. First, it is more sensitive to sampling fluctuations and statistical effects on the effective number of stars \citep{2003MNRAS.338..481C,2008A&A...491..693C} than older and redder populations, which is the reason why the SBF method is generally applied to globular clusters and elliptical galaxies. Second, theoretical predictions become somewhat unreliable at ages younger than about 3 Gyr and low metallicities \citep[see ][]{2001MNRAS.320..193B} mostly due to uncertainties in the modelling of stellar evolutionary phases such as intermediate-mass asymptotic giant branch stars and horizontal branch stars, \citep[e.g.][]{2003AJ....125.2783C}. Also, a given luminosity-weighted age does not guarantee that there are contributions from much younger populations which may also bias the measure of the sampled variance.

\begin{figure}
    \includegraphics[width=\columnwidth]{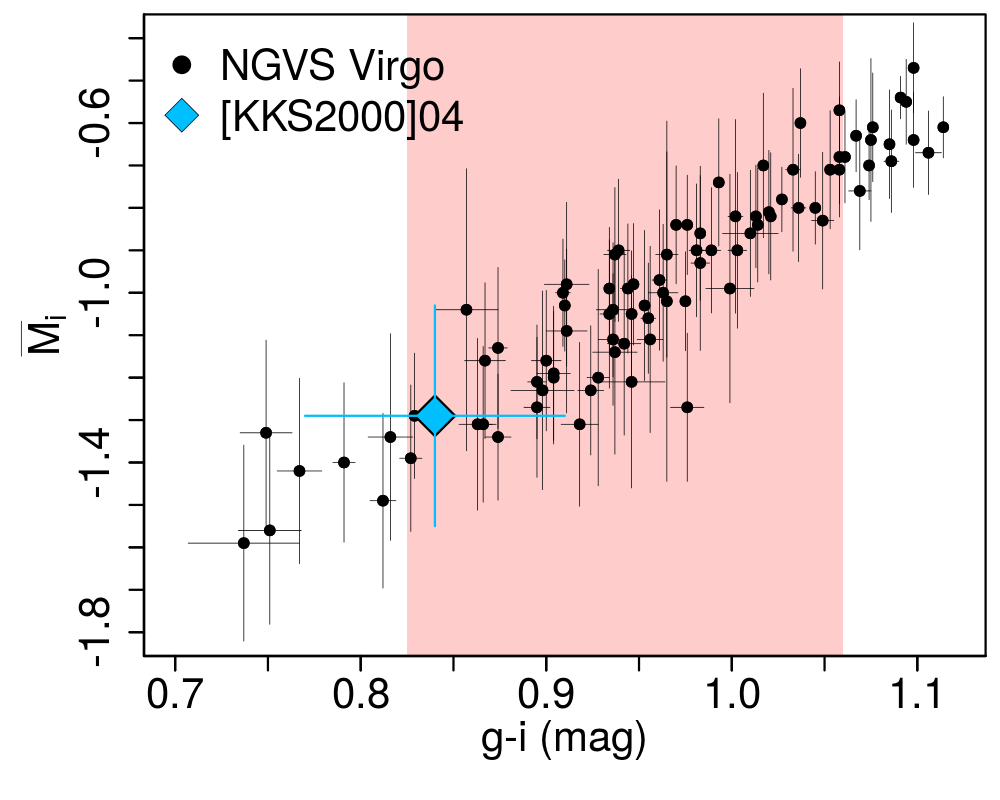}
    \caption{Absolute fluctuation magnitude in the \textsl{i} band as  a function of $g-i$ colour for the NGVS sample analysed by  \citet{2018ApJ...856..126C}. The range of validity of the calibration
    (Eq.~\ref{eq:cantiello}) is given by the pink area. \df2 (blue rhombus) is within this range, and the relation yields a distance of $14.7\pm1.7$ Mpc.}
   \label{fig:sbf4}
\end{figure}

\begin{figure}
   \includegraphics[width=\columnwidth]{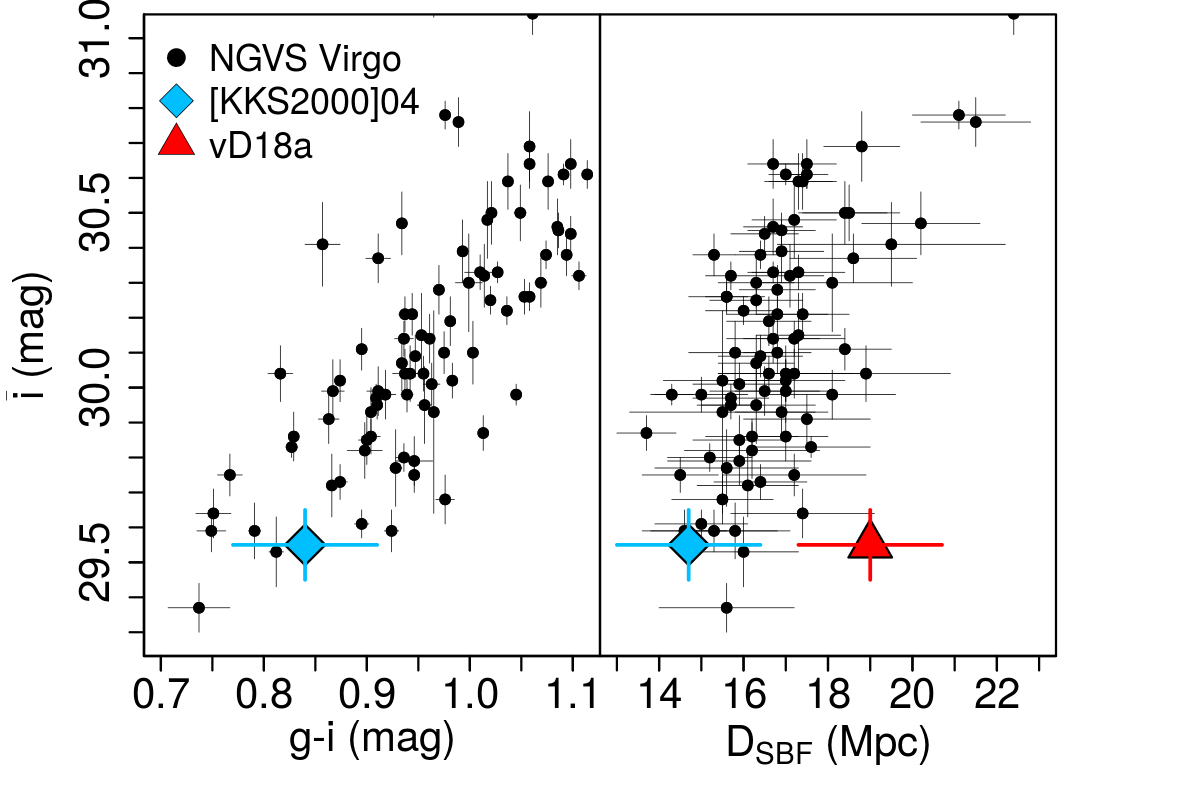}
    \caption{(Left panel) Apparent fluctuation magnitude in the \textsl{i} band as  a function of $g-i$ colour for the NGVS sample analysed by  \citet{2018ApJ...856..126C}. \df2 appears to follow the trend of the  Virgo galaxies. (Right panel) Apparent fluctuation magnitude in the \textsl{i} band as a function of distance. Once again, the distance inferred  for \df2 is well within the range expected, unlike the distance proposed by \citet{2018Natur.555..629V}.}
   \label{fig:sbf3}
\end{figure}

To bypass the uncertainties in the theoretical models, we use purely empirical SBF measures. While the SBF method has been applied to a variety of mostly red galaxies, there are very few attempts towards bluer colours \citep{2006A&A...458.1013M,2008ApJ...678..168B}, given the complications mentioned above. The only survey, as far as we are aware of, that extends SBF measurements systematically towards bluer galaxies is part of the Next Generation Virgo Cluster Survey (NGVS) \citep{2018ApJ...856..126C}\footnote{Very recently \citet{2019arXiv190107575C} have also explored  the SBF at blue colours using dwarf galaxies finding a nice agreement with \citet{2018ApJ...856..126C} in the common colour region.}. In this survey, carried out at the CFHT, the SBF analysis is performed in several filters, from \textsl{u$^*$} to \textsl{z}, and, in this first analysis to all galaxies brighter than $B=13$ mag. \citet{2018ApJ...856..126C} found several tight single-colour calibrations. The relation that uses reliable magnitudes available for \df2 is (see their Table 3):
\beq
\overline{M_i}  \; = \; (-0.93\pm0.04) + (3.25\pm0.42) \times\left[ (g-i)-0.95 \right]
\label{eq:cantiello}
\eeq
which is valid in the range $0.825 \leq (g-i) \leq 1.06$ (see Fig. \ref{fig:sbf4}). We find in this work (see Table \ref{table:sed}) $g-i$=0.84$\pm$0.07, and hence $\overline{M_i}$=-1.29$\pm$0.26. Finally, in order to estimate $\overline{M}_{814}$ we use the following correction $\overline{M}_{814}-\overline{M_i}$=-0.1 based on the models by \citet{2016MNRAS.463.3409V} for the best-fit SED. Hence, we obtain (m-M)$_0$=30.84$\pm$0.26 mag which corresponds to a distance of D$_{SBF}$=14.7$\pm$1.7 Mpc. 
%As \df2 is not detected in the \textsl{u*} and \textsl{z} bands, the dual-colour calibrations from \citet{2018ApJ...856..126C} cannot be used, but the slight improvement in precision would not lead to a significantly different distance.
Figure~\ref{fig:sbf3} shows that \df2 follows the trend traced by galaxies in the SBF analysis of the NGVS, and a distance
which is well within the observed correlation, unlike the one proposed by \citet{2018Natur.555..629V}.

An interesting outcome of this analysis is the future re-calibration of the relation between the fluctuation star count and dispersion velocity. The fluctuation star count $\overline{N}$ is a measure
of the luminosity-weighted number of stars or the total luminosity of a galaxy $L_{tot}$ in terms of the luminosity of a typical giant star \citep{2001ApJ...546..681T}, and is defined as 
\beq
\overline{N} \; = \; \overline{m} - m_{tot} \; = 
\; 2.5 \log\left( \frac{L_{tot}}{\overline{L}} \right) \, .
\eeq
Since it is measured in the same band, it has the obvious 
 advantage of being independent of extinction and zero points. Furthermore it correlates with the dispersion velocity for massive galaxies in the
 form
 \beq
 \log \sigma \; = \; 2.22 \, + \, 0.10 \times \left( \overline{N} - 20 \right) \, .
 \eeq
 In the case of \df2, the fluctuation star count in the \textsl{F814W}  filter is $\overline{N}_{814}= 13.78 \pm 0.26$, and the linearly-extrapolated dispersion velocity would be 39.6 \kms, an overestimation by a factor of less than 5 (assuming the velocity dispersion of the globular clusters of \df2 is representative of the velocity dispersion of the galaxy itself). With further measures of the dispersion velocity in low surface brightness galaxies, a new calibration of the relation can be envisioned, yielding redshift-independent distances.

\begin{figure}

\includegraphics[width=\columnwidth]{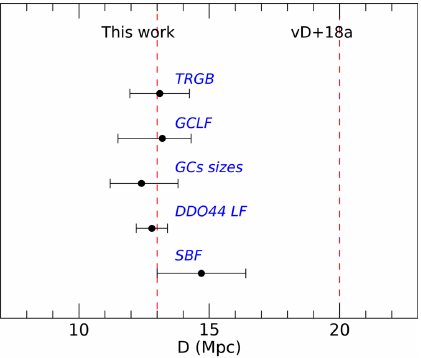}
\caption{The distance to [KKS2000]04 using five different redshift-independent methods (TRGB-the Tip of the Giant Red Branch, GCLF=the peak of the luminosity function of the GCs, GCs sizes=the effective radii of the GCs, DDO44 LF=the comparison with the LF of the stars of DDO44 and the surface brightness fluctuation (SBF)) and their uncertainties. All the determinations are compatible with a distance of $\sim$13 Mpc. The distance of 20 Mpc proposed by \citet{2018Natur.555..629V} is also indicated.}
\label{fig:distances}
\end{figure}

To summarise Section \ref{sec:distance}, up to five different distance indicators converge to a (weighted average) distance of 12.97$\pm$0.45 Mpc\footnote{Removing GC$_{new7}$ barely changes the weighted average distance to 13.03$\pm$0.45 Mpc.}  for [KKS2000]04. Figure \ref{fig:distances} summarizes the independent measurements obtained in this work and their uncertainties. In what follows we assume a distance of 13 Mpc as the most likely distance to [KKS2000]04. Under this assumption, the properties of the galaxy which depends on the distance are as follow: effective radius R$_e$=1.4$\pm$0.1 kpc, absolute V-band magnitude (M$_V$=-14.52$\pm$0.05 mag)  and total stellar mass M$_{\star}$=6.0$\pm$3.6$\times$10$^{7}$ M$_{\odot}$ (based on the (M/L)$_V$ obtained in the SED fitting). We have also fitted the surface brightness profile in the HST bands using S\'ersic models. The total magnitudes we retrieve using those fittings are compatible with the ones obtained using aperture photometry. With the new estimate for the size of the galaxy, the object would not longer belong to the category of UDGs but it would classify as a regular dwarf spheroidal galaxy.

\section{The velocity field of the galaxies in the line-of-sight of [KKS2000]04}
\label{sec:velfield}

\begin{figure*}
    \centering
    \includegraphics[width=\textwidth]{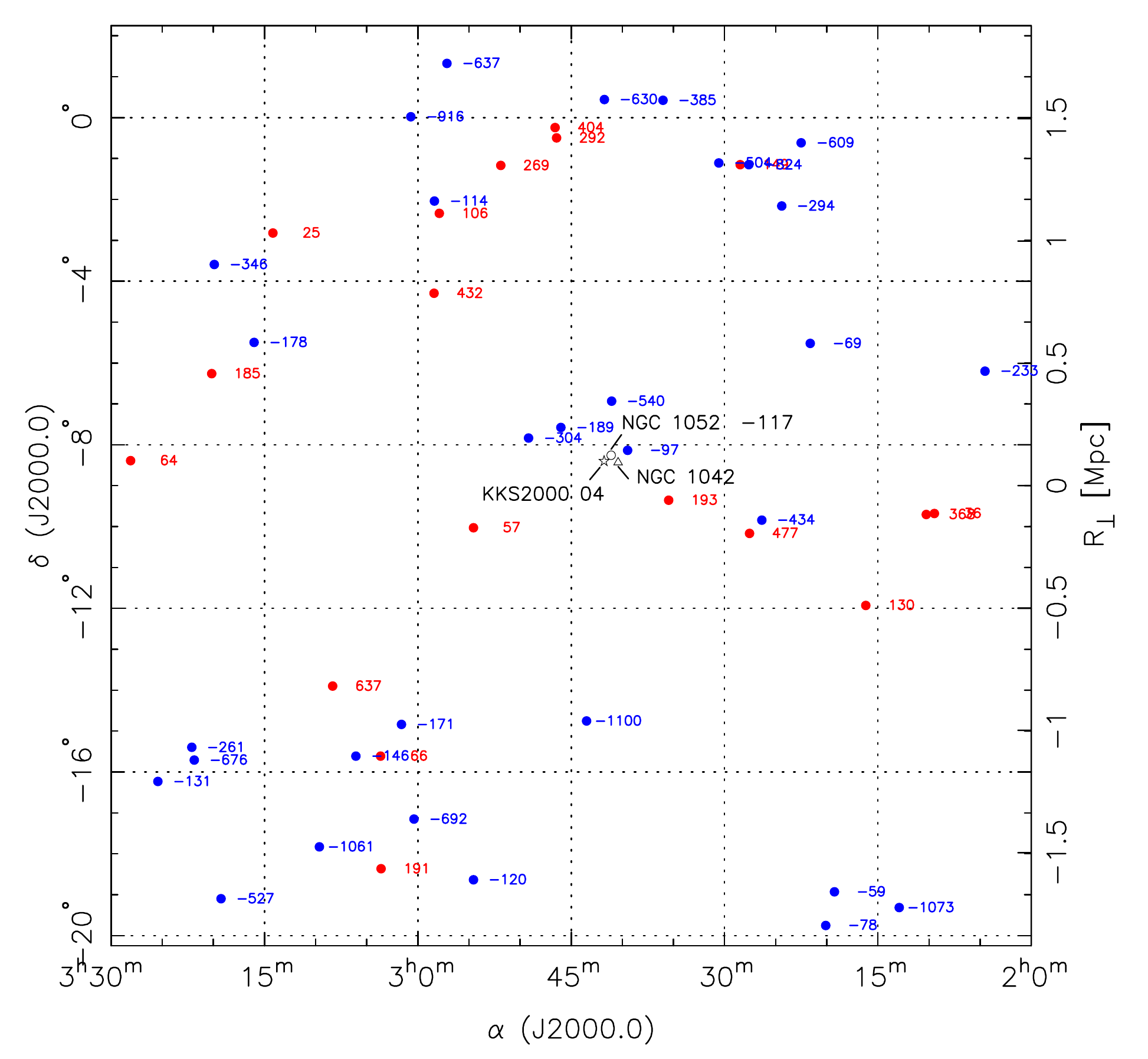}
    \caption{The peculiar velocity field around \df2, as traced by galaxies with heliocentric velocities in
the range $ 500 \kms < v_{hel} < 3000 \kms$. \df2, NGC~1042 and NGC~1052 are shown as star, triangle and open circle, respectively. Filled red circles give galaxies with positive peculiar
velocities,  while filled blue circles are galaxies with negative peculiar
velocities. The amplitude and sign of the peculiar velocity is given by the side of each galaxy. Note the large shear in the peculiar velocity field, as traced by galaxies with large positive peculiar velocities lying close
to galaxies with large negative peculiar velocities. 
 For reference, the right vertical axis provides the physical scale of the projected distance at 10 Mpc.}
    \label{fig:vpecsky}
\end{figure*}

At first sight, a distance of 13 Mpc might imply an implausible large peculiar velocity,  given the observed heliocentric velocity of $v_{hel} = 1803 \pm 2 \kms$ \citep[the mean velocity of the 10 GCs detected by][]{2018Natur.555..629V}. \citet{2018Natur.555..629V} argue that a distance as short as 8 Mpc would imply a large peculiar velocity of 1200 \kms assuming a velocity of $1748\pm 16\kms$ once corrected from the Virgocentric infall and the effects of the Shapley supercluster and Great Attractor. Given the uncertainties in the assumed corrections, the way to deal properly with peculiar velocities is to use the CMB reference frame \citep{1996ApJ...473..576F}, that is, 
\beq
v_{pec} \; = \; v_{CMB} \, - \, H_0 \cdot d
\eeq
with $d$ is the distance and $H_0$ is the Hubble constant, for which we adopt the value of 73$\pm2$~\hubble \citep{2012ApJ...745..156R,2016ApJ...826...56R}. In the CMB reference frame, the velocity of [KKS2000]04 is 1587 km s$^{-1}$, and its peculiar
velocity  640$\pm 25$ km s$^{-1}$, a seemingly large value but (as we will show) not unusually so. Interestingly, the field of galaxy velocities around [KKS2000]04 is rather complex. 
Figure~\ref{fig:vpecsky} shows the distribution of peculiar velocities for galaxies
with heliocentric velocities in the range $ 500 \kms < v_{hel} < 3000 \kms$, as given in the compilation
by the Extragalactic Distance Database \citep[EDD, ][]{2009AJ....138..323T}\footnote{\url{http://edd.ifa.hawaii.edu}},  and
in particular its latest version \texttt{cosmicflows-3}  \citep{2016AJ....152...50T}.

\begin{figure}
    \centering
    \includegraphics[width=\columnwidth]{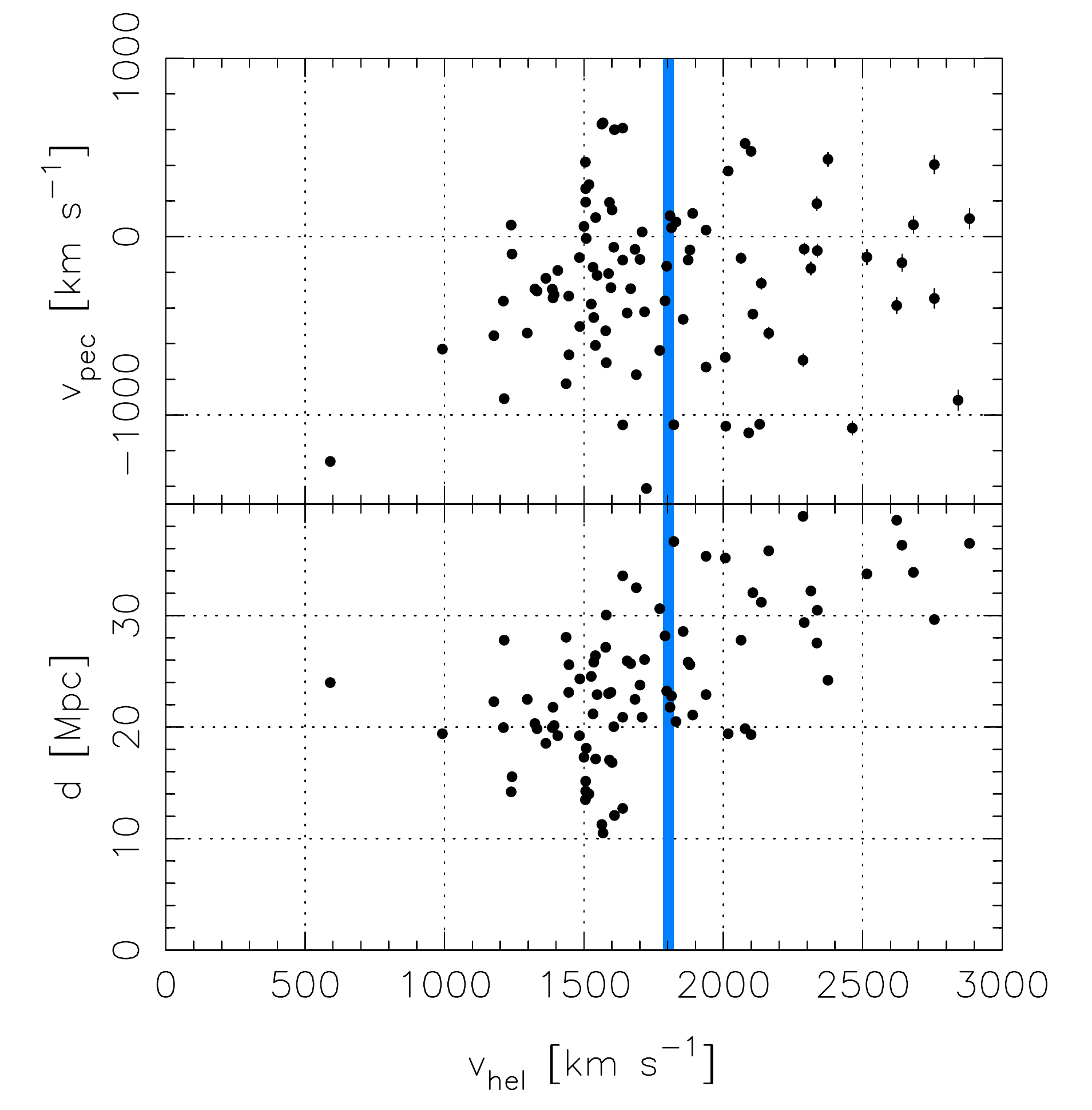}
    \caption{(Top) The peculiar velocity field in the area given in
    Fig.~\ref{fig:vpecsky} appears to show a very perturbed Hubble flow centred around $v_{pec} = -230~\kms$  superimposed on 
    infalling filaments and small-scale voids which give rise
    to an unusually wide distribution of peculiar velocities for
    a given heliocentric velocity. (Bottom) The measured
    distances to the galaxies in this field show no signature of
    large-scale Virgocentric-like flows. The vertical blue line gives the observed heliocentric velocity of \df2, which corresponds to a wide range
    of possible peculiar velocities and distances.}
    \label{fig:vpec2}
\end{figure}

Within a projected radius of 1.75 Mpc (at a distance of 10 Mpc), the peculiar velocity field shows huge variations, with galaxies with large infalling velocities in excess of $-500 \kms$ being angularly close to galaxies with large outflowing velocities, a reflection of strong gradients in the tidal tensor. There is a large-scale filament crossing the field with typical heliocentric velocities around $v_{hel} \sim 1600 \kms$, and another more extended structure around $v_{hel} \sim 2200 \kms$ 
with a local minimum around $v_{hel} \sim 1800 \kms$. This is also illustrated in Figure~\ref{fig:vpec2} which shows the distribution of measured peculiar velocities (top) and
distances (bottom) as a function of the observed heliocentric velocity, in the range of interest. Overall, the peculiar velocity field  shows an average peculiar velocity of $-230 \kms$, significantly larger than the expected null value, and a dispersion much larger than the standard pairwise velocity dispersion on these scales. The bottom panel shows no tell-tale S-shaped signature of a cluster 
(e.g., Virgocentric infall). This complex anisotropic pattern can be accounted for if several filaments (as traced by a number of groups present in this volume, see \S~\ref{sec:group}) and small voids coexist in this volume, giving raise to large amplitudes in the tidal tensor on small scales. The vertical blue line in Fig.~\ref{fig:vpec2} gives the observed heliocentric velocity of \df2, and shows that a wide range of peculiar velocities, from around $-1400$~\kms~ to +700~\kms~ is possible, although
only distances larger than 20 Mpc appear to be present. This, however, could be misleading, as the peculiar velocity field is not well-sampled and relies on the existence of redshift-independent distances in the catalog by \citet{2009AJ....138..323T}. The presence of large positive peculiar velocities around +600 \kms, and galaxies as close as 11 Mpc, within a small angular range and around the observed heliocentric velocity makes the likelihood of the expected peculiar velocity of \df2 much larger than anticipated.

\begin{figure}
    \centering
    \includegraphics[width=\columnwidth]{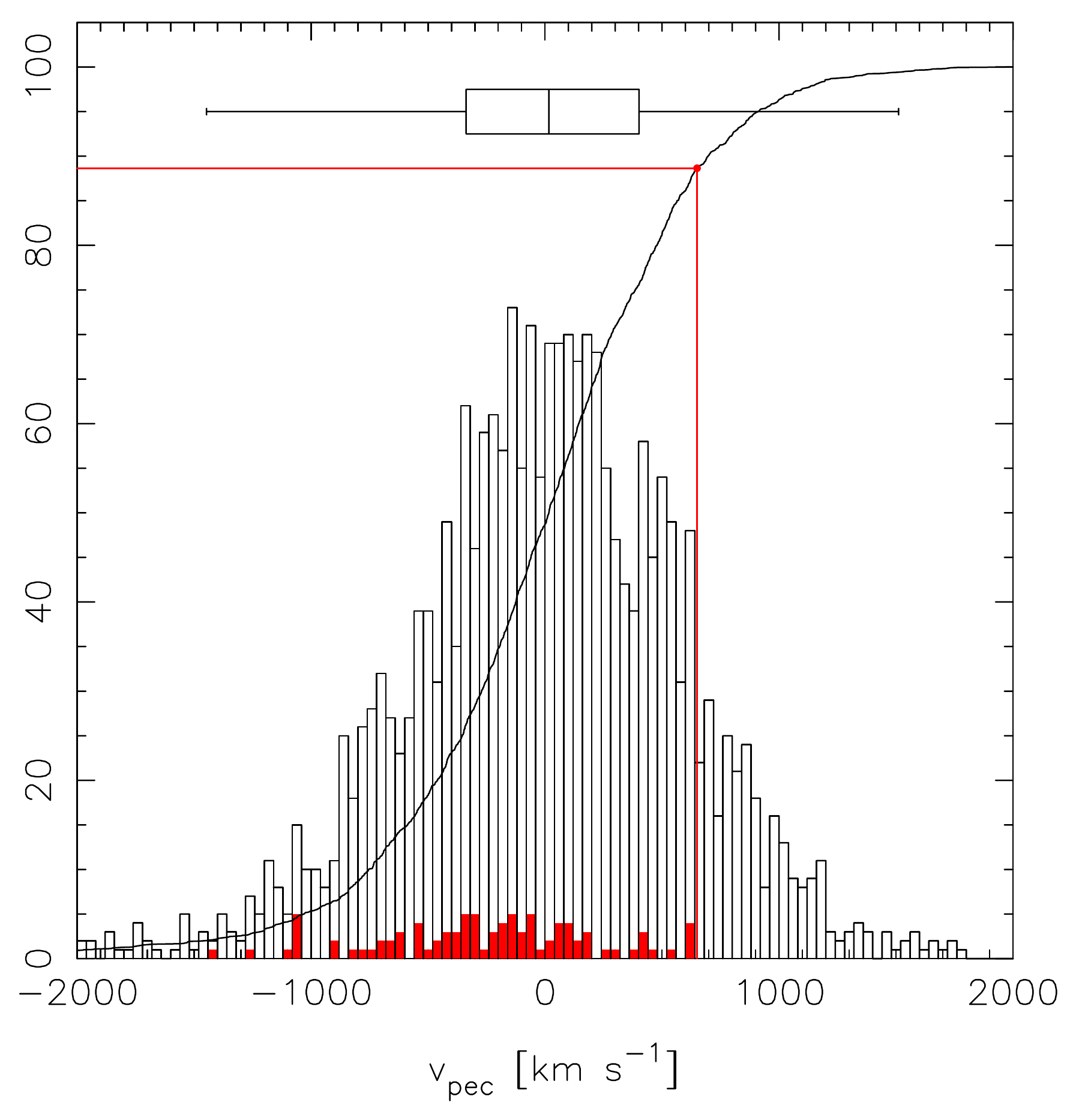}
    \caption{The distribution of peculiar velocities in the
    area given in Fig.~\ref{fig:vpecsky} (red histogram) and in
    the shell $500 \kms < v_{hel} < 3000 \kms$ (black histogram), as probed
    by the \texttt{cosmicflows-3} catalogue \citep{2016AJ....152...50T}. The
    cumulative distribution function (solid line) shows that
    a peculiar velocity of 640~\kms~ is rather likely (88\% quantile).
     The top box-whisker plot
    gives the median and interquartile range (IQR=Q3-Q1; upper/lower hinges), along with the
  lengths of the  whiskers at $Q_{1,3} \pm 1.5 \cdot IQR$. Q$_3$ and Q$_1$ are the upper and lower quartiles corresponding to the 75th and 25th percentiles of the distribution.} 
    \label{fig:vpechist}
\end{figure}

To quantify this, given the sparse sampling of the \texttt{cosmicflows-3} catalogue in this area, we compute the probability that the inferred peculiar velocity of \df2 could arise  in similar volumes of  the nearby universe. Figure~\ref{fig:vpechist} shows the distribution of peculiar velocities for the field at hand (red histogram), and the one for all \texttt{cosmicflows-3} galaxies in the range $ 500 \kms < v_{hel} < 3000 \kms$. The average peculiar velocity in this spherical shell is, as expected, very small $-24$~\kms, while its dispersion is 671~\kms, which would place the expected
 peculiar velocity of \df2 well within 1$\sigma$ (68\% confidence level) of the velocity distribution. The distribution, however, is not Gaussian, and a more robust measure of the dispersion is required.  Figure~\ref{fig:vpechist}  shows the box-whisker plot with the median (16~\kms), first ($Q_1$) and third ($Q_3$) quartiles at $-354$~\kms~ and 385~\kms~ respectively. For a Gaussian distribution the interquartile range (IQR) would be 905~\kms, while the observed one is much smaller at 739~\kms. 
Another robust measure of the dispersion is the median absolute deviation, MAD, given by  median$\left( \left| v_{pec, i} - \mathrm{median}(v_{pec,i})\right|\right)$, which is 369 \kms~ in this shell, in perfect agreement with the average of the first and third quantiles. Independently of the estimator of the dispersion, a peculiar velocity of $\sim$640~\kms~ appears to be rather likely in both the local universe and in the local volume around \df2, and is certainly expected within the context of the $\Lambda$CDM paradigm \cite[see][]{2017MNRAS.470..445N,2017MNRAS.467.2787H}.

\begin{figure}
    \centering
   \includegraphics[width=\columnwidth]{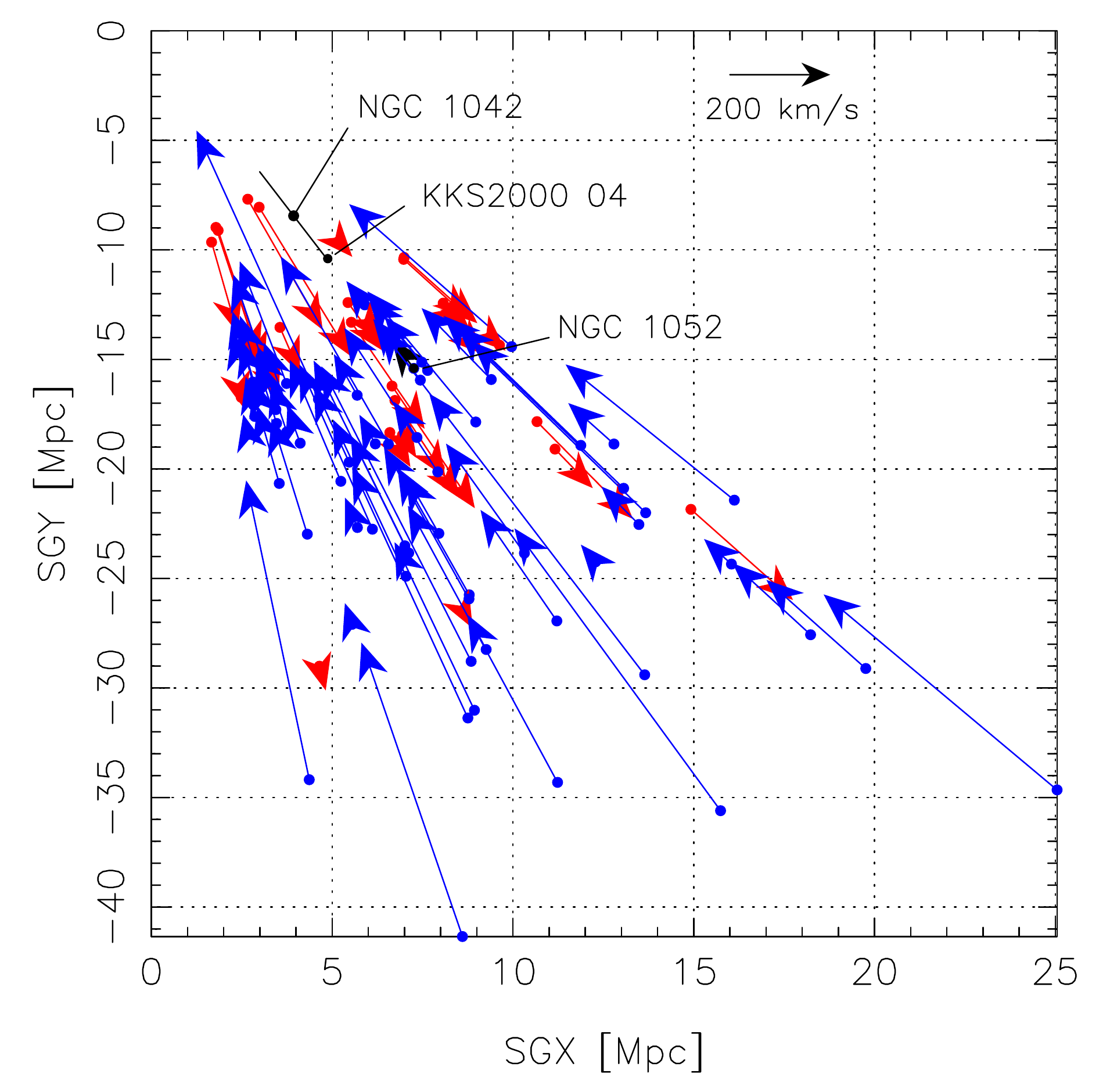}
    \caption{The peculiar velocity field of the sky area shown in Fig.~\ref{fig:vpecsky} projected on the super-galactic plane. The length of the arrow gives the scale of the peculiar velocity.}
    \label{fig:vpecSG}
\end{figure}

Another feature of Fig.~\ref{fig:vpecsky} is the large shear in peculiar velocities at small scales, as also illustrated in Fig.~\ref{fig:vpecSG}, where the peculiar velocity field is projected onto the super-galactic plane. The usual Wiener reconstruction of the field \citep[e.g.][]{2013AJ....146...69C} shows that this volume straddles a local void and the outskirts of the Fornax cluster, although the resolution of the reconstructed fields is not high enough at these small scales.

Still, the large differences between the peculiar velocities of \df2 and NGC~1052 or NGC~1042, which are so close by in projection, might be an issue. One may ask how likely is it that a large positive peculiar velocity of 640~\kms~ is found close to a galaxy with a large negative peculiar velocity, such as, say, NGC1052 (located at $\sim$20 Mpc) at $-117$~\kms?\footnote{Note that the peculiar velocity of NGC1042 is rather uncertain as its distance has not been well settled in the literature, with a likely range from 8 to 13 Mpc; see  discussion in Sec. \ref{sec:group}.}. These large differences are actually not that strange in that field, as there are many more extreme cases. For example, PGC013684 has $v_{pec} = -908~\kms$ and lies at 15\arcmin~ of  PGC013612 which has $v_{pec} = +630~\kms$.  Likewise NGC936 and NGC 941 are within 12\farcmin6 and yet their peculiar velocities differ by more than 970 \kms ($-824$ and 149 \kms respectively).

\begin{figure}
    \centering
    \includegraphics[width=\columnwidth]{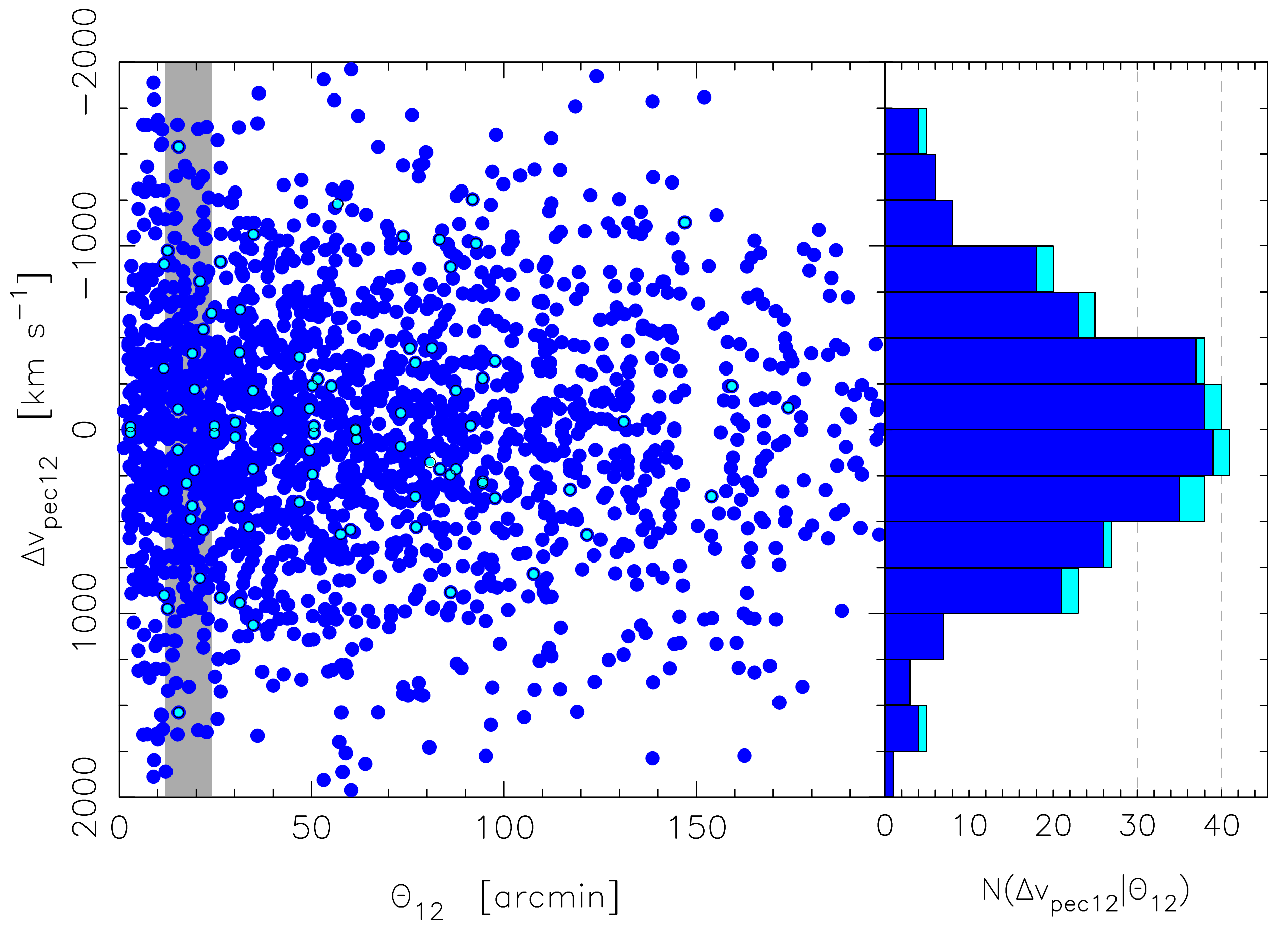}
    \caption{
    The difference in peculiar velocity of nearest
    neighbour galaxies, separated by an angular distance $\theta_{12}$, in the surveyed area (Fig.~\ref{fig:vpecsky}, blue circles) and in the shell $500 \kms < v_{hel} < 3000 \kms$ (filled circles). 
    The grey rectangle gives the range of angular distance
    between \df2 and its nearest neighbours (13\farcmin7 to NGC 1052 and 20\farcmin8 to NGC 1042), and shows that
    differences as large as $\pm 600~\kms$ are rather 
    common, as quantified by the histogram on the right hand side. }
    \label{fig:vpec12}
\end{figure}

To go beyond the classical estimation of the average line-of-sight dispersion \citep{1999ApJ...515L...1F}, we show in Figure~\ref{fig:vpec12} the difference in peculiar velocities 
between a galaxy and its nearest neighbour at an angular distance of $\theta_{12}$ as $\Delta v_{pec 12} = v_{pec 1} - v_{pec 2}$. In the shaded area which corresponds to the nearest (catalogued) galaxies around \df2, very large differences are found, ranging from $-1600~\kms$ to $+1520~\kms$. As the volume probed is small, Figure~\ref{fig:vpec12} also shows the distribution in the control sample. Once again, differences as large as $\pm600~\kms$ are well within 1$\sigma$ in that shell.

In summary, all the properties of the expected large peculiar velocity of \df2, $v_{pec} \sim 640~\kms$, appear to be not only rather normal in that particular  field, but also in the nearby universe, as traced in the state-of-the-art \texttt{cosmicflows-3} catalogue. Hence the distance of 13 Mpc cannot be ruled out on the basis of the inferred peculiar velocity. 

%Radial velocity (cz) with respect to the CMB radiation:
%NGC1042: 1155+-1 km/s
%NGC1052: 1268+-6 km/s
%[KKS2000]04 (NGC1052-DF2): 1588+-2 km/s

\section{[KKS2000]04 a member of the NGC~988 group?}
%----------------------------------------
\label{sec:group}

\begin{figure}

\includegraphics[width=\columnwidth]{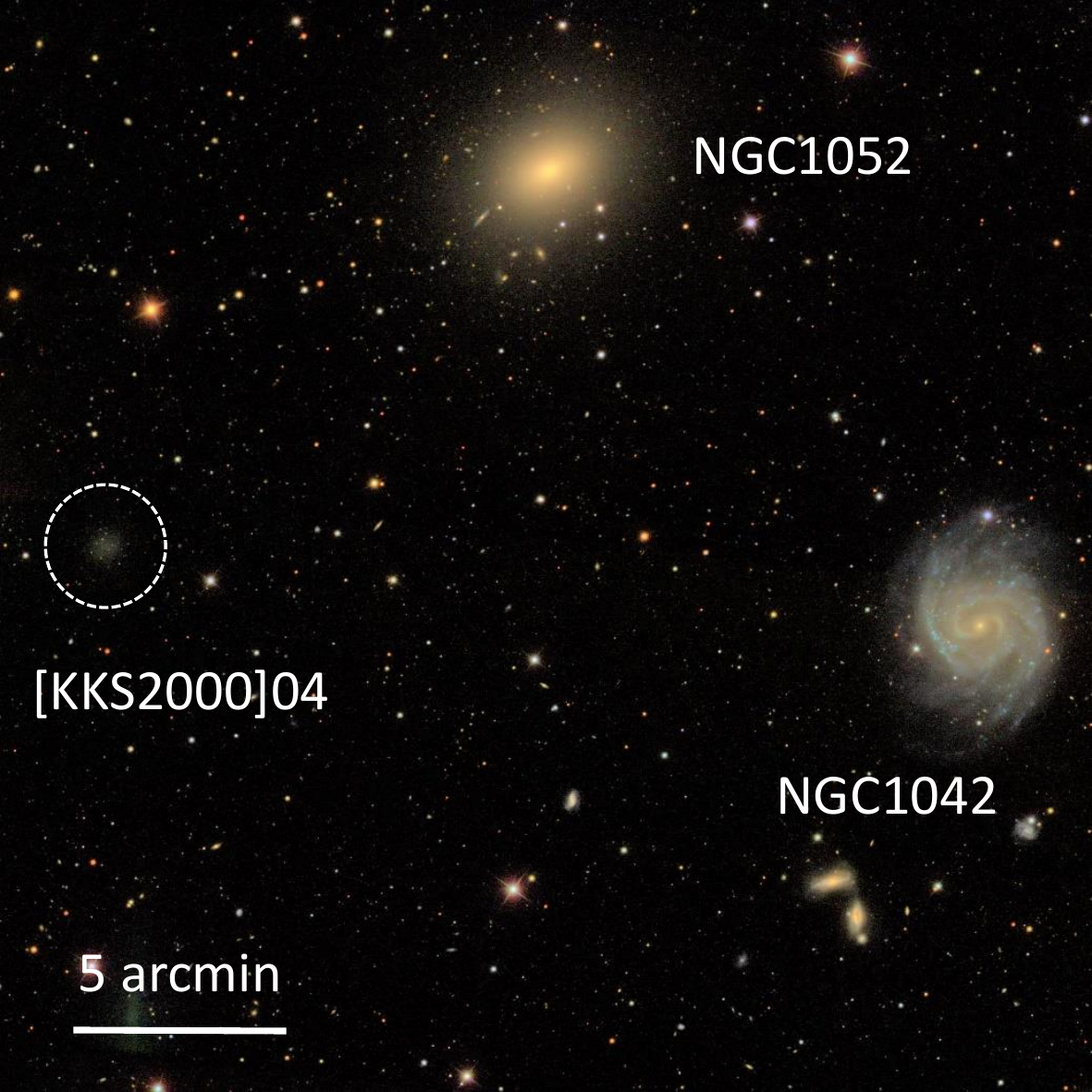}

\caption{Field of view showing the two closest (in projection) large galaxies around [KKS2000]04: NGC1052 (13\farcmin7) and NGC1042 (20\farcmin8). The colour image is taken from SDSS. The location of [KKS2000]04 has been highlighted with a dashed circle.}
\label{fig:ngc1042}
\end{figure}

The two most prominent galaxies in the line of sight of [KKS2000]04 are NGC1052 and NGC1042 (see Fig. \ref{fig:ngc1042}). \citet{2018Natur.555..629V} have used the vicinity (13\farcmin7 in projection) of [KKS2000]04  to NGC1052 to support their claim that [KKS2000]04 is at 20 Mpc\footnote{The  redshift-independent distance measurement to NGC1052 is $\sim$20 Mpc \citep[see e.g.][]{2001ApJ...546..681T,2001MNRAS.327.1004B}.}. If this were the case, [KKS2000]04 would be a satellite located at $\sim$80 kpc from NGC1052 \citep{2018Natur.555..629V}. However, these authors do not explore the possibility that the system could be physically linked to the also very close in projection (20\farcmin8) NGC1042 (cz=1371 km s$^{-1}$).

The distance to NGC1042 has been estimated in the literature following different approaches. Making use of the Tully-Fisher relation, a distance estimation independent of the redshift, several authors have found that the galaxy NGC1042 is located at $\sim$8 Mpc (8.4 Mpc \citet{1992ApJS...80..479T}; 7.8 \citet{2007A&A...465...71T}; 8 Mpc \citet{2016ApJ...823...85L}). Interestingly, \citet{2007A&A...465...71T}  obtained a distance of 13.2 Mpc reconstructing the proper motions of the galaxies in the local universe, which includes considering both the peculiar velocities of our Galaxy and NGC1042. Consequently, if [KKS2000]04  were a satellite of NGC1042 its distance would be within a range from $\sim$8-13 Mpc\footnote{It is worth noting that based on the Tully-Fisher relation, \citet{2008ApJ...676..184T} claimed that NGC1042 is located as close as 4.2 Mpc. This distance was based on the inclination of 57 $\deg$ they measured for this galaxy using the \textsl{B}-band. However, measurements both in the near infrared using 2MASS as well as kinematical measurements find a more likely inclination for the system of 37-38 $\deg$ \citep{2016ApJ...823...85L}, which moves the galaxy to a distance of 8 Mpc.}. If this were the case, the projected distance of [KKS2000]04 to NGC1042 would be between 49 to 80 kpc (i.e. physically closer than if the object were a satellite of NGC1052). Still, even if the NGC1042 were located at 13.2 Mpc a physical association with [KKS2000]04 could be debatable considering their difference in heliocentric velocities are $\sim$400 km s$^{-1}$. In Sec. \ref{sec:velfield} we argue that a difference in velocities as large as 400 km s$^{-1}$ are not implausibly high considering the small-scale shear in that region. Since the distance estimate to NGC1042 is still rather uncertain (8-13 Mpc), it is worth exploring whether there are other associations of galaxies in the line-of-sight of [KKS2000]04 and at a distance around 13 Mpc. 

As diffuse galaxies are normally found in clusters \citep[see e.g.][]{2003AJ....125...66C,2015ApJ...798L..45V,2015ApJ...807L...2K,2015ApJ...809L..21M,2015ApJ...813L..15M,2017A&A...608A.142V} and groups \citep{2017MNRAS.468..703R,2017MNRAS.468.4039R,2017ApJ...836..191T} and much more rarely in the field \citep{1997AJ....114..635D,2016AJ....151...96M,2017MNRAS.467.3751B,2017ApJ...842..133L} the probability that [KKS2000]04 belongs to a group should be high. 

Given that the first turnaround radii of groups of galaxies have typical sizes of a few Mpc \citep{2017ApJ...843...16K}, which also corresponds to the present scale  (Fig.~\ref{fig:vpecsky}), could  [KKS2000]04 be a member of a known group of galaxies?  The nearest
group is that containing NGC~988 which is also associated with NGC~1032 and NGC~1068 \citep{2011MNRAS.412.2498M}, and whose center \citep[R.A.(2000)=39.5716 and Dec(2000)=-8.4943;][]{2017ApJ...843...16K} is only 57\farcmin5 away from [KKS2000]04.  Interestingly, galaxies in this group and its associations span a wide range in redshift from  1160 km s$^{-1}$$<$v$_{hel}$$<$2750  km s$^{-1}$ and enclose a 
total mass of about $\sim$10$^{13}$ M$_{\odot}$ \citep{2011MNRAS.412.2498M}. The rich 
NGC~988 group (22 members with measured radial velocities) lies at a distance of 15.1 Mpc and has
$<$v$_{hel}$$>$~= 1550 km s$^{-1}$ \citep{2016AJ....152...50T} with a dispersion velocity of 143 km s$^{-1}$ \citep{2017ApJ...843...16K}. NGC988 has a peculiar velocity of +193  km s$^{-1}$. The difference in velocity with [KKS2000]04 is less than 450 km s$^{-1}$, and as shown in Fig.~\ref{fig:vpec12} differences in peculiar velocities as large as $\pm$450 km s$^{-1}$ at angular separations of order of the degree 
are very frequent in the nearby universe.

For [KKS2000]04 to be bound to the NGC~988 group itself, its total energy must be negative:
\beq
\frac{1}{2} \, \frac{ \Delta v_{12}^2 \; \Delta R_{12} }{G \, (M_1 + M_2)} \; < \; 1 \, ,
\label{eq:binding}
\eeq
where $\Delta v_{12}$ and $\Delta R_{12}$ are the differences in velocity and separation, respectively. M$_1$ and M$_2$ are the mass of the system and the group and G is the gravitational constant. Observationally, we have the projected separation, $R_{\perp} = 215 $ kpc (assuming a distance of 13 Mpc), the difference in radial velocities, $\Delta v_{pec 12}$ = 450 km s$^{-1}$, and also the group mass $M_{\mathrm{NGC\, 988}}$=7$\times$10$^{12}$ M$_{\odot}$ \citep{2017ApJ...843...16K}.  Since the mass of \df2 is much smaller by some three orders of magnitude,  we can safely assume that $M_{\mathrm{NGC\, 988}} \, + \, M_{\mathrm{KKS2000 \, 04}} \approx M_{\mathrm{NGC\, 988}} \approx 7 \times 10^{12} \, \Msun$. In terms of these  quantities, Eq.~\ref{eq:binding} becomes \citep[see Eq. (3) in][]{2011MNRAS.412.2498M}: 

\beq
\left( \frac{ R_{\perp} }{215 \, \mathrm{kpc} } \right) \;
\left( \frac{ \Delta v_{pec 12} }{450 \kms} \right)^2 \;  \lesssim \; 1.4 \; 
\left( \frac{ M_{\mathrm{NGC\, 988}}  }{7 \cdot 10^{12} \, \Msun} \right) \; , \label{eq1} 
\eeq
which indeed is fulfilled. This is not surprising given
that the projected separation of  $R_{\perp} = 215 $ kpc is smaller than the mean harmonic radius (a good measure of the effective radius of the gravitational potential of the system) of the NGC~988 group which is 379 kpc \citep{2011MNRAS.412.2498M}. In this sense, [KKS2000]04 could be bound to the NGC~988 group.

A second condition for [KKS2000]04 to belong to this group is that their separation must remain at least within the zero-velocity sphere \citep{1986ApJ...307....1S}, a condition which translates into the following inequality, again in terms of observable quantities \citep[see Eq. (4) in][]{2011MNRAS.412.2498M}: 
\beq
\left( \frac{ R_{\perp} }{215 \, \mathrm{kpc} }\right)^3 \; \lesssim \; 
1452 \; \left( \frac{ M_{\mathrm{NGC\, 988}} }{7 \cdot 10^{12} \, \Msun} \right) \; \left( \frac{H_0}{73 \hubble} \right)^{-2}
 \; .
\eeq
This relation also appears to be fulfilled, and provides tantalizing evidence that [KKS2000]04 might indeed belong to this galaxy group associated to NGC~988.

\section{Revisiting the total mass of [KKS2000]04}
\label{sec:mass}

On the basis of the adopted distance to [KKS2000]04 of 13 Mpc, we revisit how this new distance affects the estimate of the total mass of the system. The total mass of the system is derived from its dynamical mass M$_{\rm dyn}$.

To derive the dynamical mass of [KKS2000]04 based on its GCs, we used the Tracer Mass Estimator (TME) of \citet{2010MNRAS.406..264W}. The TME was also used by \citet{2018Natur.555..629V} and was first employed for UDGs in \citet{2016ApJ...819L..20B}. For [KKS2000]04 we have line-of-sight radial velocities and projected radii and so we use Eq. (26) of \citet{2010MNRAS.406..264W}. Three parameters enter into the TME prefactor: $\alpha$,  the power law index of the host potential, $\beta$, the velocity anisotropy parameter and $\gamma$, the power law slope of the volume density profile.

\begin{figure*}
\includegraphics[width=\textwidth]{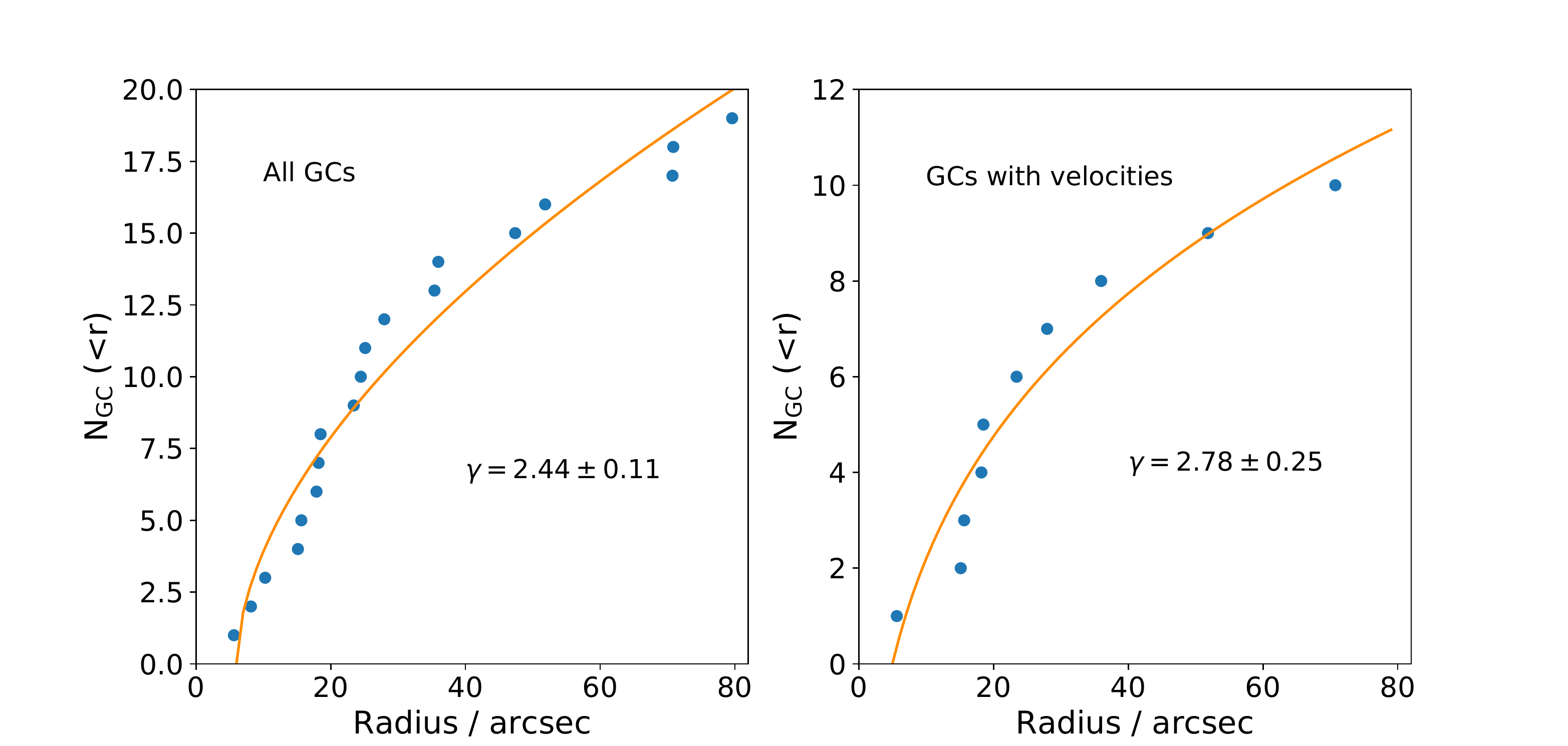}
    \caption{Cumulative radial distributions of GCs in [KKS2000]04 for the full sample of 19 GCs (left panel), and for the 10 GCs with radial velocities (right panel). The solid lines indicate fits to these data with the values of $\gamma$ indicated. The uncertainties given are the statistical uncertainties from linear, least-squares fits.}
    \label{fig:perfs}
\end{figure*}

Simulations constrain $\alpha$ to lie in the range between 0 and 0.6 \citep{2012MNRAS.423.1883D,2018arXiv180411348W} and we fix $\beta=0$ (which corresponds to isotropic orbits for the GCs). We measured $\gamma$ by fitting a 
power-law function of the form $n(r) \propto r^{3-\gamma}$ \citep{2010MNRAS.406..264W}
to the cumulative radial distributions of the GCs (Fig. \ref{fig:perfs}). We did this for the combined sample of 19 GCs, and also for the 10 GCs with velocities. For the sample of 19 GCs we obtain $\gamma=2.44\pm0.11$\footnote{Removing GC$_{new7}$ from the analysis changes only slightly the value of  $\gamma$, i.e. $\gamma=2.39\pm0.11$.}. We obtain consistent, though slightly higher values for the reduced sample of 10 GCs ($\gamma=2.78\pm0.25$). This approach is preferable to measuring $\gamma$ from a differential volume density plot since it is not dependent on the choice of binning. We performed linear least-squares fits to the linearised power-law function. Since the distribution is cumulative and the points are not independent, the
formal uncertainties on the fit are generally underestimated. To account for this we performed bootstrap sampling with replacement where we randomly
sampled the observed cumulative distribution function and performed linear regression. The quoted uncertainties are the 16 and 84 percentiles of the
distributions. We note that the cumulative spatial distributions of the [KKS2000]04 GCs  are not particularly well described by a single power-law, however this
is the form required by the \citet{2010MNRAS.406..264W} mass estimator \citep[see on this point][]{2018MNRAS.481L..59H}.

From the same sample of 10 GCs, \citet{2018Natur.555..629V} measured $\gamma=0.9\pm0.3$ which is inconsistent with our measurements. For a fixed $\alpha$ and $\beta$, the inferred mass from the TME scales linearly with $\gamma$. Therefore, our TME mass measurements are higher than \citet{2018Natur.555..629V} even though we place [KKS2000]04 at 13 rather than 20 Mpc (see below).

\begin{figure}
\includegraphics[width=\columnwidth]{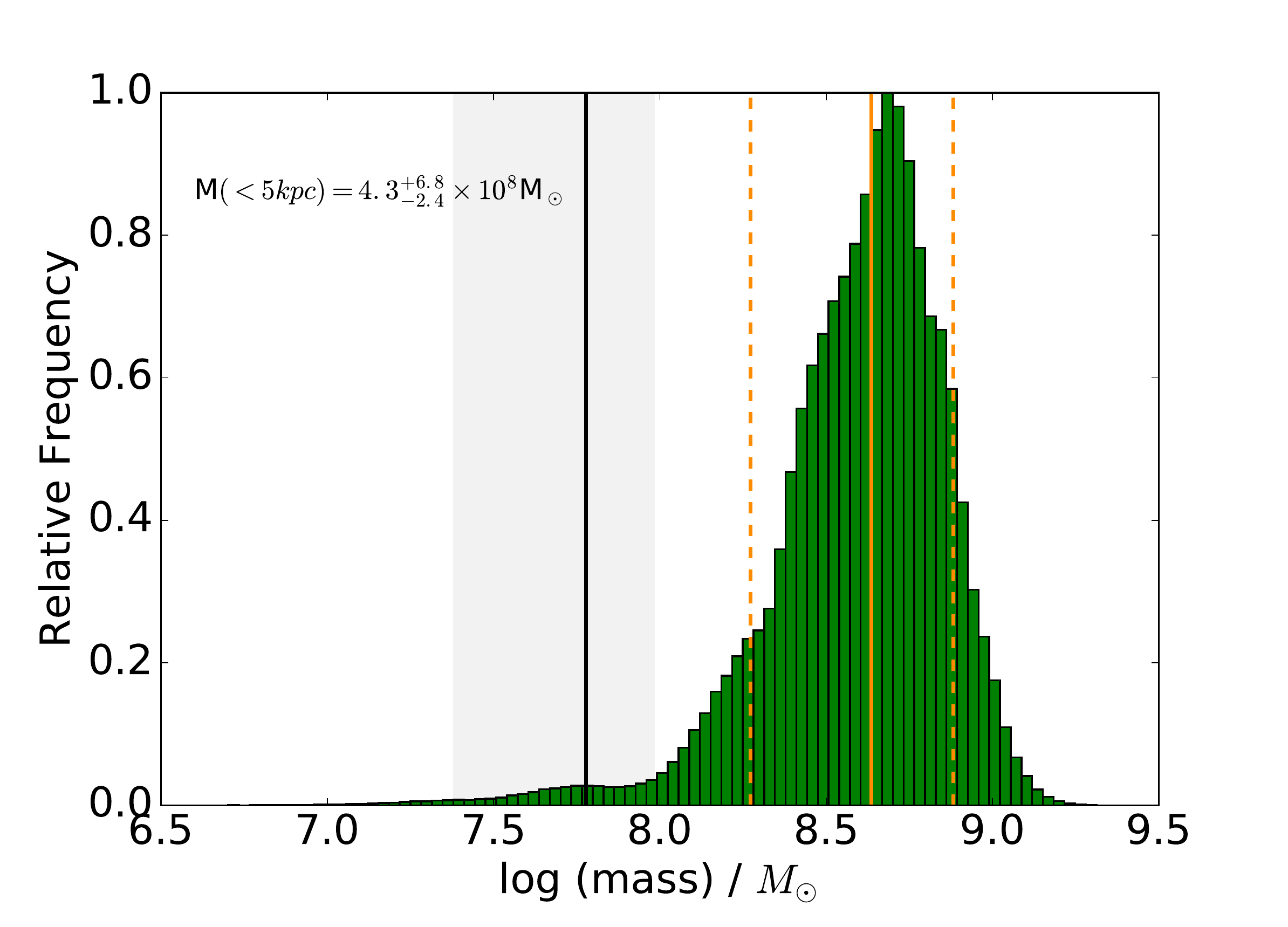}
    \caption{Distribution of TME (i.e. dynamical) masses for [KKS2000]04 from Monte Carlo simulations for D = 13 Mpc. The solid orange line indicates the median mass of the distribution, the dashed lines indicate the 10, 90 percentiles of the distribution. The vertical black line indicates the stellar mass of [KKS2000]04 obtained in this work and the gray shaded region indicates the range of  uncertainties on the stellar mass.}
    \label{fig:mass_monte}
\end{figure}

We explored the range of inferred masses by performing Monte Carlo simulations where $\alpha$ was allowed to vary between 0 and 0.6, and $\gamma$ vary between 2.33 and 2.55 (the range in our measured profile slopes for the full sample of 19 GCs). GC velocities and projected radii were selected randomly with replacement and the TME mass recalculated each time. The results of these simulations are shown in Fig. \ref{fig:mass_monte}. We obtain a 
median mass within 5 kpc, M$_{\rm dyn}=4.3^{+6.8}_{-2.4}\times10^8\rm M_{\odot}$, as constrained from the 10 GCs with velocities\footnote{\citet{2018arXiv180604685V} have very recently remeasured the velocity of GC98. With the new estimation, our dynamical mass within 5 kpc changes slightly: M$_{\rm dyn}=2.9^{+4.5}_{-1.3}\times10^8\rm M_{\odot}$.}. The uncertainties are the 10, 90 percentiles of the distribution of masses. Also indicated in Fig. \ref{fig:mass_monte} is the stellar mass and its uncertainty for [KKS2000]04. The figure shows that TME masses are generally higher than the stellar mass of [KKS2000]04. Less than 3\% of the realizations overlap with the permitted range of stellar masses. The results of the obtained dynamical masses as a function of distance from galactic center are presented in Tab. \ref{tab:TME_mass_radius}.

\begin{table}
\centering
\caption{Dynamical (median) mass (M$_{dyn}$) of [KKS2000]04 computed within different radial distances from the galaxy center,  using an increasing number of GCs (N$_{GC}$($<$r)). The second and fourth column represent the 10th and 90th percentile of the distribution of TME, as presented in Fig. \ref{fig:mass_monte}.}
\label{tab:TME_mass_radius} 
\begin{tabular}{ccccc}
\hline
\hline
r (kpc) & 10$\%$ile & $\rm M_{\rm dyn}(<r)$[$10^{8}$M$_{\odot}$]   & 90$\%$ile & N$_{\rm GC}$(<r) \\
\hline
1.47  & 0.29 & 1.00 &	 	1.52 &   4\\ 
2.23  & 0.95 & 2.59 &	 	4.56 &	 6\\
3.27  & 1.23 & 2.88 &  	    5.02 &	 8\\
4.46  & 1.36 & 3.14 &       5.49 &   9 \\
5.02  & 2.39 & 4.32 &	  	6.83 &	10\\
\hline
\hline
\end{tabular}
\end{table}

\begin{figure*}
\includegraphics[width=\textwidth]{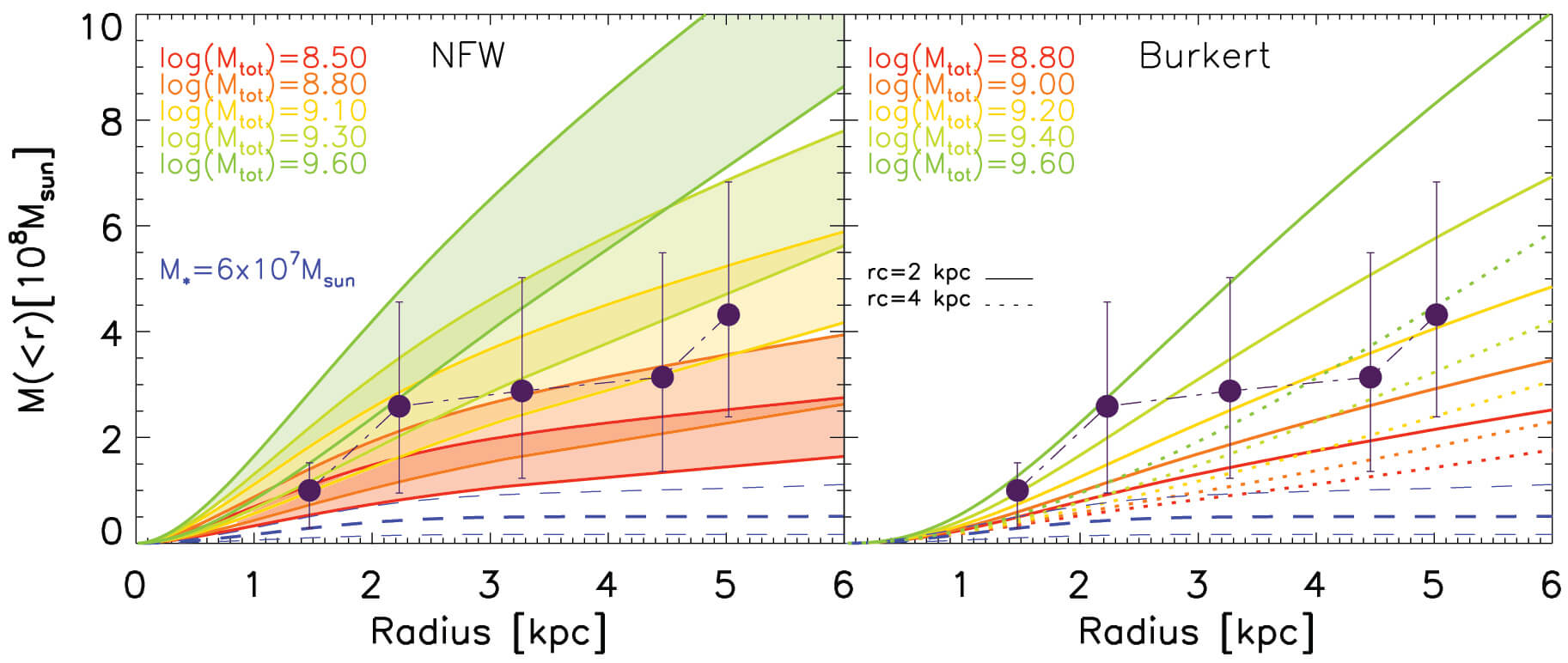}
    \caption{The dynamical mass of [KKS2000]04 vs. distance from the galaxy center. Each dark circle corresponds to measurements of the dynamical mass at different radial distances, between the effective radius  and the radius of the outer-most GC ($\sim$5 kpc). The data-points are correlated as indicated by the connecting dashed line. Vertical lines indicate the 10th and 90th percentile in the derivation of M$_{\rm dyn}$. The contribution from the stellar surface density is indicated as dashed blue lines, including the uncertainty in the (M/L)$_V$. \textit{Left panel:} NFW  mass  profiles for halos of increasing masses are shown as  colored shaded areas. For each DM halo, we indicated the upper and lower bound in their M($<$r), taking into account the c-M relation scatter and the uncertainties in the stellar component. The best NFW halo  that describes the dynamical mass of [KKS2000]04 for increasing radii has a total mass of M$_{\rm tot}\sim10^{9.1}$M$_{\odot}$. \textit{Right panel:} Burkert cored mass profiles for haloes of increasing masses, using a core radius $r_c$ of 2 (solid lines) and 4 kpc (dashed lines) respectively. The largest the core the more massive is the DM halo that can fit [KKS2000]04 data. The best halo describing  [KKS2000]04, at its outermost measured point, is M$_{\rm tot}\simeq10^{9.25}$M$_{\odot}$, for a core radius of 2 kpc, and M$_{\rm tot}\simeq10^{9.6}$M$_{\odot}$, for $r_c$= 4 kpc.}
    \label{fig:M_r}
\end{figure*}

To estimate the dark matter (DM) halo mass, in Fig. \ref{fig:M_r} we show as dark points the dynamical masses of [KKS2000]04 as a function of radius, from the effective radius, R$_{e}=$1.4$\pm$0.1 kpc, to the position of the outer-most globular cluster, R$_{out}$=5 kpc, constrained using the above mentioned \citet{2010MNRAS.406..264W} TME. Error bars indicate  the 10, 90 percentiles of the distribution of masses, as shown in Fig. \ref{fig:mass_monte}, and should  therefore not be confused with the 1$\sigma$ uncertainties. The mass profiles of dark matter haloes of different masses  are overplotted as colored shaded areas, taking into account  the cosmic baryon fraction. In the left panel, we assume a NFW profile \citep{1997ApJ...490..493N} and the Planck cosmology concentration-mass (c-M) relation from \citet{2014MNRAS.441.3359D}. Scatter in the c-M relation has been included, using a value of $\sigma$=0.11 dex. In the right panel we instead employed a Burkert profile \citep{1995ApJ...447L..25B}, that well describes the matter density distribution in the presence of a central core. We used two values for the free parameter $r_c$, i.e. core radius, namely 2 and 4 kpc\footnote{Note that the value $r_c$ represents roughly the radius at which the Burkert density profile becomes shallower than a NFW model, such that the actual core, defined as the position where the log slope of the profile is zero, it is  smaller than $r_c$.}.

The reason for using both a NFW as well as a cored profile, is that analytic calculations have shown that even in a  DM halo as small as 10$^9$ M$_{\odot}$ it is possible to form a central shallow core, proven that there is enough energy from stellar feedback \citep{2012ApJ...759L..42P,2012MNRAS.421.3464P}. In \citet{2014MNRAS.437..415D} this point was made explicitly by using cosmological simulations, showing that the core formation efficiency depends ultimately on the ratio M$_{\star}$/M$_{\rm halo}$, rather than on the halo mass itself: if a galaxy has a larger-than-expected amount of stars compared to abundance matching results, it will be able to flatten its central matter density. More recently, \cite{2016MNRAS.459.2573R} showed a similar example by simulating a dwarf living in a 10$^9$ M$_{\odot}$  DM halo and having as many as 3.5$\times$10$^6$M$_{\odot}$ in stars: with more stars than what predicted by abundance matching relations for a similar halo mass, their dwarf was able to create a central shallow core of size $\sim$1 kpc.  For this reason, it is reasonable to think that a galaxy such as [KKS2000]04, with one order of magnitude more stars than the simulated \cite{2016MNRAS.459.2573R} example, could easily be able to form a 1 kpc or a even larger core via SNae feedback.

To estimate the total mass of [KKS2000]04 we need to account for the contribution to the mass provided by the stellar mass. The stellar density profile and its   upper and lower bounds, coming from uncertainties in the derivation of the  mass-to-light ratio,  are shown  as  dashed blue lines in Fig. \ref{fig:M_r}. To derive the surface stellar density profile we followed the method described in \citet[][see Eq.\,(1)]{2008ApJ...683L.103B}. We have used the surface brightness profile obtained in the band F606W (corrected by Galactic extinction) as a proxy for estimating the surface brightness profiles in the V-band. Following the SED fitting presented in Sec. \ref{sec:sed}, we used $(M/L)_{V}$ = 1.07$^{+0.80}_{-0.54}$. The uncertainties from sky noise and the intrinsic variation on the elliptical aperture were calculated by using Monte Carlo and bootstrapping simulations. Finally we combine these errors with those from the mass-to-light ratio $(M/L)_{V}$, which dominate the final uncertainties of the stellar mass density. As expected, the total stellar mass for [KKS2000]04 using the surface brightness profile (M$_{\star}$=5.0$^{+3.9}_{-2.5}$$\times$10$^7$M$_{\odot}$) is in agreement with the one using aperture photometry.

To constrain [KKS2000]04's highest (lowest) allowed mass in the case of NFW DM haloes, including the contribution of the stars, we added the upper (lower) bounds of each DM halo to the upper (lower) bounds of the stellar distributions: in this way, each colour contour  represents the highest and lowest possible mass profile associated to a specific DM halo, once the  1$\sigma$ error in the c-M relation and the uncertainties in the derivation of the stellar mass profile have been considered.

The  best  NFW halo mass describing the dynamical mass of [KKS2000]04  at every radius is  M$_{\rm tot}$$\sim$10$^9$ M$_{\odot}$. Specifically, a NFW halo of mass M$_{\rm tot}=1.26\times10^{9}$M$_{\odot}$ best fits  the outermost dynamical mass measured at $\sim$5 kpc. Altogether, the  range of halo masses spanned by the derived dynamical mass of [KKS2000]04, including its uncertainties, varies between M$_{\rm tot}$$\simeq$10$^{8.5}$M$_{\odot}$ and M$_{\rm tot}$$\simeq$10$^{9.6}$M$_{\odot}$. Similarly, we constrained the best halo mass  for [KKS2000]04 to be  M$_{\rm tot}$=$10^{9.25}$M$_{\odot}$ and M$_{\rm tot}$=$10^{9.6}$M$_{\odot}$ in the case of assuming a Burkert profile with a 2 or 4 kpc scale-radius, respectively. Clearly, the larger the core radius the higher the halo mass that can fit the data.

\begin{figure*}
	\includegraphics[width=\textwidth]{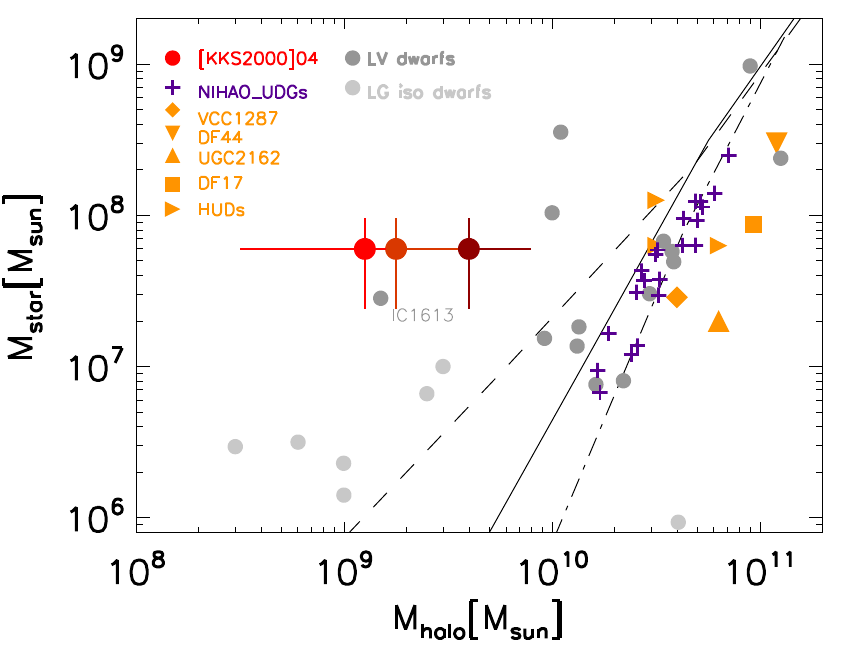}
    \caption{The stellar mass-halo mass relation of [KKS2000]04, shown as red circle with error bars. From dark red to bright red, we show our best mass estimates assuming a Burkert profile with core radius of $r_c$=4 and 2 kpc, and a NFW profile. Other known UDGs for which a constraint on halo mass exists are shown as orange symbols (see text for more details), while simulated UDGs from the NIHAO project \citep{2015MNRAS.454...83W, 2017MNRAS.466L...1D} are indicated as purple crosses. Expectations from several abundance matching relations are shown as solid \citep{2013MNRAS.428.3121M}, dashed \citep{2013ApJ...770...57B} and dot-dashed \citep{2014ApJ...784L..14B} lines. Isolated LG dwarfs \citep{2015MNRAS.450.3920B} and Local volume dwarfs \citep{2016MNRAS.460.3610O} with extended rotation curves are shown as light and dark  gray points, respectively. Both datasets assume an underlying NFW profile. We have highlighted the location of IC1613, a galaxy with some characteristics similar to [KKS2000]04.}
    \label{fig:AM_rel}
\end{figure*}

In Fig. \ref{fig:AM_rel} we show the expected stellar mass-halo mass relation from abundance matching techniques, together with the derived M$_{\star}$-M$_{\rm halo}$ of [KKS2000]04, as well as other UDGs and local dwarfs. We show three different abundance matching results as  solid \citep{2013MNRAS.428.3121M}, dashed \citep{2013ApJ...770...57B} and dot-dashed \citep{2014ApJ...784L..14B} lines. Note that while the \citet{2014ApJ...784L..14B} relation has been constrained using Local Group data, and reaches M$_{\star}\sim3\times10^6$ M$_{\odot}$, the other relations are simply extrapolated to such low masses. It is worth noting that the effect of the environment could play an important role on this relation at such low masses \citep[see e.g.][]{2017MNRAS.467.2019R}.

Different estimates of the total mass of [KKS2000]04  are shown as red circles. Specifically, from dark to light red we indicate the estimated mass assuming a Burkert profile with $r_c$=4 and 2 kpc, and a NFW profile (lowest mass estimate, bright red point). Uncertainties in  stellar  and halo mass are indicated as red error bars: the uncertainty in the halo mass represents the highest and lowest possible M$_{\rm tot}$ compatible with the highest and lowest mass estimates of [KKS2000]04 at $\sim$5 kpc, as in Fig. \ref{fig:M_r}. Other known UDGs for which a constraint on halo mass exists are shown as orange symbols. Data for VCC1287, UGC2162, DF44 and HUDs  (HI-bearing Ultra Diffuse sources) are taken from \citet{2016ApJ...819L..20B}, \citet{2017ApJ...836..191T}, \citet{2016ApJ...828L...6V} and \citet{2017ApJ...842..133L}, respectively. For these objects, the best fit halo masses have been determined (or reassessed) by comparing data with mass profiles of NIHAO hydrodynamical simulations, as in Fig. 2 of \citet{2017MNRAS.466L...1D}.
The mass of DF17, from \citet{2016ApJ...822L..31P} and \citet{2016ApJ...830...23B},  is instead based on GC counts. Simulated UDGs from the NIHAO  project \citep{2015MNRAS.454...83W,2017MNRAS.466L...1D} are shown as purple crosses.
Light gray circles represent Local Group isolated galaxies whose masses have been estimated from their stellar velocity dispersions at R$_e$ as in \cite{2015MNRAS.450.3920B}, assuming a NFW profile,  while dark gray circles are Local Volume dwarfs with HI rotation curves extending to at least two disk scale lengths, as presented in \citet{2016MNRAS.460.3610O}, and for which we have re-computed the expected halo mass in a Planck cosmology.

While [KKS2000]04 appears to be an outlier in the M$_{\star}$-M$_{\rm halo}$ relation, its inferred dark matter mass is clearly much larger than that obtained in \citet{2018Natur.555..629V}, making this galaxy a DM dominated object. [KKS2000]04 is not the only example of galaxy showing a lower than expected halo mass: IC1613, highlighted in the plot, is an example of galaxy with a similar DM content, as already noted in \citet{2016MNRAS.460.3610O} (see, however, \cite{2015MNRAS.450.3920B} for a larger  estimation of the total mass of IC1613, assuming an underlying cored profile). While IC1613's rotation curve seems  difficult to reconcile within $\Lambda$CDM expectations,  a possible explanation could be that the inclination errors have been underestimated or even that the inclination itself has been incorrectly measured \citep[see e.g.][]{2016MNRAS.462.3628R}, which compromises the  inferred mass profile. In the same way, caution should be taken when interpreting the derived mass of [KKS2000]04: the low number of GCs on which the TME is based might provide biased results (see \cite{2018ApJ...859L...5M,2018MNRAS.tmp.2765L} for a recent discussion).

We find that the amount of dark matter in \df2 is not consistent with zero, as claimed by \citet{2018Natur.555..629V}. We can show this one more time estimating the actual fraction of dark matter
inside its 3D half-light radius $r_{1/2}$ following the same procedure as \citet{2016ApJ...828L...6V}. First, we estimate the
stellar mass from the empirical correlation found in the GAMA survey \citep[see Eq. 8 in ][]{2011MNRAS.418.1587T}, namely
\beq
log M_{\star} \; = \; 1.15 \, + \, 0.70\times(g-i) \, - \, 0.4 M_i \; , \label{eq:gama}
\eeq
where $M_{\star}$ is in solar units and $M_i$ is the absolute AB magnitude in the \textsl{i} band. Our
accurate Gemini \textsl{g} magnitude combined with the SDSS \textsl{i} magnitude (see Table~\ref{table:sed}) yields $g-i=0.84\pm0.07$, while the absolute magnitude at the revised distance of 13 Mpc becomes $M_i=-15.03\pm0.07$. From equation~\ref{eq:gama}, the stellar mass of \df2 is $M_{\star}$ =5.6$^{+1.3}_{-1.0}$ \,$\times$ 10$^7$ $M_{\odot}$, where we have adopted a 1$\sigma$ accuracy of $\sim$0.1 dex \citep{2011MNRAS.418.1587T}. This independent estimate agrees with the mass inferred from the SB profile and from the fitting of the SED we have derived in this paper.

Second, we can estimate the mass enclosed within the 3D half-mass radius assuming that \df2 has an
approximately isothermal profile. In this case, we assume $\sigma_{\star} \approx \sigma_{GC} \sim 8 \kms$, and, as argued by \citet{2010MNRAS.406.1220W} the deprojected 3D circularised half-mass radius is\footnote{Note that assuming an isothermal profile implies that the mass we are deriving here is a lower limit of the total mass, as we would expect that $\sigma_{\star}$ should be higher than  $\sigma_{GC}$ under this assumption.}
\beq
r_{1/2} \; \approx \; \frac{4}{3} \, R_{e} \, \sqrt{\frac{b}{a}} \,
\eeq
where $R_{e}$ is the effective radius from the observed SB profile along the semi-major axis, $b/a$ the axis ratio. Adopting $R_{e}$=1.4 kpc and $b/a=0.85$ yields $r_{1/2} = 1.72$ kpc. The mass enclosed within the 3D half-light radius \citep[see Eq. 2 in ][]{2010MNRAS.406.1220W} becomes
\beq
M\left(r<r_{1/2}\right) \; \gtrsim \;  10^8 \,M_{\odot}  \, \left( \frac{\sigma_{\star}}{8 \, \kms}\right)^2 \,\left( \frac{ r_{1/2} }{ 1.72 kpc} \right) \; .
\eeq
Hence the dark matter fraction within $r_{1/2}$ is $f_{DM} = (M\left(r<r_{1/2}\right) -0.5M_{\star})/M\left(r<r_{1/2}\right) \gtrsim$ 75\%. It is worth noting that using the $\sigma_{\star}$ value measured by \citet{2018arXiv181207345E}, the dark matter fraction for this galaxy rises to $\gtrsim$ 93\%. We conclude that \df2 is a dark matter dominated galaxy.

This result is in stark contrast with the claim that the galaxy lacks dark matter, and the difference arises from two factors. First, while the measured distance (i.e. 13 Mpc vs 20 Mpc) makes a factor (20/13)$\sim$1.5 smaller the dynamical mass, the change on the stellar mass associated to the distance is larger 10$^{0.4\times(31.505-30.569)}$$\sim$2.37. Second,
our estimated stellar mass ranges from 5.0$^{+3.9}_{-2.5}\,10^7\Msun$ (surface brightness fitting), through $5.6\pm 1.3 \,10^7\Msun$ (GAMA relations) to $6.0\pm3.6 \,10^7\Msun$ (SED fitting), that is, another factor $\sim$1.5 smaller in the (M$_{\star}$/L) than the value adopted by \citet{2018Natur.555..629V}. The combination of these two factors yields a galaxy which is, far and large, dominated by dark matter. 

\section{Is [KKS2000]04 an anomalous galaxy?}
\label{sec:discussion}

In this paper we have concentrated our attention on the distance to the galaxy [KKS2000]04, leaving aside the ongoing discussion \citep[see e.g.][]{2018ApJ...859L...5M,2018MNRAS.tmp.2765L} about the reliability of the velocity dispersion of the compact sources of the system. Taking into account the velocity of the most discrepant GC-98 source, \citet{2018Natur.555..629V} claim the following 90\% confidence limit interval for the intrinsic velocity dispersion of the system\footnote{Very recently \citet{2018arXiv180604685V} readdressing the velocity of GC98 obtains $\sigma_{int}$=7.8$^{+5.2}_{-2.2}$ km s$^{-1}$.}: 8.8$<$$\sigma_{int}$$<$10.5 km s$^{-1}$. In the following discussion, we use this interval of velocity dispersion, together with our estimations of the stellar mass, dynamical mass and effective radius to compare the galaxy with other low mass galaxies. 

\subsection{[KKS2000]04 compared to other Local Group galaxies}

\begin{figure*}

\includegraphics[width=\textwidth]{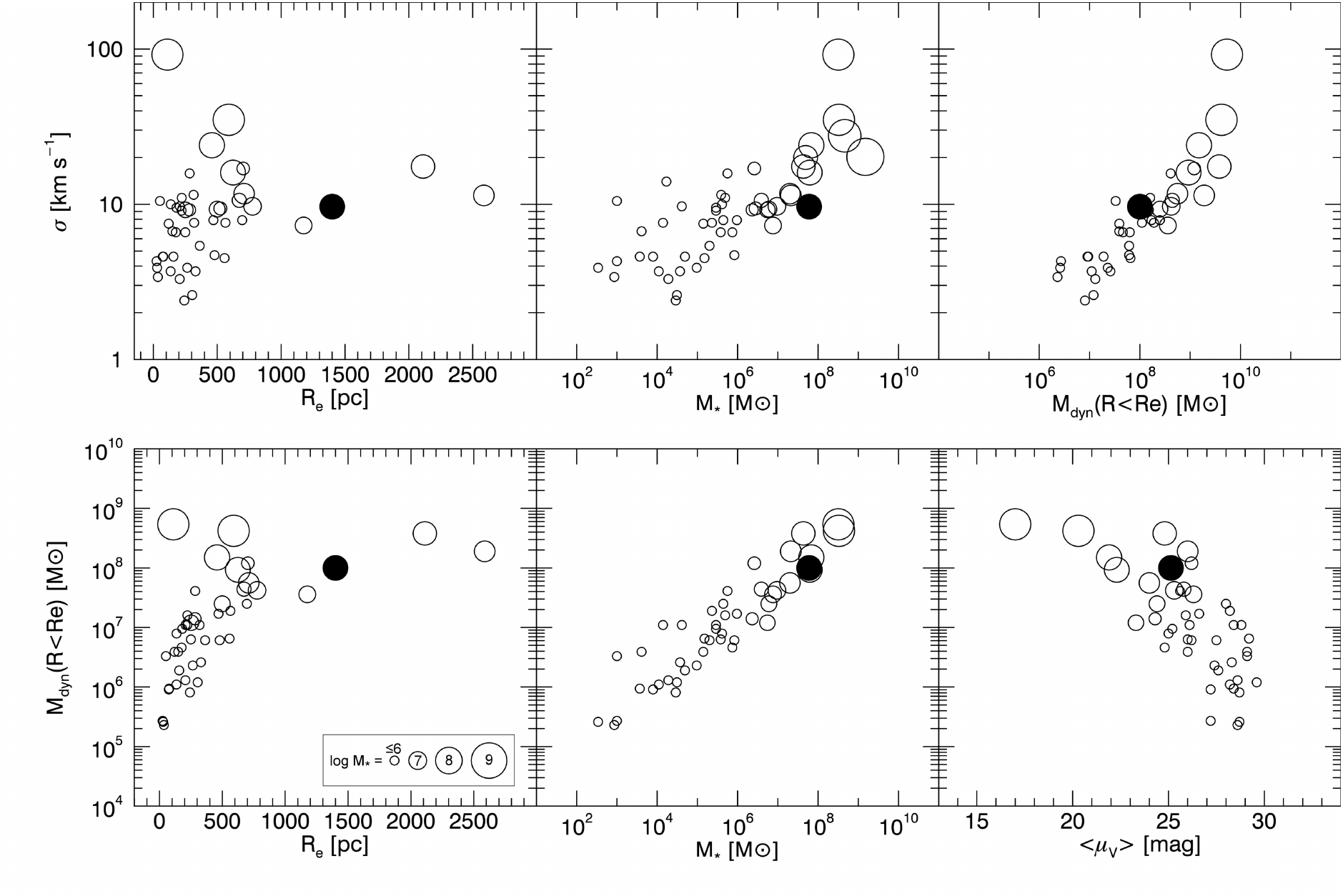}
  \caption{{\it Upper panels.} Velocity dispersion versus effective radius, stellar mass and dynamical mass within 1 R$_e$ for the Local Group galaxies \citep{2012AJ....144....4M}. {\it Lower panels.} Same but using the dynamical mass within 1 R$_e$. The figure illustrates the location of [KKS2000]04  (black filled circle) in comparison with the Local Group galaxies. At 13 Mpc distance, [KKS2000]04  properties follows the relations traced by dwarf galaxies in the Local Group.}
\label{fig:sigmare}
\end{figure*}

In Fig. \ref{fig:sigmare} we plot the velocity dispersion and the dynamical mass within 1 R$_e$ versus other properties of the object. The galaxy is shown together with other Local Group galaxies taken from the compilation by \citet{2012AJ....144....4M}. Fig. \ref{fig:sigmare} shows that [KKS2000]04 is not anomalous in relation to other low mass galaxies in the Local Group.

[KKS2000]04 presents some characteristics which are close to the isolated object IC1613 \citep{2014MNRAS.439.1015K} located in our Local Group at a distance of $\sim$760 kpc. This object has R$_e$$\sim$1 kpc, $\sigma_v$=10.8$\pm$1 km s$^{-1}$ and 10$^8$ L$_\odot$. Interestingly, this galaxy also has a similar amount of dark matter within the effective radius (M/L$_V$)$_{1/2}$=2.2$\pm$0.5 $\Upsilon$$_\odot$ (in the case of \df2 this value is 4.3$\pm$2.0 $\Upsilon$$_\odot$).

\subsection{Is [KKS2000]04 a tidal dwarf galaxy?}

The existence of galaxies without dark matter have been suggested previously in the literature. Tidal Dwarf Galaxies (TDGs) have little to no dark-matter \citep[e.g. ][]{2015A&A...584A.113L,2015MNRAS.447.2512P,2018MNRAS.474..580P}. These objects,  due to their surface brightness, could resemble closely the properties of [KKS2000]04.

The question then arises as to whether [KKS2000]04  could be a TDG. This interpretation seems unlikely. [KKS2000]04  does not fit in with the expected properties of TDGs. TDGs are formed from the tidal debris of gas-rich merging galaxies \citep[e.g. ][]{1992A&A...256L..19M,1998A&A...333..813D}.  Thus, they inherit many of their properties from the parent galaxies, namely, TDGs tend to be gas rich with relatively higher  metallicity for their low stellar mass \citep[$10^7$ -- $10^8\,M_\odot$, ][]{2012MNRAS.419...70K}, and have young stellar populations. They are born embedded in tidally stripped material, so, they often connect to the central galaxy through tidal tails. They are thought to be short lived although fossil TDGs are known with ages around 4 Gyr \citep{2014MNRAS.440.1458D, 2016ApJ...830...66K}. [KKS2000]04 does not fulfill these characteristics. It lacks neutral gas \citep[from the HIPASS][limit we estimate $M_{\rm HI}/M_\star < 1$\,\%]{2004MNRAS.350.1195M}\footnote{See a confirmation of this result by \citet{2019MNRAS.482L..99C} and \citet{2019arXiv190107586S}.}, and has low stellar metallicity consistent with its stellar mass and its age is compatible with being older than 4 Gyr. No obvious stellar stream connects [KKS2000]04 with the galaxies in the field, despite the fact the Gemini \textsl{g}-band image is deep enough to reveal tidal features  ($\sim$30 mag\,arcsec$^{-2}$; 3$\sigma$ in 10$\times$10 arcsec boxes). Last but not least, TDGs are not expected to exhibit significant GC systems. Thus, it is unlikely that [KKS2000]04 is a fossil TDG.  

Although [KKS2000]04 is not a TDG according to the properties of these objects described in the literature. One could argue that these properties are observationally biased towards the youngest brightest TDGs. Really old TDGs may be different. They may have lost their gaseous content through stellar feedback and interaction with the halo of the parent galaxy. Moreover, the primitive merger from which they were formed may have involved metal poor galaxies, producing TDGs of low metallicity. However, these hypothetical fossil TDGs, with characteristics closer to [KKS2000]04, are disfavoured by numerical simulations. Even if purely stellar systems  can survive for long in the halo of a host galaxy, their stars get stripped by tidal forces so that the final remnant conserves only a small fraction of the original mass. If 1\% of the initial mass survives after 7-9 Gyr \citep[e.g.,][]{1998ApJ...498..143K}, the progenitor of  [KKS2000]04 should have had a stellar mass  comparable to 10$^{10}$ M$_\odot$. This stellar mass is close to the putative host galaxy NGC1052  \citep[$\sim10^{11}\,M_\odot$, e.g.,][]{2011MNRAS.411L..21F}. This similarity in stellar mass makes [KKS2000]04 unlikely to be a debris from the merger that gave rise to NGC1052. This is because newborn TDGs are typically less than a 100 times fainter and less massive than the progenitor galaxy
\citep[e.g.][]{2000A&A...358..819W,2001AJ....121.2524M}. In other words, if [KKS2000]04 was created as a TDG 9 Gyr ago, it would have been an extraordinary massive TDG compared to the TDGs currently forming. Finally, the disruption of TDGs in relatively short time-scales explains why fossil TDGs do not appear in cosmological numerical simulations of galaxy formation \citep{2018MNRAS.474..580P}.

\section{Conclusions}
\label{sec:conclusions}

With the revised distance to \df2 of 13.0$\pm$0.4 Mpc, all the claimed anomalies of the galaxy and its system of globular clusters disappear. The revised distance comes from five different redshift-independent distance indicators: the TRGB, the luminosity of its GCs, the size of its GCs, the comparison of the luminosity function of its stars with the very similar galaxy DDO44 and the surface brightness fluctuations. \df2 is a rather ordinary low surface brightness galaxy (R$_e$=1.4$\pm$0.1 kpc, M$_\star$$\sim$6$\times$10$^7$M$_{\odot}$, M$_{\rm tot}$$\gtrsim$10$^9$ M$_{\odot}$ and M$_{halo}$/M$_\star$$>$20), surrounded by globular clusters whose properties are very similar to those found in other systems associated with dwarf galaxies. Moreover, with this new distance estimate, the ratio between the GC mass and the total galaxy mass is low enough to prevent, by dynamical friction, the orbital decay of the GCs in a short  period of time. A serious problem already pointed out by \citet{2018ApJ...863L..17N}.

%% Note that the \setcounter and \renewcommand are needed here because
%% this example is using a mix of deluxetable and tabular.  Here the
%% deluxetable counters are set with \tablenum but the situation is a bit
%% more complex for tabular.  Use the first command to set the Table number
%% to ONE LESS than it should be.  The next command will auto increment it
%% to the desired number.

%% Putting eqnarrays or equations inside the mathletters environment groups
%% the enclosed equations by letter. For instance, the eqnarray below, instead
%% of being numbered, say, (4) and (5), would be numbered (4a) and (4b).
%% LaTeX the paper and look at the output to see the results.

\section{acknowledgments}

We thank the referee, Nicolas Martin, for his constructive report. We thank Juan E. Betancort-Rijo for interesting comments on the redshift distribution of the galaxies in the large scale structure. We also thank Gabriella Raimondo and Michele Cantiello for useful advice on the Teramo/SPoT SBF database. Miguel Cervi\~no gave us interesting comments on the SBF method and Sebastian Hidalgo on the use of DOLPHOT.  We thank Fran\c{c}oise Combes for useful remarks on the $M/L$ ratio. I.T. acknowledges financial support from the European Union's Horizon 2020 research and innovation programme under Marie Sklodowska-Curie grant agreement No 721463 to the SUNDIAL ITN network. MAB acknowledges support from the Severo Ochoa Excellence program (SO FEYG 16).
This research has been partly supported by the Spanish Ministry of Economy and Competitiveness (MINECO) under grants AYA2016-77237-C3-1-P and AYA2016-79724-C4-2-P. A.DC. acknowledges financial support from a Marie-Sk\l{}odowska-Curie Individual Fellowship grant, H2020-MSCA-IF-2016 Grant agreement 748213 DIGESTIVO. M.E.F. gratefully acknowledges the financial support of the ''Funda\c c\~ao para a Ci\^encia e Tecnologia'' (FCT -- Portugal), through the research grant SFRH/BPD/107801/2015. 

Funding for the Sloan Digital Sky Survey (SDSS) has been provided by the Alfred P. Sloan Foundation, the Participating Institutions, the National Aeronautics and Space Administration, the National Science Foundation, the U.S. Department of Energy, the Japanese Monbukagakusho, and the Max Planck Society. The SDSS Web site is http://www.sdss.org/. The SDSS is managed by the Astrophysical Research Consortium (ARC) for the Participating Institutions. The Participating Institutions are The University of Chicago, Fermilab, the Institute for Advanced Study, the Japan Participation Group, The Johns Hopkins University, Los Alamos National Laboratory, the Max-Planck-Institute for Astronomy (MPIA), the Max-Planck-Institute for Astrophysics (MPA), New Mexico State University, University of Pittsburgh, Princeton University, the United States Naval Observatory, and the University of Washington. GALEX (Galaxy Evolution Explorer) is a NASA Small Explorer, launched in April 2003. We gratefully acknowledge NASA's support for construction, operation, and science analysis for the GALEX mission, developed in cooperation with the Centre National d'Etudes Spatiales (CNES) of France and the Korean Ministry of Science and Technology. This publication makes use of data products from the Wide-field Infrared Survey Explorer (WISE), which is a joint project of the University of California, Los Angeles, and the Jet Propulsion Laboratory/California Institute of Technology, funded by the National Aeronautics and Space Administration.

Based on observations obtained at the Gemini Observatory (acquired through the Gemini Observatory Archive), which is operated by the Association of Universities for Research in Astronomy, Inc., under a cooperative agreement with the NSF on behalf of the Gemini partnership: the National Science Foundation (United States), the National Research Council (Canada), CONICYT (Chile), Ministerio de Ciencia, Tecnolog\'ia e Innovaci\'on Productiva (Argentina), and Minist\'erio da Ci\^{e}ncia, Tecnologia e Inova\c{c}\~{a}o (Brazil).

This work was partly done using GNU Astronomy Utilities (Gnuastro) version 0.5. Gnuastro is a generic package for astronomical data manipulation and analysis which was initially created and developed for research funded by the Monbukagakusho (Japanese government) scholarship and ERC advanced grant 339659-MUSICOS.

\appendix

\section{The Fundamental Plane distance}

Since the work of \citet{1987ApJ...313...59D}  and \citet{1987ApJ...317....1D}  it has been known that pressure supported structures (i.e., ellipticals galaxies and bulges of spiral galaxies) occupy a well-defined plane determined by a relation between their global properties. The Fundamental Plane (FP), as it is known, is a direct consequence of the virial theorem, which relates the potential and kinetic energy of a galaxy in equilibrium \citep{1987gady.book.....B}. The FP can be expressed as a relation among the velocity dispersion $\sigma_{e}$, the mean surface brightness within the effective radius $<\mu_{e}>$ and the effective radius $R_{e}$

\begin{equation}
\log R_{e} = a \, \log \sigma_{e} + b <\mu_{e}> + c
\end{equation}

\noindent where $a$, $b$ and $c$ are the FP coefficients. $R_{e}$ is expressed in kpc, $\sigma_{e}$ in km s$^{-1}$ and $<\mu_{e}>$ in mag arcsec$^{-2}$. 

\begin{figure}
\includegraphics[width=\columnwidth]{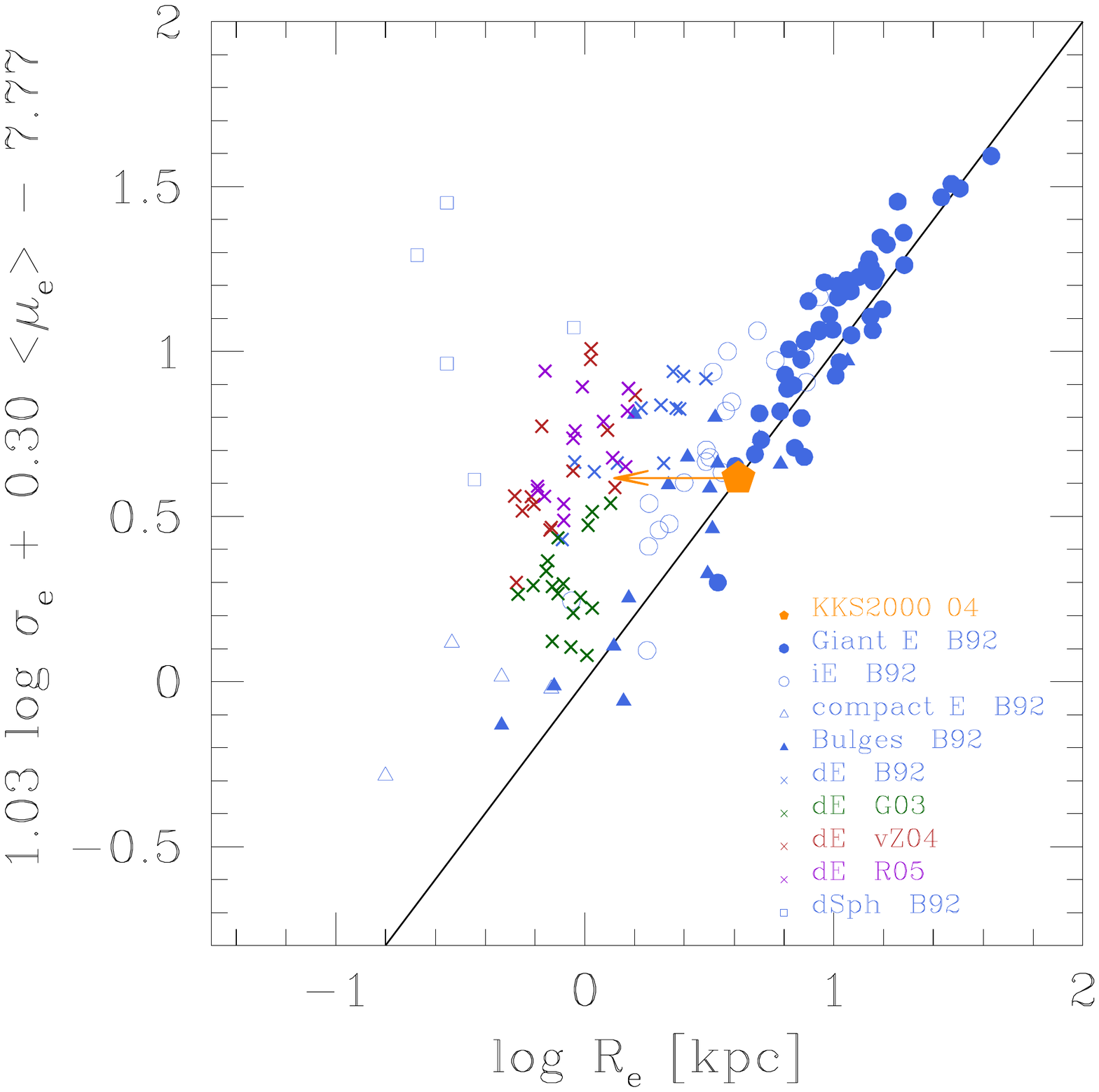}

\vspace{-2 cm}

\caption{An edge-on view of the FP projected onto the plane defined by $\log R_{e}$ and 1.03$\log\sigma_e$+0.30$<\mu_e>$-7.77. E = ellipticals, iE = intermediate luminosity ellipticals, dE = dwarf ellipticals, and dSph = dwarf spheroidals. B92 = \citet{1992ApJ...399..462B}, G03 = \citet{2003AJ....126.1794G} (Virgo cluster), vz04 = \citet{2004AJ....128..121V} (Virgo cluster) and R05 \citet{2005A&A...438..491D}.}
\label{fig:fundplane}
\end{figure}

\citet[][see also \citet{2003AJ....125.1866B}]{2013A&A...557A..21S} provide a compilation of literature FP coefficients and a calibration of the FP using 93000 elliptical galaxies from the SDSS DR8. For the $r$-band (their Table~1) the FP coefficients are $a$ = 1.034, $b$ = 0.3012 (assuming $\log I_{e}$ = $-<\mu_{e}>/2.5$; their Eq.~[17]) and $c$ = -7.77. The work of \citet[][ their Fig.~2 left]{2005A&A...438..491D} show that dwarf ellipticals (dE) and dwarf spheroidals (dSph), and presumably the case of [KKS2000]04, lie above the FP projection; for the same $\sigma_{e}$ and $<\mu_{e}>$, dEs and dSphs have smaller physical effective radii than bulges and bright and intermediate luminosity ellipticals (Fig. \ref{fig:fundplane}). Discrepancies can be larger than two orders of magnitudes in  physical effective radius for the more extreme dSph cases.

As $\sigma_{e}$ and $<\mu_{e}>$ are distance-independent quantities, the FP plane can be used to derive the physical effective radius, and, hence, the distance, when the observed effective radius is measured. In the case of [KKS2000]04, not being a massive elliptical, the FP will provide an upper limit to the physical effective radius, and, hence, an upper limit to the distance \citep{2005A&A...438..491D}. Adopting the \citet{2013A&A...557A..21S} $r$-band coefficients  \citep[which is the closest match for the V$_{606}$ data of ][]{2018Natur.555..629V}, assuming $<\mu_{e}>$ = 24.9 mag arcsec$^{-2}$ \citep[the central surface brightness in \textsl{V$_{606}$} is 24.4 mag arcsec$^{-2}$; ][]{2018Natur.555..629V}, adding the colour correction \textsl{r-F606W}=-0.18 (obtained from Table \ref{table:sed}), and using a representative intrinsic central velocity dispersion of $\sigma_{\rm e}$ = 8 km s$^{-1}$ \citep{2018Natur.555..629V} provides an effective radius upper limit of $R_{e}$ = 4.1 kpc (Fig.~\ref{fig:fundplane}; orange pentagon). Using the observed effective radius provided by \citet{2018Natur.555..629V}, $R_{e}$ = 22.6 arcsec, the previous R$_e$=4.1 kpc would imply a distance upper limit of 37.8 Mpc. For the same $\sigma_{e}$ and $<\mu_{e}>$ as [KKS2000]04, dEs (Fig. \ref{fig:fundplane}; blue open squares) show effective radii that are approximately one order of magnitude smaller than that predicted from the FP relation. Hence, if a conservative 0.5 dex decrease in effective radius is applied (i.e. assuming [KKS2000]04 is among the largest dEs or UDGs), this provides a distance of 12 Mpc for [KKS2000]04 (Fig. \ref{fig:fundplane}; orange arrow). Considering an uncertainty of 0.1 dex, this will imply the following distance range: 12$\pm$3 Mpc. So, although not very conclusive,  a mean FP distance of 12$\pm$3 Mpc for [KKS2000]04 also fits well with the FP properties of the galaxy assuming this object is an extended dE. As the FP distance estimate depends on the assumed decrease in the effective radius, we prefer not to use it on our final estimations of the distance to [KKS2000]04. Future work, including more and more UDG galaxies will be able to provide a much better constraint on the distance to this object based on the Fundamental Plane.

\vspace{1 cm}

% Don't change these lines
\bsp	% typesetting comment
\label{lastpage}
\end{document}